%% file: main.tex
\documentclass[acmtog, nonacm]{acmart}
\usepackage{amsthm}
\usepackage{graphicx}
\usepackage{enumitem}
\newtheorem{theorem}{Theorem}

\usepackage[final]{pdfpages}
\begin{document}


\author{Dong Xiao}
\email{xiaodong@ustc.edu.cn}
    \affiliation{%
  \department{School of Mathematical Sciences}
  \institution{University of Science and Technology of China}
  \country{China}
}
\author{Renjie Chen}
\authornote{Corresponding author}
\email{renjiec@ustc.edu.cn}
\affiliation{%
  \department{School of Mathematical Sciences}
  \institution{University of Science and Technology of China}
 \country{China}
}
\author{Bailin Deng}
\email{dengb3@cardiff.ac.uk}
\affiliation{%
  \department{School of Computational and Mathematical Sciences}
  \institution{Cardiff University}
  \country{UK}
}




\title{Anisotropic Green Coordinates}



\begin{abstract}
  We live in a world filled with anisotropy, a ubiquitous characteristic of both natural and engineered systems. In this study, we concentrate on space deformation and introduce \textbf{Anisotropic Green Coordinates (AGC)}, which provide versatile effects for cage-based and variational deformations in both two and three dimensions. The AGC are derived from the anisotropic Laplace equation $\nabla\cdot(\mathbf{A}\nabla u)=0$, where $\mathbf{A}$ is a symmetric positive definite (SPD) matrix. This equation belongs to the class of constant-coefficient second-order elliptic equations, exhibiting properties analogous to the Laplace equation but incorporating the matrix $\mathbf{A}$ to characterize anisotropic behavior. Based on this equation, we establish the boundary integral formulation, which is subsequently discretized to derive the deformation coordinates defined on the vertices and normals of oriented simplicial cages. Our method satisfies basic properties such as linear reproduction and translation invariance, and possesses closed-form expressions for both 2D and 3D scenarios. We also give an intuitive geometric interpretation of the approach, demonstrating that our method can generate a quasi-conformal mapping. We demonstrate both theoretically and empirically that the deformation effect is more pronounced when the normal of the cage face aligns with the eigenvector corresponding to the larger eigenvalue of $\mathbf{A}$. This indicates that anisotropy amplifies the deformation sensitivity along this direction, enabling more targeted cage design and matrix selection. Furthermore, we derive the gradients and Hessians of the deformation coordinates and employ the local-global optimization framework to facilitate variational shape deformation, enabling flexible shape manipulation while achieving as-rigid-as-possible (ARAP) shape deformation. Experimental results demonstrate that AGC offer versatile and diverse deformation options, providing artists with enhanced flexibility and introducing a novel perspective on spatial deformation. The source code is available at~\url{https://github.com/Submanifold/AGC}.
\end{abstract}



\maketitle

\input{sections/sec1}
\input{sections/sec2}
\input{sections/sec3}
\input{sections/sec4}
\input{sections/sec5}
\input{sections/sec6}
\input{sections/sec7}
\input{sections/sec8}

\bibliographystyle{ACM-Reference-Format}
\bibliography{sample-base}

\clearpage


\section*{Supplementary material}

# Supplementary Material of *Anisotropic Green Coordinates*

DONG XIAO, University of Science and Technology of China, China

RENJIE CHEN*, University of Science and Technology of China, China

BAILIN DENG, Cardiff University, UK

## A  Mathematical Details of Important Lemmas

In the main text, our derivation of the anisotropic Green coordinates relies on several lemmas. This section provides detailed derivations of these lemmas. Although many of these derivations are not mathematically advanced, they are non-trivial and may not be immediately obvious within the context of this research topic. Therefore, they are included here for clarity and completeness.

LEMMA A.1. *Let* $\mathbf{x}, \mathbf{y} \in \mathbb{R}^d$, *the matrix* $\mathbf{B} \in \mathbb{R}^{d \times d}$ *is an invertible matrix such that* $\mathbf{y} = \mathbf{B}\mathbf{x}$. *Then, for a* $C^2$ *scalar function* $f$ *and a* $C^2$ *vector-valued function* $F$, *we have* $\nabla_{\mathbf{x}} f = \mathbf{B}^\top \nabla_{\mathbf{y}} f$, $\mathbf{H}_{\mathbf{x}}(f) = \mathbf{B}^\top \mathbf{H}_{\mathbf{y}}(f)\mathbf{B}$ *and* $\nabla_{\mathbf{x}} \cdot \mathbf{F} = \nabla_{\mathbf{y}} \cdot (\mathbf{B}\mathbf{F})$. *Here,* $\nabla_{\mathbf{x}} f$ *and* $\mathbf{H}_{\mathbf{x}}(f)$ *denote the gradient and Hessian of* $f$ *with respect to* $\mathbf{x}$, *respectively, and* $\nabla_{\mathbf{x}} \cdot \mathbf{F}$ *is the divergence of* $\mathbf{F}$ *with respect to* $\mathbf{x}$.

PROOF. Since $\mathbf{y} = \mathbf{B}\mathbf{x}$, we have $y_j = \sum_{k=1}^{d} B_{jk} x_k$ and $\frac{\partial y_j}{\partial x_i} = B_{ji}$. Therefore,

$$\frac{\partial f}{\partial x_i} = \sum_{j=1}^{d} \frac{\partial f}{\partial y_j} \frac{\partial y_j}{\partial x_i} = \sum_{j=1}^{d} \frac{\partial f}{\partial y_j} B_{ji}, \tag{1}$$

which is equivalent to $\nabla_{\mathbf{x}} f = \mathbf{B}^\top \nabla_{\mathbf{y}} f$.

Calculate the second derivative, we have

$$\frac{\partial^2 f}{\partial x_k \partial x_i} = \frac{\partial^2 f}{\partial x_i \partial x_k} = \frac{\partial}{\partial x_i} \left( \sum_{j=1}^{d} \frac{\partial f}{\partial y_j} B_{jk} \right) = \sum_{j=1}^{d} B_{jk} \frac{\partial}{\partial x_i} \left( \frac{\partial f}{\partial y_j} \right) = \sum_{j=1}^{d} B_{jk} \sum_{l=1}^{d} \frac{\partial^2 f}{\partial y_j \partial y_l} \frac{\partial y_l}{\partial x_i} = \sum_{j,l=1}^{d} B_{jk} \frac{\partial^2 f}{\partial y_j \partial y_l} B_{li}. \tag{2}$$

This implies that $\mathbf{H}_{\mathbf{x}}(f) = \mathbf{B}^\top \mathbf{H}_{\mathbf{y}}(f)\mathbf{B}$.

Moreover, let $\mathbf{F} = (F_1, \dots, F_d)^\top$, we have

$$\nabla_{\mathbf{x}} \cdot \mathbf{F} = \sum_{i=1}^{d} \frac{\partial F_i}{\partial x_i} = \sum_{i=1}^{d} \sum_{j=1}^{d} \frac{\partial F_i}{\partial y_j} \frac{\partial y_j}{\partial x_i} = \sum_{i=1}^{d} \sum_{j=1}^{d} \frac{\partial F_i}{\partial y_j} B_{ji} = \sum_{j=1}^{d} \frac{\partial}{\partial y_j} \left( \sum_{i=1}^{d} B_{ji} F_i \right) = \nabla_{\mathbf{y}} \cdot (\mathbf{B}\mathbf{F}), \tag{3}$$

completing the proof. □

LEMMA A.2. *Let* $f \in \mathcal{D}'(\mathbb{R}^d)$ *be a scalar-valued distribution and* $\mathbf{F} = (F_1, \dots, F_d)^\top \in (\mathcal{D}'(\mathbb{R}^d))^d$ *be a vector-valued distribution. Consider* $f$ *and* $\mathbf{F}$ *as distributions in the variables* $\mathbf{x}$ *and* $\mathbf{y}$ *via the invertible linear change of variables* $\mathbf{y} = \mathbf{B}\mathbf{x}$. *Then,* $\nabla_{\mathbf{x}} f = \mathbf{B}^\top \nabla_{\mathbf{y}} f$ *and* $\nabla_{\mathbf{x}} \cdot \mathbf{F} = \nabla_{\mathbf{y}} \cdot (\mathbf{B}\mathbf{F})$, *where* $\nabla_{\mathbf{x}} f$ *denotes the gradient of* $f$ *with respect to* $\mathbf{x}$, *and* $\nabla_{\mathbf{x}} \cdot \mathbf{F}$ *is the divergence of* $\mathbf{F}$ *with respect to* $\mathbf{x}$. *All derivatives are interpreted in the distributional sense.*

---

*Corresponding author

Authors' Contact Information: Dong Xiao, xiaodong@ustc.edu.cn, School of Mathematical Sciences, University of Science and Technology of China, China; Renjie Chen, renjiec@ustc.edu.cn, School of Mathematical Sciences, University of Science and Technology of China, China; Bailin Deng, dengb3@cardiff.ac.uk, School of Computer Science and Informatics, Cardiff University, UK.

PROOF. Let $\varphi \in C_c^\infty(\mathbb{R}_x^d)$ be an arbitrary test function, where $C_c^\infty(\mathbb{R}_x^d)$ denotes the space of smooth functions with compact support in $\mathbb{R}^d$. For a distribution $f$, its partial derivative $\partial_{x_i} f$ is defined by

$$\langle \partial_{x_i} f, \varphi \rangle = -\langle f, \partial_{x_i} \varphi \rangle \text{ for all } \varphi \in C_c^\infty(\mathbb{R}_x^d). \tag{4}$$

Define $\psi(\mathbf{y}) = \varphi(\mathbf{B}^{-1}\mathbf{y}) \in C_c^\infty(\mathbb{R}_y^d)$. According to the distribution theory [Schwartz 1966], $\langle u, \varphi \rangle = |\det \mathbf{B}|^{-1}\langle u, \psi \rangle$ for any distribution $u \in \mathcal{D}'(\mathbb{R}^d)$. By Lemma A.1 and the smoothness of $\varphi$ and $\psi$, we have

$$\langle \partial_{x_i} f, \varphi \rangle = -\langle f, \partial_{x_i} \varphi \rangle = -|\det \mathbf{B}|^{-1}\langle f, (\partial_{x_i} \varphi) \circ \mathbf{B}^{-1} \rangle = -|\det \mathbf{B}|^{-1}\langle f, \partial_{x_i} \psi \rangle = -|\det \mathbf{B}|^{-1}\langle f, \sum_{j=1}^d B_{ji} \partial_{y_j} \psi \rangle$$

$$= -\sum_{j=1}^d B_{ji} |\det \mathbf{B}|^{-1}\langle f, \partial_{y_j} \psi \rangle = \sum_{j=1}^d B_{ji} |\det \mathbf{B}|^{-1}\langle \partial_{y_j} f, \psi \rangle = \sum_{j=1}^d B_{ji} \langle \partial_{y_j} f, \varphi \rangle. \tag{5}$$

which implies $\partial_{x_i} f = \sum_{j=1}^d B_{ji} \partial_{y_j} f$ in $\mathcal{D}'(\mathbb{R}^d)$ and hence $\nabla_{\mathbf{x}} f = \mathbf{B}^\top \nabla_{\mathbf{y}} f$.

$\nabla_{\mathbf{x}} \cdot \mathbf{F}$ is the divergence of the vector-valued distribution $\mathbf{F} = (F_1, F_2, ..., F_d)^\top$ if

$$\langle \nabla_{\mathbf{x}} \cdot \mathbf{F}, \varphi \rangle = -\sum_{i=1}^d \langle F_i, \partial_{x_i} \varphi \rangle \text{ for all } \varphi \in C_c^\infty(\mathbb{R}^d). \tag{6}$$

We then have

$$\langle \nabla_{\mathbf{x}} \cdot \mathbf{F}, \varphi \rangle = -\sum_{i=1}^d \langle F_i, \partial_{x_i} \varphi \rangle = -\sum_{i=1}^d |\det \mathbf{B}|^{-1}\langle F_i, (\partial_{x_i} \varphi) \circ \mathbf{B}^{-1} \rangle = -|\det \mathbf{B}|^{-1}\sum_{i=1}^d \langle F_i, \sum_{j=1}^d B_{ji} \partial_{y_j} \psi \rangle$$

$$= -|\det \mathbf{B}|^{-1}\sum_{j=1}^d \langle \sum_{i=1}^d B_{ji} F_i, \partial_{y_j} \psi \rangle = |\det \mathbf{B}|^{-1}\langle \nabla_{\mathbf{y}} \cdot (\mathbf{B}\mathbf{F}), \psi \rangle = \langle \nabla_{\mathbf{y}} \cdot (\mathbf{B}\mathbf{F}), \varphi \rangle. \tag{7}$$

Therefore, $\nabla_{\mathbf{x}} \cdot \mathbf{F} = \nabla_{\mathbf{y}} \cdot (\mathbf{B}\mathbf{F})$ in $\mathcal{D}'(\mathbb{R}^d)$, which completes the proof. □

LEMMA A.3. *Let* $\mathbf{x} \in \mathbb{R}^d$, $\mathbf{B} \in \mathbb{R}^{d \times d}$ *be an invertible matrix and* $\delta(\mathbf{x})$ *be the Dirac delta distribution satisfying* $\delta(\mathbf{x}) = 0$ *for* $\mathbf{x} \neq \mathbf{0}$ *and* $\int_{\mathbb{R}^d} \delta(\mathbf{x}) = 1$. *Then,*

$$\delta(\mathbf{B}\mathbf{x}) = \frac{1}{|\det \mathbf{B}|} \delta(\mathbf{x}). \tag{8}$$

PROOF. Let $\mathbf{y} = \mathbf{B}\mathbf{x}$, we have $d\mathbf{y} = |\det \mathbf{B}| \, d\mathbf{x}$. Moreover, let $\varphi \in C_c^\infty(\mathbb{R}^d)$ be an arbitrary smooth test function with compact support. Then,

$$\int_{\mathbb{R}^d} \delta(\mathbf{B}\mathbf{x})\varphi(\mathbf{x}) \, d\mathbf{x} = \int_{\mathbb{R}^d} \delta(\mathbf{y})\varphi(\mathbf{B}^{-1}\mathbf{y})\frac{1}{|\det \mathbf{B}|} \, d\mathbf{y} = \frac{1}{|\det \mathbf{B}|}\varphi(\mathbf{0}) = \frac{1}{|\det \mathbf{B}|}\int_{\mathbb{R}^d} \delta(\mathbf{x})\varphi(\mathbf{x}) \, d\mathbf{x}. \tag{9}$$

Therefore, for all test functions $\varphi$:

$$\int_{\mathbb{R}^d} \delta(\mathbf{B}\mathbf{x})\varphi(\mathbf{x}) \, d\mathbf{x} = \frac{1}{|\det \mathbf{B}|}\int_{\mathbb{R}^d} \delta(\mathbf{x})\varphi(\mathbf{x}) \, d\mathbf{x}. \tag{10}$$

This implies that $\delta(\mathbf{B}\mathbf{x}) = \frac{1}{|\det \mathbf{B}|}\delta(\mathbf{x})$. □

LEMMA A.4. *Let* $\mathbf{B}$ *be an invertible matrix, and let* $\mathbf{n}$ *be the normal vector of* $C$, *where* $C$ *is a non-degenerate 1-simplex in* $\mathbb{R}^2$ *or 2-simplex in* $\mathbb{R}^3$. *If* $C$ *is transformed by* $\mathbf{B}$, *i.e., all vertices* $\mathbf{v}_i$ *of* $C$ *are mapped to* $\mathbf{B}\mathbf{v}_i$. *Then, the normal vector with the same orientation of the transformed simplex* $C'$ *is given by* $\frac{\mathbf{B}^{-\top}\mathbf{n}}{||\mathbf{B}^{-\top}\mathbf{n}||}$.

Supplementary Material of *Anisotropic Green Coordinates* 3

PROOF. Let $\mathbf{v}_1, \mathbf{v}_2 \in \mathbb{R}^2$. By definition, $\mathbf{n} \cdot (\mathbf{v}_2 - \mathbf{v}_1) = 0$. Then,

$$(\mathbf{B}^{-\top}\mathbf{n}) \cdot (\mathbf{B}\mathbf{v}_2 - \mathbf{B}\mathbf{v}_1) = \mathbf{n}^\top \mathbf{B}^{-1}\mathbf{B}(\mathbf{v}_2 - \mathbf{v}_1) = \mathbf{n} \cdot (\mathbf{v}_2 - \mathbf{v}_1) = 0. \tag{11}$$

Therefore, $\frac{\mathbf{B}^{-\top}\mathbf{n}}{||\mathbf{B}^{-\top}\mathbf{n}||}$ is the normal vector of $C'$. Moreover, if $\mathbf{p}$ in $\mathbb{R}^2$ satisfies $\mathbf{n} \cdot (\mathbf{p} - \mathbf{v}_1) > 0$, we also have $(\mathbf{B}^{-\top}\mathbf{n}) \cdot (\mathbf{B}\mathbf{p} - \mathbf{B}\mathbf{v}_1) > 0$, which proves the orientation consistency.

Let $\mathbf{v}_1, \mathbf{v}_2, \mathbf{v}_3 \in \mathbb{R}^3$. By definition, $\mathbf{n} \cdot (\mathbf{v}_2 - \mathbf{v}_1) = 0$ and $\mathbf{n} \cdot (\mathbf{v}_3 - \mathbf{v}_1) = 0$. Similarly to the 2D case, we have $(\mathbf{B}^{-\top}\mathbf{n}) \cdot (\mathbf{B}\mathbf{v}_2 - \mathbf{B}\mathbf{v}_1) = 0$ and $(\mathbf{B}^{-\top}\mathbf{n}) \cdot (\mathbf{B}\mathbf{v}_3 - \mathbf{B}\mathbf{v}_1) = 0$. Therefore, $\frac{\mathbf{B}^{-\top}\mathbf{n}}{||\mathbf{B}^{-\top}\mathbf{n}||}$ is the normal vector of $C'$. Moreover, if $\mathbf{p}$ in $\mathbb{R}^3$ satisfies $\mathbf{n} \cdot (\mathbf{p} - \mathbf{v}_1) > 0$, we also have $(\mathbf{B}^{-\top}\mathbf{n}) \cdot (\mathbf{B}\mathbf{p} - \mathbf{B}\mathbf{v}_1) > 0$, which proves the orientation consistency. □

LEMMA A.5. *Let $C$ be a regular curve in $\mathbb{R}^2$ parameterized by $\mathbf{r}(u)$ with the length element $\mathrm{d}s(u) = \|\mathbf{r}_u\| \, \mathrm{d}u$. $C'$ is the curve obtained by applying an invertible linear transformation $\mathbf{B}$ to $C$, i.e., $C'$ is parameterized by $\mathbf{s}(u) = \mathbf{B}\mathbf{r}(u)$. Then the length element $\mathrm{d}s'(u)$ of $C'$ at $u$ is given by:*

$$\mathrm{d}s'(u) = \|\mathbf{B}\mathbf{t}(u)\| \, \mathrm{d}s(u), \tag{12}$$

*where $\mathbf{t}(u)$ is the unit tangent vector to $C$ at $\mathbf{r}(u)$. In particular, if $\mathbf{B}$ is symmetric, we also have*

$$\mathrm{d}s'(u) = |\det \mathbf{B}| \, \|\mathbf{B}^{-1}\mathbf{n}(u)\| \, \mathrm{d}s(u), \tag{13}$$

*where $\mathbf{n}(u)$ is the unit normal vector to $C$ at $\mathbf{r}(u)$.*

PROOF. The tangent vector of $C'$ is $\mathbf{s}_u = \mathbf{B}\mathbf{r}_u$. The length element

$$\mathrm{d}s'(u) = \|\mathbf{s}_u\| \, \mathrm{d}u = \|\mathbf{B}\mathbf{r}_u\| \, \mathrm{d}u = \|\mathbf{B}\mathbf{t}(u)\|\|\mathbf{r}_u\| \, \mathrm{d}u = \|\mathbf{B}\mathbf{t}(u)\| \, \mathrm{d}s(u). \tag{14}$$

where $\mathbf{t}(u)$ is the unit tangent vector to $C$ at $\mathbf{r}(u)$. This completes the proof of the first part.

Without loss of generality, we can choose an orthonormal coordinate system at $\mathbf{r}(u)$ such that $\mathbf{t}(u) = (1, 0)^\top$ and $\mathbf{n}(u) = (0, 1)^\top$. When $\mathbf{B}$ is symmetric, it can be expressed as $\mathbf{B} = \begin{pmatrix} a & b \\ b & c \end{pmatrix}$. We also have $\mathbf{B}^{-1} = \frac{1}{ac-b^2}\begin{pmatrix} c & -b \\ -b & a \end{pmatrix}$. Therefore, $\|\mathbf{B}\mathbf{t}(u)\| = \sqrt{a^2 + b^2}$. Given that $|\det \mathbf{B}| = |ac - b^2|$, we have

$$\|\mathbf{B}^{-1}\mathbf{n}(u)\| = \|\frac{1}{ac-b^2}(b, -a)\| = \frac{\sqrt{a^2+b^2}}{|\det \mathbf{B}|} = \frac{\|\mathbf{B}\mathbf{t}(u)\|}{|\det \mathbf{B}|}. \tag{15}$$

This completes the proof of the second part. □

LEMMA A.6. *Let $C$ be a regular surface in $\mathbb{R}^3$ parameterized by $\mathbf{r}(u, v)$ with the area element $\mathrm{d}s(u, v) = \|\mathbf{r}_u \times \mathbf{r}_v\| \, \mathrm{d}u\mathrm{d}v$. $C'$ is the surface obtained by applying an invertible linear transformation $\mathbf{B}$ to $C$, i.e., $C'$ is parameterized by $\mathbf{s}(u, v) = \mathbf{B}\mathbf{r}(u, v)$. Then the area element $\mathrm{d}s'(u, v)$ of $C'$ at $(u, v)$ is given by:*

$$\mathrm{d}s'(u, v) = |\det \mathbf{B}| \, \|\mathbf{B}^{-\top}\mathbf{n}(u, v)\| \, \mathrm{d}s(u, v), \tag{16}$$

*where $\mathbf{n}(u, v)$ is the unit normal vector to $C$ at $\mathbf{r}(u, v)$.*

PROOF. The tangent vectors of $C'$ are $\mathbf{s}_u = \mathbf{B}\mathbf{r}_u$ and $\mathbf{s}_v = \mathbf{B}\mathbf{r}_v$. The area element on $C'$ is given by $\mathrm{d}s'(u, v) = \|\mathbf{s}_u \times \mathbf{s}_v\| \, \mathrm{d}u\mathrm{d}v$. We first show that $(\mathbf{B}\mathbf{a}) \times (\mathbf{B}\mathbf{b}) = (\det \mathbf{B})\mathbf{B}^{-\top}(\mathbf{a} \times \mathbf{b})$ for arbitrary vectors $\mathbf{a}$ and $\mathbf{b}$ in $\mathbb{R}^3$. Consider another vector $\mathbf{c} \in \mathbb{R}^3$ and the triple scalar product:

$$\mathbf{c} \cdot [(\mathbf{B}\mathbf{a}) \times (\mathbf{B}\mathbf{b})] = \det(\mathbf{B}\mathbf{a}, \mathbf{B}\mathbf{b}, \mathbf{c}) = \det(\mathbf{B})\det(\mathbf{a}, \mathbf{b}, \mathbf{B}^{-1}\mathbf{c})$$
$$= \det(\mathbf{B})(\mathbf{a} \times \mathbf{b}) \cdot (\mathbf{B}^{-1}\mathbf{c}) = \det(\mathbf{B})(\mathbf{c}^\top \mathbf{B}^{-\top}(\mathbf{a} \times \mathbf{b})) = \det(\mathbf{B})(\mathbf{B}^{-\top}(\mathbf{a} \times \mathbf{b})) \cdot \mathbf{c}. \tag{17}$$

Since $\mathbf{c}$ is also arbitrary, we conclude that

$$(\mathbf{B}\mathbf{a}) \times (\mathbf{B}\mathbf{b}) = (\det \mathbf{B})\mathbf{B}^{-\top}(\mathbf{a} \times \mathbf{b}). \tag{18}$$

Therefore,

$$\mathbf{s}_u \times \mathbf{s}_v = (\mathbf{B}\mathbf{r}_u) \times (\mathbf{B}\mathbf{r}_v) = (\det \mathbf{B})\mathbf{B}^{-\top}(\mathbf{r}_u \times \mathbf{r}_v). \tag{19}$$

Taking the norms, we obtain

$$\begin{aligned}
\mathrm{d}s'(u,v) = \|\mathbf{s}_u \times \mathbf{s}_v\| \, \mathrm{d}u\mathrm{d}v &= |\det \mathbf{B}| \|\mathbf{B}^{-\top}(\mathbf{r}_u \times \mathbf{r}_v)\| \, \mathrm{d}u\mathrm{d}v \\
&= |\det \mathbf{B}| \|\mathbf{B}^{-\top}\mathbf{n}(u,v)\| \, \|\mathbf{r}_u \times \mathbf{r}_v\| \, \mathrm{d}u\mathrm{d}v = |\det \mathbf{B}| \|\mathbf{B}^{-\top}\mathbf{n}(u,v)\| \, \mathrm{d}s(u,v).
\end{aligned} \tag{20}$$

The third equality holds because $\mathbf{n}(u,v) = \frac{\mathbf{r}_u \times \mathbf{r}_v}{\|\mathbf{r}_u \times \mathbf{r}_v\|}$. This completes the proof. □

LEMMA A.7. *Let $v_0, v_1, \ldots, v_k$ be affinely independent points in $\mathbb{R}^d$ with $d \geq k$, forming a non-degenerate $k$-simplex $C$. Then, the barycentric coordinates of any point $p$ in $C$ are unique with respect to $v_0, v_1, \ldots, v_k$.*

PROOF. Suppose there exist two sets of barycentric coordinates $(\lambda_0, \lambda_1, \ldots, \lambda_k)$ and $(\mu_0, \mu_1, \ldots, \mu_k)$ such that:

$$p = \sum_{i=0}^{k} \lambda_i v_i = \sum_{i=0}^{k} \mu_i v_i, \tag{21}$$

where $\lambda_i, \mu_i \geq 0$ and $\sum_{i=0}^{k} \lambda_i = \sum_{i=0}^{k} \mu_i = 1$. Subtracting the two representations gives

$$\sum_{i=0}^{k} (\lambda_i - \mu_i)v_i = 0. \tag{22}$$

Define $\delta_i = \lambda_i - \mu_i$, it follows that:

$$\sum_{i=0}^{k} \delta_i v_i = 0 \quad \text{and} \quad \sum_{i=0}^{k} \delta_i = 0. \tag{23}$$

Using $v_0$ as a reference point, we have

$$\sum_{i=0}^{k} \delta_i v_i = \sum_{i=0}^{k} \delta_i(v_i - v_0) + v_0 \sum_{i=0}^{k} \delta_i. \tag{24}$$

According to Eq. (23) and (24), we obtain $\sum_{i=1}^{k} \delta_i(v_i - v_0) = 0$. By the affine independence of the vertices, the vectors $v_1 - v_0, \ldots, v_k - v_0$ are linearly independent. Hence, $\delta_i = 0$ for all $i = 1, \ldots, k$. Since $\sum_{i=0}^{k} \delta_i = 0$, we have $\delta_0 = 0$ as well. Therefore, $\lambda_i = \mu_i$ for all $i = 0, 1, \ldots, k$. □

## B  Closed-Form Expressions of Isotropic Green Coordinates and Their Derivatives

In this section, we present detailed derivations of closed-form expressions for the classical isotropic Green coordinates, including their gradients and Hessians, for both 2D and 3D oriented simplicial cages. These formulations form the foundation for the development of anisotropic techniques. Although these results have appeared fragmentarily in previous literature [Ben-Chen et al. 2009; Lipman and Levin 2009; Lipman et al. 2008; Michel et al. 2025; Urago 2000], to our knowledge, a systematic derivation and summary is still lacking. In particular, many publications in this field do not provide thorough derivations for the gradient and Hessian in the 3D scenario.

Mathematically, we denote the cage (2D or 3D) as $P = (\mathbb{V}, \mathbb{T})$, where $\mathbb{V}$ denote the collection of the vertices and $\mathbb{T}$ represent the collections of the face normals. Isotropic Green coordinates reproduce arbitrary point $\eta$ within the cage

using the cage vertices $\{\mathbf{v}_i | i \in I_V\}$ and face normals $\{\mathbf{n}_j | j \in I_T\}$ as follows:

$$\eta = F(\eta, P) = \sum_{i \in I_V} \phi_i(\eta)\mathbf{v}_i + \sum_{j \in I_T} \psi_j(\eta)\mathbf{n}_j, \ \ \eta \in \Omega, \tag{25}$$

where

$$\phi_i(\eta) = \int_{\xi \in N\{\mathbf{v}_i\}} \Gamma_i(\xi) \frac{\partial G}{\partial \mathbf{n}}(\xi, \eta) \, d\sigma_\xi, \, i \in I_V, \tag{26}$$

$$\psi_j(\eta) = -\int_{\xi \in t_j} G(\xi, \eta) \, d\sigma_\xi, \, j \in I_T. \tag{27}$$

Here, $G(\xi, \eta)$ is the fundamental solution of the Laplacian equation, and $\Gamma_i(\xi)$ is the hat-function, which equals one at vertex $\mathbf{v}_i$, zero at other vertices, and varies linearly over the one-ring neighborhood of $\mathbf{v}_i$. Our goal is to derive closed-form expressions for $\phi_i(\eta), \psi_j(\eta)$, along with their gradients $\nabla_\eta \phi_i(\eta), \nabla_\eta \psi_j(\eta)$ and Hessians $\mathbf{H}_\eta \phi_i(\eta), \mathbf{H}_\eta \psi_j(\eta)$. For brevity and clarity, we will omit the explicit dependence on $\eta$ in the gradient and Hessian notations when no ambiguity arises. In the following, we primarily adopt column vector notation, meaning that row vectors will be denoted with the transpose symbol $\top$.

### B.1 Isotropic Green coordinates for 2D

We begin by deriving the formulation for 2D, which mainly use the methods established in [Lipman and Levin 2009] and [Michel et al. 2025]. Fig. 1 provides an illustration of a 2D cage $P$ along with its mathematical notations. We denote

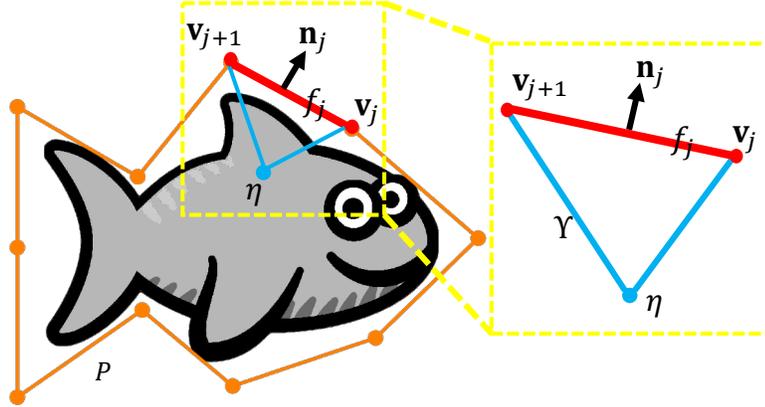

Fig. 1. Illustration of a 2D cage and its mathematical notations.

$f_j = \overrightarrow{\mathbf{v}_j \mathbf{v}_{j+1}}$ as one face of $P$. We first focus on computing $\psi_j(\eta)$, which is defined as the integral of $-G(\xi, \eta)$ over $f_j$. We define $\mathbf{a}_j = \mathbf{v}_{j+1} - \mathbf{v}_j$ and $\mathbf{b}_j = \mathbf{v}_j - \eta$. Consider the parametric curve $\gamma(t) = \mathbf{v}_j + t\mathbf{a}_j$, where $t \in [0, 1]$, we have:

$$\begin{aligned}
\psi_j(\eta) &= -\int_{t_j} G(\xi, \eta) \, d\sigma_\xi = -\frac{1}{2\pi} \|\mathbf{a}_j\| \int_{t=0}^{1} \log \|\mathbf{b}_j + t\mathbf{a}_j\| \, dt \\
&= -\frac{\|\mathbf{a}_j\|}{4\pi} \int_{t=0}^{1} \log(\|\mathbf{a}_j\|^2 t^2 + 2(\mathbf{a}_j \cdot \mathbf{b}_j)t + \|\mathbf{b}_j\|^2) \, dt \\
&= -\frac{\|\mathbf{a}_j\|}{4\pi} (Q(1) - Q(0)),
\end{aligned} \tag{28}$$

where $Q(t)$ is the antiderivative of $\log(\|\mathbf{a}_j\|^2 t^2 + 2(\mathbf{a}_j \cdot \mathbf{b}_j)t + \|\mathbf{b}_j\|^2)$, which has the following expression [Lipman and Levin 2009]:

$$Q(t) = \left(t + \frac{\mathbf{a}_j \cdot \mathbf{b}_j}{\|\mathbf{a}_j\|^2}\right) \log\left(\|\mathbf{a}_j\|^2 t^2 + 2(\mathbf{a}_j \cdot \mathbf{b}_j)t + \|\mathbf{b}_j\|^2\right)$$
$$- \left(\frac{\|\mathbf{a}_j\|^2 + (\mathbf{a}_j \cdot \mathbf{b}_j)}{\|\mathbf{a}_j\|^2 \sqrt{\|\mathbf{a}_j\|^2 \|\mathbf{b}_j\|^2 - (\mathbf{a}_j \cdot \mathbf{b}_j)^2}}\right) \arctan\left(\frac{\|\mathbf{a}_j\|^2 t + (\mathbf{a}_j \cdot \mathbf{b}_j)}{\sqrt{\|\mathbf{a}_j\|^2 \|\mathbf{b}_j\|^2 - (\mathbf{a}_j \cdot \mathbf{b}_j)^2}}\right). \tag{29}$$

We now focus on the Dirichlet term $\phi_{\mathbf{v}_j}(\eta)$, which is defined on cage vertices $\mathbf{v}_j$. We denote $\phi_{\mathbf{v}_j, f_j}(\eta)$ as the contribution of face $f_j$ to $\phi_{\mathbf{v}_j}(\eta)$. The value of $\phi_{\mathbf{v}_j}(\eta)$ is obtained by summing the contributions from all faces incident to $\mathbf{v}_j$. Recall that the parametric equation of $f_j$ is $\gamma(t) = \mathbf{v}_j + t\mathbf{a}_j$, where $t \in [0, 1]$. Therefore,

$$\phi_{\mathbf{v}_j, f_j}(\eta) = \int_0^1 (1-t) \left(\frac{\mathbf{a}_j t + \mathbf{b}_j}{2\pi \|\mathbf{a}_j t + \mathbf{b}_j\|^2} \cdot \mathbf{n}_j\right) \|\mathbf{a}_j\| \, dt$$
$$= \frac{\|\mathbf{a}_j\|(\mathbf{b}_j \cdot \mathbf{n}_j)}{2\pi} \int_0^1 \frac{(1-t) \, dt}{\|\mathbf{a}_j\|^2 t^2 + 2t(\mathbf{a}_j \cdot \mathbf{b}_j) + \|\mathbf{b}_j\|^2}$$
$$= -\frac{\|\mathbf{a}_j\|(\mathbf{b}_j \cdot \mathbf{n}_j)}{2\pi}(W(1) - W(0)), \tag{30}$$

where $W(t)$ is the antiderivative of $\frac{t-1}{\|\mathbf{a}_j\|^2 t^2 + 2t(\mathbf{a}_j \cdot \mathbf{b}_j) + \|\mathbf{b}_j\|^2}$, which is given by

$$W(t) = \frac{1}{2\|\mathbf{a}_j\|^2} \log(\|\mathbf{a}_j\|^2 t^2 + 2t(\mathbf{a}_j \cdot \mathbf{b}_j) + \|\mathbf{b}_j\|^2) - \left(\frac{\|\mathbf{a}_j\|^2 + (\mathbf{a}_j \cdot \mathbf{b}_j)}{\|\mathbf{a}_j\|^2 \sqrt{\|\mathbf{a}_j\|^2 \|\mathbf{b}_j\|^2 - (\mathbf{a}_j \cdot \mathbf{b}_j)^2}}\right) \arctan\left(\frac{\|\mathbf{a}_j\|^2 t + (\mathbf{a}_j \cdot \mathbf{b}_j)}{\sqrt{\|\mathbf{a}_j\|^2 \|\mathbf{b}_j\|^2 - (\mathbf{a}_j \cdot \mathbf{b}_j)^2}}\right). \tag{31}$$

We now proceed to compute the derivatives of $\phi_{\mathbf{v}_j}(\eta)$ and $\psi_j(\eta)$. The core idea involves expressing these values in terms of integrals of rational functions with even-powered denominators with respect to the parameter $t$. The expressions are then subjected to partial fraction decomposition, and the integrals are evaluated using recurrence relations [Michel et al. 2025]. We denote $\gamma_\eta(t) = \gamma(t) - \eta$, where $\gamma(t) = \mathbf{v}_j + t\mathbf{a}_j$. Then we have

$$\nabla_\eta \phi_{\mathbf{v}_j, f_j}(\eta) = -\frac{\|\mathbf{a}_j\|}{2\pi} \int_{t=0}^1 (1-t) \left[\frac{\mathbf{n}_j}{\|\gamma_\eta(t)\|^2} - 2\frac{(\gamma_\eta(t) \cdot \mathbf{n}_j) \, \gamma_\eta(t)}{\|\gamma_\eta(t)\|^4}\right] dt, \tag{32}$$

$$\mathbf{H}_\eta \phi_{\mathbf{v}_j, f_j}(\eta) = -\frac{\|\mathbf{a}_j\|}{2\pi} \int_{t=0}^1 (1-t) \left[2\frac{\mathbf{n}_j \gamma_\eta(t)^\top + \gamma_\eta(t)\mathbf{n}_j^\top}{\|\gamma_\eta(t)\|^4} + 2\frac{(\gamma_\eta(t) \cdot \mathbf{n}_j) \, \mathbf{I}}{\|\gamma_\eta(t)\|^4} - 8\frac{(\gamma_\eta(t) \cdot \mathbf{n}_j) \, \gamma_\eta(t)\gamma_\eta(t)^\top}{\|\gamma_\eta(t)\|^6}\right] dt, \tag{33}$$

$$\nabla_\eta \psi_j(\eta) = \frac{\|\mathbf{a}_j\|}{2\pi} \int_{t=0}^1 \frac{\gamma_\eta(t)}{\|\gamma_\eta(t)\|^2} \, dt, \tag{34}$$

$$\mathbf{H}_\eta \psi_j(\eta) = -\frac{\|\mathbf{a}_j\|}{2\pi} \int_{t=0}^1 \frac{\|\gamma_\eta(t)\|^2 \, \mathbf{I} - 2\gamma_\eta(t)\gamma_\eta(t)^\top}{\|\gamma_\eta(t)\|^4} \, dt. \tag{35}$$

Therefore, the remaining work is to compute integrals of the following form $F_{K,n}^c = \int_0^1 \frac{t^n \, dt}{P_\gamma(t)^K}$, where $P_\gamma(t) = \|\gamma(t) - \eta\|^2$ is a polynomial of degree 2. The discriminant of $P_\gamma(t) = 0$ is non-positive, and it equals zero only when the cage is concave and $\eta$ lies on the line containing this edge. In such cases, further decomposition is unnecessary. Otherwise, $P_\gamma(t) = 0$ have two complex roots satisfying $\omega_1 = \omega_2^*$, and $P_\gamma(t)$ can be factored as

$$P_\gamma(t) = A(t - \omega_1)(t - \omega_2). \tag{36}$$

Then we obtain

$$F_{K,n}^c = \frac{1}{A^K} \int_0^1 \frac{t^n}{(t - \omega_1)^K (t - \omega_2)^K} \, dt. \tag{37}$$

We can denote $Q = (t - \omega_1)^K (t - \omega_2)^K$. Then,

$$\frac{1}{Q} = \sum_{k=1}^{K} \frac{\alpha_k}{(t - \omega_1)^k} + \sum_{k=1}^{K} \frac{\beta_k}{(t - \omega_2)^k}, \text{ where}$$

$$\alpha_k = \frac{1}{(K-k)!} \frac{\mathrm{d}^{K-k}}{\mathrm{d}t^{K-k}} \left( \frac{1}{(t - \omega_2)^K} \right) \bigg|_{x = \omega_1} = (-1)^{K-k} \frac{(2K - k - 1)!}{(K-1)!(K-k)!} (\omega_1 - \omega_2)^{-(2K-k)},$$

$$\beta_k = \frac{1}{(K-k)!} \frac{\mathrm{d}^{K-k}}{\mathrm{d}t^{K-k}} \left( \frac{1}{(t - \omega_1)^K} \right) \bigg|_{x = \omega_2} = (-1)^{K-k} \frac{(2K - k - 1)!}{(K-1)!(K-k)!} (\omega_2 - \omega_1)^{-(2K-k)}.$$

(38)

The remaining terms for calculating $F_{K,n}^c$ are all of the form $G_{a,b}(\omega) = \int_0^1 t^a (t - \omega)^b \, \mathrm{d}t$, which can be computed using the following recursive formulation $G_{a,b}(\omega)$:

If $b = -1$, we have

$$G_{a,-1}(\omega) = \omega^a D(\omega) + U_a(\omega),$$

(39)

where:

$$D(\omega) = \log(1 - \omega) - \log(-\omega), \quad U_a(\omega) = \sum_{k=0}^{a-1} \frac{\omega^k}{a - k}.$$

(40)

If $a = 0$,

$$G_{0,b}(\omega) = \frac{1}{b+1} \left[ (1 - \omega)^{b+1} - (-\omega)^{b+1} \right].$$

(41)

If $a > 0$ and $b \neq -1$, the recurrence relation is given by:

$$G_{a,b}(\omega) = \frac{1}{b+1} \left[ (1 - \omega)^{b+1} - a G_{a-1,b+1}(\omega) \right].$$

(42)

## B.2 Isotropic Green coordinates for 3D

We now proceed to derive the isotropic Green's coordinates and their derivatives in three dimensions. This part is a central focus of this section, as to our knowledge, previous studies in the field [Ben-Chen et al. 2009], including foundational works such as [Urago 2000], have primarily provided final results while offering relatively limited elaboration on the derivation process. Consider the tetrahedron depicted in Fig. 2, which is formed by a query point $\eta$ and a triangle $t_j = \triangle \mathbf{v}_{j_1} \mathbf{v}_{j_2} \mathbf{v}_{j_3}$ of the 3D cage, along with other relevant mathematical notations. We compute the Neumann term $\psi_j(\eta)$ associated with the face $t_j$, as well as the Dirichlet term $\phi_{\mathbf{v}_{j_1}}(\eta)$ associate with the vertex $\mathbf{v}_{j_1}$, together with their gradients and Hessians. We also use $\phi_{\mathbf{v}_{j_1}, t_j}(\eta)$ to denote the contribution of the face $t_j$ to $\phi_{\mathbf{v}_{j_1}}(\eta)$. In most cases, we focus on the computation of $\phi_{\mathbf{v}_{j_1}, t_j}(\eta)$. However, some terms cancel out when summing over the one-ring neighbors of $\mathbf{v}_{j_1}$ when computing $\phi_{\mathbf{v}_{j_1}}(\eta)$. Such instances will be explicitly indicated.

To ensure clarity in subsequent expressions, we adopt the following notation. The tetrahedron formed by $t_j$ and $\eta$ is denoted as $\Upsilon$. The ourward normal of $t_j$ is denoted as $\mathbf{n}_t$. Additionally, we define $\mathbf{e}_1 = \mathbf{v}_{j_1} - \eta$, $\mathbf{e}_2 = \mathbf{v}_{j_2} - \eta$, $\mathbf{e}_3 = \mathbf{v}_{j_3} - \eta$, and their corresponding unit vectors as $\bar{\mathbf{e}}_1 = \mathbf{e}_1 / \|\mathbf{e}_1\|$, $\bar{\mathbf{e}}_2 = \mathbf{e}_2 / \|\mathbf{e}_2\|$ and $\bar{\mathbf{e}}_3 = \mathbf{e}_3 / \|\mathbf{e}_3\|$. We also define $R_1 = \|\mathbf{e}_2\| + \|\mathbf{e}_3\|$, $R_2 = \|\mathbf{e}_3\| + \|\mathbf{e}_1\|$, $R_3 = \|\mathbf{e}_1\| + \|\mathbf{e}_2\|$, $\mathbf{d}_1 = \mathbf{v}_{j_2} - \mathbf{v}_{j_3}$, $\mathbf{d}_2 = \mathbf{v}_{j_3} - \mathbf{v}_{j_1}$, $\mathbf{d}_3 = \mathbf{v}_{j_1} - \mathbf{v}_{j_2}$, $\mathbf{Z}_1 = -\mathbf{e}_2 \times \mathbf{e}_3$, $\mathbf{Z}_2 = -\mathbf{e}_3 \times \mathbf{e}_1$ and $\mathbf{Z}_3 = -\mathbf{e}_1 \times \mathbf{e}_2$. Additionally, $\mathbf{z}_i = \mathbf{Z}_i / \|\mathbf{Z}_i\|$ denotes the unit outward normal vector of the side faces $S_i$ for $i = 1, 2, 3$. In subsequent computations, the following quantities will also be frequently used:

$$C_i = \frac{1}{4\pi \|\mathbf{d}_i\|} \log \left( \frac{R_i + \|\mathbf{d}_i\|}{R_i - \|\mathbf{d}_i\|} \right), \quad \overline{C}_i = \frac{1}{2\pi (R_i + \|\mathbf{d}_i\|)(R_i - \|\mathbf{d}_i\|)}, \quad i = 1, 2, 3,$$

(43)

$$P_t = \mathbf{n}_t \times \sum_{i=1}^{3} C_i \mathbf{d}_i - \omega_t \mathbf{n}_t,$$

(44)

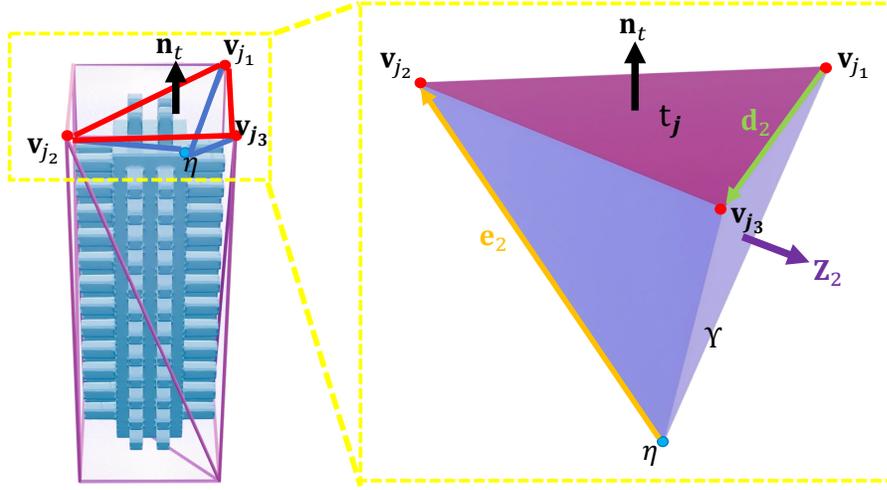

Fig. 2. Illustration of a tetrahedron formed by a query point $\eta$ and a triangle $t_j = \triangle \mathbf{v}_{j_1} \mathbf{v}_{j_2} \mathbf{v}_{j_3}$ of the 3D cage, along with other mathematical notations for important variables.

where $4\pi\omega_t$ is the signed solid angle of $t_j$ toward $\eta$. We also define $A_t$ as the area of $t_j$, and $vol_t$ as the signed volume of $\Upsilon$. Moreover, $H_t$ is the signed height of $\eta$ to $t_j$. In our formulation, when $\eta$ is in the direction opposite to $\mathbf{n}_j$, as shown in Fig. 2, we define $\omega_t$, $H_t$ and $vol_t$ as positive.

**1. Derivation of $\psi_j(\eta)$.** We first focus on the calculation of $\psi_j(\eta)$, which is the integral of $-G(\xi, \eta)$ over $t_j$ as

$$\psi_j(\eta) = -\int_{t_j} G(\xi, \eta) \, d\sigma_\xi, \tag{45}$$

we first begin with the divergence theorem of the tetrahedron $\Upsilon$ as

$$\int_\Upsilon \nabla_\xi \cdot (G(\xi, \eta)\mathbf{n}_j) \, dV_\xi = \int_{t_j + S_1 + S_2 + S_3} G(\xi, \eta) \, (\mathbf{n}_t \cdot \mathbf{n}(\xi)) \, d\sigma_\xi$$

$$= \int_{t_j} G(\xi, \eta) \, d\sigma_\xi + (\mathbf{n}_t \cdot \mathbf{z}_1) \int_{S_1} G(\xi, \eta) \, d\sigma_\xi + (\mathbf{n}_t \cdot \mathbf{z}_2) \int_{S_2} G(\xi, \eta) \, d\sigma_\xi + (\mathbf{n}_t \cdot \mathbf{z}_3) \int_{S_3} G(\xi, \eta) \, d\sigma_\xi. \tag{46}$$

The left-hand side of the above equation can be calculated as:

$$\int_\Upsilon \nabla_\xi \cdot (G(\xi, \eta)\mathbf{n}_t) \, dV_\xi = \int_\Upsilon \nabla_\xi G(\xi, \eta) \cdot \mathbf{n}_t \, dV_\xi = \int_{h=0}^{H_t} \left( \int_{S(h)} \nabla_\xi G(\xi, \eta) \cdot \mathbf{n}_t \, dS \right) dh$$

$$= \omega_t \int_{h=0}^{H_t} dh = \omega_t H_t = \frac{3\omega_t \, vol_t}{A_t}, \tag{47}$$

Then we consider the integration of $G(\xi, \eta)$ over the side face $\triangle \eta \mathbf{v}_{j_1} \mathbf{v}_{j_2}$, which is equivalent to the integration over $\triangle \mathbf{O} \mathbf{v}'_{j_1} \mathbf{v}'_{j_2}$, where $\mathbf{O}$ is the origin, $\mathbf{v}'_{j_1} = \mathbf{v}_{j_1} - \eta$ and $\mathbf{v}'_{j_2} = \mathbf{v}_{j_2} - \eta$. In Fig. 3, the region of integration is represented using polar coordinates $R(\theta)$. We also use $\alpha$, $\beta$ and $\theta$ to denote key angular variables. According to the law of sines, we have

$$R(\theta) = \frac{\|\mathbf{e}_1\| \sin(\alpha)}{\sin(\pi - \alpha - \theta)}. \tag{48}$$

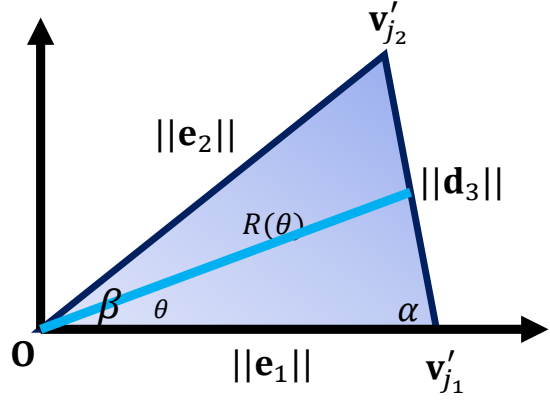

Fig. 3. Illustration of the integral of $G(\xi, \eta)$ over the side face, along with other mathematical notations for important variables.

Then, the integral of $G(\xi, \eta)$ over $\triangle O\mathbf{v}'_{j_1}\mathbf{v}'_{j_2}$ can be computed as:

$$
\int_{\triangle O\mathbf{v}'_{j_1}\mathbf{v}'_{j_2}} G(\xi, \eta)\, \mathrm{d}\sigma_\xi = -\frac{1}{4\pi} \int_{\theta=0}^{\beta} \int_{r=0}^{R(\theta)} \frac{1}{r} \times r\, \mathrm{d}r \mathrm{d}\theta = -\frac{1}{4\pi} \int_{\theta=0}^{\beta} R(\theta)\, \mathrm{d}\theta = -\frac{1}{4\pi} \int_{\theta=0}^{\beta} \frac{\|\mathbf{e}_1\| \sin(\alpha)}{\sin(\alpha + \theta)}\, \mathrm{d}\theta.
$$

$$
= -\frac{\|\mathbf{e}_1\| \sin(\alpha)}{4\pi} \int_{\theta=\alpha}^{\alpha+\beta} \csc\theta\, \mathrm{d}\theta = -\frac{\|\mathbf{e}_1\| \sin(\alpha)}{8\pi} \log\left|\frac{1-\cos\theta}{1+\cos\theta}\right|_{\alpha}^{\alpha+\beta}.
$$

(49)

Additionally, we have

$$
\cos(\alpha+\beta) = -\cos(\pi-\alpha-\beta) = \frac{\|\mathbf{e}_1\|^2 - \|\mathbf{e}_2\|^2 - \|\mathbf{d}_3\|^2}{2\|\mathbf{e}_2\|\|\mathbf{d}_3\|}, \quad \cos\alpha = \frac{\|\mathbf{d}_3\|^2 + \|\mathbf{e}_1\|^2 - \|\mathbf{e}_2\|^2}{2\|\mathbf{e}_1\|\|\mathbf{d}_3\|}.
$$

(50)

Therefore,

$$
\log\left|\frac{1-\cos\theta}{1+\cos\theta}\right|_{\alpha}^{\alpha+\beta} = \log \frac{(\|\mathbf{e}_2\| + \|\mathbf{d}_3\|)^2 - \|\mathbf{e}_1\|^2}{\|\mathbf{e}_1\|^2 - (\|\mathbf{e}_2\| - \|\mathbf{d}_3\|)^2} - \log \frac{(\|\mathbf{e}_1\| + \|\mathbf{d}_3\|)^2 - \|\mathbf{e}_2\|^2}{\|\mathbf{e}_2\|^2 - (\|\mathbf{e}_1\| - \|\mathbf{d}_3\|)^2}
$$

$$
= 2\log \frac{\|\mathbf{e}_1\| + \|\mathbf{e}_2\| + \|\mathbf{d}_3\|}{\|\mathbf{e}_1\| + \|\mathbf{e}_2\| - \|\mathbf{d}_3\|}.
$$

(51)

Thus,

$$
\int_{\triangle O\mathbf{v}'_{j_1}\mathbf{v}'_{j_2}} G(\xi, \eta)\, \mathrm{d}\sigma_\xi = -\frac{\|\mathbf{e}_1\| \sin(\alpha)}{8\pi} \log\left|\frac{1-\cos\theta}{1+\cos\theta}\right|_{\alpha}^{\alpha+\beta} = -\frac{\|\mathbf{e}_1 \times \mathbf{e}_2\|}{4\pi\|\mathbf{d}_3\|} \log \frac{\|\mathbf{e}_1\| + \|\mathbf{e}_2\| + \|\mathbf{d}_3\|}{\|\mathbf{e}_1\| + \|\mathbf{e}_2\| - \|\mathbf{d}_3\|}.
$$

(52)

The integral of $G(\xi, \eta)$ over $\triangle O\mathbf{v}'_{j_2}\mathbf{v}'_{j_3}$ and $\triangle O\mathbf{v}'_{j_3}\mathbf{v}'_{j_1}$ can be evaluated using the same approach. Then, we can obtain the closed-form formula for $\psi_j(\eta)$ as

$$
\psi_j(\eta) = -\int_{t_j} G(\xi, \eta)\, \mathrm{d}\sigma_\xi = -\frac{1}{4\pi} \Big( \frac{\mathbf{e}_1 \times \mathbf{e}_2}{\|\mathbf{d}_3\|} \log \frac{\|\mathbf{e}_1\| + \|\mathbf{e}_2\| + \|\mathbf{d}_3\|}{\|\mathbf{e}_1\| + \|\mathbf{e}_2\| - \|\mathbf{d}_3\|}
$$

$$
+ \frac{\mathbf{e}_2 \times \mathbf{e}_3}{\|\mathbf{d}_1\|} \log \frac{\|\mathbf{e}_2\| + \|\mathbf{e}_3\| + \|\mathbf{d}_1\|}{\|\mathbf{e}_2\| + \|\mathbf{e}_3\| - \|\mathbf{d}_1\|} + \frac{\mathbf{e}_3 \times \mathbf{e}_1}{\|\mathbf{d}_2\|} \log \frac{\|\mathbf{e}_3\| + \|\mathbf{e}_1\| + \|\mathbf{d}_2\|}{\|\mathbf{e}_3\| + \|\mathbf{e}_1\| - \|\mathbf{d}_2\|} \Big) \cdot \mathbf{n}_t - \frac{3}{A_t} \omega_t vol_t.
$$

(53)

Therefore,

$$\psi_t(\eta) = -\sum_{i=1}^{3} C_i (\mathbf{Z}_i \cdot \mathbf{n}_t) - \frac{3}{A_t} \omega_t vol_t. \tag{54}$$

**2. Derivation of** $\phi_{\mathbf{v}_{j_1}}(\eta)$**.** We begin by extending the hat-function $\Gamma_{\mathbf{v}_{j_1}, t_j}(\xi)$ from face $t_j$ to the entire volume of $\Upsilon$. Recall that $\Gamma_{\mathbf{v}_{j_1}, t_j}(\xi)$ takes the value 1 at $\mathbf{v}_{j_1}$, 0 at its adjacent vertices, and varies linearly within the 1-neighborhood of $\mathbf{v}_{j_1}$. Mathematically, for $\xi \in \Upsilon$, $\Gamma_{\mathbf{v}_{j_1}, t_j}(\xi)$ is given by 1.0 minus the ratio of the volumes of the tetrahedron $\xi, \mathbf{v}_{j_2}, \mathbf{v}_{j_3}, \eta$ and $\mathbf{v}_{j_1}, \mathbf{v}_{j_2}, \mathbf{v}_{j_3}, \eta$. We define $\mathbf{q}$ as the foot of the perpendicular from $\mathbf{v}_{j_1}$ to the plane spanned by $\mathbf{v}_{j_2}, \mathbf{v}_{j_3}$ and $\eta$. Then,

$$\Gamma_{\mathbf{v}_{j_1}, t_j}(\xi) = 1 - \frac{(\xi - \mathbf{v}_{j_1}) \cdot (\mathbf{q} - \mathbf{v}_{j_1})}{\|\mathbf{q} - \mathbf{v}_{j_1}\|^2}. \tag{55}$$

Based on the relationships between the cross product, scalar triple product, and their geometric interpretations in terms of area and volume, the gradient of $\Gamma_{\mathbf{v}_{j_1}, t_j}(\xi)$ can be computed as

$$\nabla_\xi \Gamma_{\mathbf{v}_{j_1}, t_j}(\xi) = \frac{\mathbf{v}_{j_1} - \mathbf{q}}{\|\mathbf{v}_{j_1} - \mathbf{q}\|^2} = \frac{\mathbf{e}_2 \times \mathbf{e}_3}{\|\mathbf{e}_2 \times \mathbf{e}_3\|} \cdot \frac{\|\mathbf{e}_2 \times \mathbf{e}_3\|}{|\mathbf{e}_1 \cdot (\mathbf{e}_2 \times \mathbf{e}_3)|} = \frac{\mathbf{e}_2 \times \mathbf{e}_3}{|\mathbf{e}_1 \cdot (\mathbf{e}_2 \times \mathbf{e}_3)|} = -\frac{\mathbf{Z}_1}{6 \operatorname{vol}_t}. \tag{56}$$

According to the divergence theorem,

$$\int_{\partial \Upsilon} \Gamma_{\mathbf{v}_{j_1}, t_j}(\xi) \frac{\partial G}{\partial \mathbf{n}}(\xi, \eta) \, d\sigma_\xi = \int_\Upsilon \nabla_\xi \cdot (\Gamma_{\mathbf{v}_{j_1}, t_j}(\xi) \nabla_\xi G(\xi, \eta)) \, dV_\xi. \tag{57}$$

Recall that $\mathbf{z}_i = \mathbf{Z}_i / \|\mathbf{Z}_i\|$ is unit normal vector for $S_i$. We then observe that $\nabla_\xi G(\xi, \eta) \cdot \mathbf{z}_i = 0$ in $S_i$ for $i = 1, 2, 3$. Therefore, computing $\phi_{\mathbf{v}_{j_1}, t_j}(\eta)$ is equivalent to evaluating the right-hand side of Eq. (57). Moreover,

$$\nabla_\xi \cdot (\Gamma_{\mathbf{v}_{j_1}, t_j}(\xi) \nabla_\xi G(\xi, \eta)) = \Gamma_{\mathbf{v}_{j_1}, t_j}(\xi) \Delta_\xi G(\xi, \eta) + \nabla_\xi \Gamma_{\mathbf{v}_{j_1}, t_j}(\xi) \cdot \nabla_\xi G(\xi, \eta). \tag{58}$$

Since $\nabla_\xi G(\xi, \eta) = \delta(\xi - \eta)$, the integration of the first term in the equation above yields only $\Gamma_{\mathbf{v}_{j_1}, t_j}(\eta)$, which is vanish at $\eta$. Thus, we only need to compute

$$\int_\Upsilon \nabla_\xi G(\xi, \eta) \, \nabla_\xi \Gamma_{\mathbf{v}_{j_1}, t_j}(\xi) \, dV_\xi = -\frac{\mathbf{Z}_1}{6 \operatorname{vol}_t} \cdot \int_\Upsilon \nabla_\xi G(\xi, \eta) \, dV_\xi. \tag{59}$$

According to the component-wise application of the divergence theorem, we have

$$\int_\Upsilon \nabla_\xi G(\xi, \eta) \, dV_\xi = \int_{\partial \Upsilon} G(\xi, \eta) \mathbf{n}(\xi) \, d\sigma_\xi = \mathbf{n}_t \int_{t_j} G(\xi, \eta) \, d\sigma_\xi + \mathbf{z}_1 \int_{S_1} G(\xi, \eta) \, d\sigma_\xi + \mathbf{z}_2 \int_{S_2} G(\xi, \eta) \, d\sigma_\xi + \mathbf{z}_3 \int_{S_3} G(\xi, \eta) \, d\sigma_\xi. \tag{60}$$

According to previous calculations, $\mathbf{z}_i \int_{S_i} G(\xi, \eta) \, d\sigma_\xi = -C_i \mathbf{Z}_i$ for $i = 1, 2, 3$ (Eq 52). If we denote $\mathbf{L} = \int_\Upsilon \nabla G(\xi, \eta) \, dV_\xi$, we have

$$\mathbf{L} = -\psi_t(\eta) \, \mathbf{n}_t - \sum_{i=1}^{3} C_i \mathbf{Z}_i. \tag{61}$$

Taking the dot product with $\mathbf{Z}_1$ gives

$$\begin{aligned}
\mathbf{Z}_1 \cdot \mathbf{L} &= -\left( -\sum_{i=1}^{3} C_i (\mathbf{Z}_i \cdot \mathbf{n}_t) - \frac{3}{A_t} \omega_t \operatorname{vol}_t \right) (\mathbf{Z}_1 \cdot \mathbf{n}_t) - \sum_{i=1}^{3} C_i (\mathbf{Z}_1 \cdot \mathbf{Z}_i) \\
&= -\sum_{i=1}^{3} C_i \left( \mathbf{Z}_1 \cdot \mathbf{Z}_i - (\mathbf{Z}_1 \cdot \mathbf{n}_t)(\mathbf{Z}_i \cdot \mathbf{n}_t) \right) + \frac{3}{A_t} \omega_t \operatorname{vol}_t \, (\mathbf{Z}_1 \cdot \mathbf{n}_t).
\end{aligned} \tag{62}$$

We will now establish the coordinate system to prove the following important equation:

$$\mathbf{Z}_i - (\mathbf{Z}_i \cdot \mathbf{n}_t)\,\mathbf{n}_t = H_t\,(\mathbf{n}_t \times \mathbf{d}_i), \quad i = 1, 2, 3, \tag{63}$$

where $H_t = \frac{3\,\text{vol}_t}{A_t}$ is the height from $\eta$ to $t_j$. We place $t_j$ in the plane $z = 0$ with outward unit normal $\mathbf{n}_t = (0, 0, 1)$. Then, we denote $\mathbf{v}_{j_k} = (x_i, y_i, 0)$ for $k = 1, 2, 3$. We also locate $\eta$ on z-axis and denote $\eta = (0, 0, -H_t)$. Then, $\mathbf{e}_i = \mathbf{v}_{j_i} - \eta = (x_i, y_i, H_t)$, $\mathbf{d}_i = \mathbf{v}_{j_{i+1}} - \mathbf{v}_{j_{i+2}} = (x_{i+1} - x_{i+2}, y_{i+1} - y_{i+2}, 0)$. Note that indices greater than 3 should be taken modulo 3. Let $a = x_{i+1} - x_{i+2}, b = y_{i+1} - y_{i+2}$, we have

$$\mathbf{Z}_i = -\mathbf{e}_{i+1} \times \mathbf{e}_{i+2} = -\mathbf{e}_{i+1} \times (\mathbf{e}_{i+1} - \mathbf{d}_i) = \mathbf{e}_{i+1} \times \mathbf{d}_i = (-H_t b, H_t a, x_{i+1} b - y_{i+1} a). \tag{64}$$

Therefore, we have $\mathbf{Z}_i - (\mathbf{Z}_i \cdot \mathbf{n}_t)\mathbf{n}_t = (-H_t b, H_t a, 0)$. On the other hand,

$$H_t(\mathbf{n}_t \times \mathbf{d}_i) = H_t \cdot ((0, 0, 1) \times (a, b, 0)) = (-H_t b, -H_t a, 0). \tag{65}$$

We complete the proof of the equation $\mathbf{Z}_i - (\mathbf{Z}_i \cdot \mathbf{n}_t)\mathbf{n}_t = H_t\,(\mathbf{n}_t \times \mathbf{d}_i)$. Then, we have

$$\begin{aligned}
\mathbf{Z}_1 \cdot \mathbf{L} &= -\sum_{i=1}^{3} C_i \Big(\mathbf{Z}_1 \cdot \mathbf{Z}_i - (\mathbf{Z}_1 \cdot \mathbf{n}_t)(\mathbf{Z}_i \cdot \mathbf{n}_t)\Big) + \frac{3}{A_t}\,\omega_t\,\text{vol}_t\,(\mathbf{Z}_1 \cdot \mathbf{n}_t) \\
&= -\sum_{i=1}^{3} C_i \Big(\mathbf{Z}_1 \cdot (\mathbf{Z}_i - (\mathbf{Z}_i \cdot \mathbf{n}_t)\mathbf{n}_t)\Big) + \frac{3}{A_t}\,\omega_t\,\text{vol}_t\,(\mathbf{Z}_1 \cdot \mathbf{n}_t) \\
&= -\sum_{i=1}^{3} C_i \Big(\mathbf{Z}_1 \cdot H_t\,(\mathbf{n}_t \times \mathbf{d}_i)\Big) + \frac{3}{A_t}\,\omega_t\,\text{vol}_t\,(\mathbf{Z}_1 \cdot \mathbf{n}_t) \\
&= -\frac{3\,\text{vol}_t}{A_t}\Big(\mathbf{n}_t \times \sum_{i=1}^{3} C_i \mathbf{d}_i - \omega_t \mathbf{n}_t\Big) \cdot \mathbf{Z}_1.
\end{aligned} \tag{66}$$

Recall the definition $P_t = \mathbf{n}_t \times \sum_{i=1}^{3} C_i \mathbf{d}_i - \omega_t \mathbf{n}_t$, we get

$$\phi_{\mathbf{v}_{j_1}, t_j}(\eta) = -\frac{1}{6\,\text{vol}_t} \mathbf{Z}_1 \cdot \mathbf{L} = \frac{1}{2A_t} P_t \cdot \mathbf{Z}_1. \tag{67}$$

Therefore, $\phi_{\mathbf{v}_{j_1}}(\eta)$ is obtained by summing over the 1-ring neighbors of $\phi_{\mathbf{v}_{j_1}, t_j}(\eta)$ as follows:

$$\boxed{\phi_{\mathbf{v}_{j_1}}(\eta) = \sum_{t \in \mathcal{N}(\mathbf{v}_{j_1})} \frac{1}{2A_t} P_t \cdot \mathbf{Z}_{1,t}.} \tag{68}$$

**3. Derivation of** $\nabla \psi_t(\eta)$. We now show that $\mathbf{a} \cdot \nabla \psi_t(\eta) = -\mathbf{a} \cdot P_t$ for arbitrary vector $\mathbf{a} \in \mathbb{R}^3$. We first notice that $\nabla_\eta G(\xi, \eta) = -\nabla_\xi G(\xi, \eta)$. Therefore,

$$\mathbf{a} \cdot \nabla_\eta \psi_t(\eta) = -\mathbf{a} \cdot \int_{t_j} \nabla_\eta G(\xi, \eta)\,\mathrm{d}\sigma_\xi = \int_{t_j} \mathbf{a} \cdot \nabla_\xi G(\xi, \eta)\,\mathrm{d}\sigma_\xi. \tag{69}$$

The vector $\mathbf{a}$ can be decomposed into tangential parts $\mathbf{a}_\parallel$ and normal parts $\mathbf{a}_\perp$ with respect to $t_j$ as $\mathbf{a}_\perp = (\mathbf{a} \cdot \mathbf{n}_t)\mathbf{n}_t$ and $\mathbf{a}_\parallel = \mathbf{a} - \mathbf{a}_\perp$. We define a tangential vector field $\mathbf{F}(\xi)$ with respect to $\xi$ at $t_j$ as:

$$\mathbf{F}(\xi) = G(\xi, \eta)\,\mathbf{a}_\parallel. \tag{70}$$

Moreover, let $\nabla_t \cdot$ be the planar divergence operator on $t_j$, we have $\nabla_t \cdot \mathbf{F} = \nabla_t \cdot (G \, \mathbf{a}_\parallel) = \mathbf{a}_\parallel \cdot \nabla_t G$. Since $\nabla_t G$ is the tangential projection of $\nabla_\xi G$, we also have $\mathbf{a}_\parallel \cdot \nabla_t G = \mathbf{a}_\parallel \cdot \nabla_\xi G$ on $t_j$. Therefore,

$$\int_{t_j} \mathbf{a}_\parallel \cdot \nabla_\xi G \, \mathrm{d}\sigma = \int_{t_j} \nabla_t \cdot (G \, \mathbf{a}_\parallel) \, \mathrm{d}\sigma = \int_{\partial t_j} G(\xi, \eta) \, (\mathbf{a}_\parallel \cdot \nu) \, \mathrm{d}s, \tag{71}$$

where $\nu$ is the outward unit normal along $\partial t_j$ lying in the plane of $t_j$, and $\mathrm{d}s$ is the arc-length measure. Then, $\nu_i = \mathbf{n}_t \times \frac{\mathbf{d}_i}{\|\mathbf{d}_i\|}$ for $i = 1, 2, 3$, and

$$\mathbf{a}_\parallel \cdot \nu_i = \mathbf{a} \cdot \left( \mathbf{n}_t \times \frac{\mathbf{d}_i}{\|\mathbf{d}_i\|} \right) = (\mathbf{a} \times \mathbf{n}_t) \cdot \frac{\mathbf{d}_i}{\|\mathbf{d}_i\|}. \tag{72}$$

Eq. (71) yields

$$\int_{t_j} \mathbf{a}_\parallel \cdot \nabla_\xi G \, \mathrm{d}\sigma = \sum_{i=1}^{3} \int_{\mathbf{d}_i} G(\xi, \eta) \, (\mathbf{a}_\parallel \cdot \nu_i) \, \mathrm{d}s = (\mathbf{a} \times \mathbf{n}_t) \cdot \sum_{i=1}^{3} \frac{\mathbf{d}_i}{\|\mathbf{d}_i\|} \int_{\mathbf{d}_i} G(\xi, \eta) \, \mathrm{d}s. \tag{73}$$

The next step is to compute the line integral of the fundamental solution $\int_{\mathbf{d}_i} G(\xi, \eta) \, \mathrm{d}s$. Let $r_k = \|\mathbf{e}_k\| = \|\mathbf{v}_{j_k} - \eta\|$ for $k = 1, 2, 3$. Let $\mathbf{a} = \mathbf{e}_3$, $\mathbf{b} = \mathbf{e}_2 - \mathbf{e}_3 = \mathbf{d}_1$ and define $\gamma(t) = \mathbf{a} + t\mathbf{b}$ with $t \in [0, 1]$. We also have $\mathrm{d}s = \|\mathbf{b}\| \, \mathrm{d}t$. Therefore,

$$\int_{\mathbf{d}_1} G(\xi, \eta) \, \mathrm{d}s = -\frac{\|\mathbf{b}\|}{4\pi} \int_0^1 \frac{\mathrm{d}t}{\|\mathbf{a} + t\mathbf{b}\|}. \tag{74}$$

When $A > 0$, the following important integral formula holds:

$$\int \frac{\mathrm{d}t}{\sqrt{At^2 + Bt + C}} = \frac{1}{\sqrt{A}} \log \left| 2At + B + 2\sqrt{A}\sqrt{At^2 + Bt + C} \right| + \text{constant}, \tag{75}$$

where $A = \|\mathbf{b}\|^2 = \|\mathbf{d}_1\|^2$, $B = 2(\mathbf{a} \cdot \mathbf{b}) = 2(\mathbf{e}_2 \cdot \mathbf{d}_1) = \|\mathbf{e}_2\|^2 - \|\mathbf{e}_3\|^2 - \|\mathbf{d}_1\|^2$, $C = \|\mathbf{e}_3\|^2$. Then we have

$$\begin{aligned} \int_{\mathbf{d}_1} G(\xi, \eta) \, \mathrm{d}s &= -\frac{1}{4\pi} \log \frac{2\|\mathbf{d}_1\|\|\mathbf{e}_2\| + 2\|\mathbf{d}_1\|^2 + (\|\mathbf{e}_2\|^2 - \|\mathbf{e}_3\|^2 - \|\mathbf{d}_1\|^2)}{2\|\mathbf{d}_1\|\|\mathbf{e}_3\| + (\|\mathbf{e}_2\|^2 - \|\mathbf{e}_3\|^2 - \|\mathbf{d}_1\|^2)} \\ &= -\frac{1}{4\pi} \log \frac{(\|\mathbf{e}_2\| + \|\mathbf{d}_1\|)^2 - \|\mathbf{e}_3\|^2}{\|\mathbf{e}_2\|^2 - (\|\mathbf{e}_3\| - \|\mathbf{d}_1\|)^2} \\ &= -\frac{1}{4\pi} \log \frac{\|\mathbf{e}_2\| + \|\mathbf{e}_3\| + \|\mathbf{d}_1\|}{\|\mathbf{e}_2\| + \|\mathbf{e}_3\| - \|\mathbf{d}_1\|}. \end{aligned} \tag{76}$$

Recall $R_1 = \|\mathbf{e}_2\| + \|\mathbf{e}_3\| > \|\mathbf{d}_1\|$, we have

$$\int_{\mathbf{d}_1} G(\xi, \eta) \, \mathrm{d}s = -\frac{1}{4\pi} \log \frac{R_1 + \|\mathbf{d}_1\|}{R_1 - \|\mathbf{d}_1\|}, \tag{77}$$

which is actually $-\|\mathbf{d}_1\| C_1$ of Eq .(43). Moreover, we have

$$\int_{t_j} \mathbf{n}_t \cdot \nabla_\xi G(\xi, \eta) \, \mathrm{d}\sigma_\xi = \omega_t. \tag{78}$$

Recall Eqs. (73) and (78), we obtain

$$\mathbf{a} \cdot \nabla \psi_t(\eta) = (\mathbf{a}_\parallel + (\mathbf{a} \cdot \mathbf{n}_t)\mathbf{n}_t) \cdot \int_{t_j} \nabla_\xi G(\xi, \eta) \, \mathrm{d}\sigma_\xi = -(\mathbf{a} \times \mathbf{n}_t) \cdot \sum_{i=1}^{3} C_i \, \mathbf{d}_i + (\mathbf{a} \cdot \mathbf{n}_t) \, \omega_t. \tag{79}$$

Using the property that $(\mathbf{a} \times \mathbf{n}_t) \cdot \mathbf{b} = \mathbf{a} \cdot (\mathbf{n}_t \times \mathbf{b})$, we obtain

$$\mathbf{a} \cdot \nabla \psi_t(\eta) = -\mathbf{a} \cdot \left( \mathbf{n}_t \times \sum_{i=1}^{3} C_i \mathbf{d}_i - \omega_t \mathbf{n}_t \right) = \mathbf{a} \cdot (-P_t). \tag{80}$$

Since $\mathbf{a}$ is arbitrary in $\mathbb{R}^3$, we have

$$\boxed{\nabla\psi_t(\eta) = -P_t.} \tag{81}$$

**4. Derivation of $\nabla\phi_{\mathbf{v}_{j_1}}(\eta)$.** The gradient of $\phi_{\mathbf{v}_{j_1},t_j}(\eta)$ with respect to $\eta$ can be directly obtained by differentiation as

$$\nabla\phi_{\mathbf{v}_{j_1},t_j}(\eta) = \frac{1}{2A_t}(\mathbf{J}(\mathbf{Z}_1)^\top P_t + \mathbf{J}(P_t)^\top \mathbf{Z}_1). \tag{82}$$

We first compute the Jacobian matrix of $\mathbf{Z}_1 = -\mathbf{e}_2 \times \mathbf{e}_3$. Consider a perturbation $\delta\eta$ to $\eta$. The corresponding variations in $\mathbf{e}_2$ and $\mathbf{e}_3$ are $\delta\mathbf{e}_2 = -\delta\eta$ and $\delta\mathbf{e}_3 = -\delta\eta$. The variation of $\mathbf{Z}_1$ is given by:

$$\delta\mathbf{Z}_1 = -\delta(\mathbf{e}_2 \times \mathbf{e}_3) = -(\delta\mathbf{e}_2) \times \mathbf{e}_3 - \mathbf{e}_2 \times (\delta\mathbf{e}_3) = (\delta\eta) \times \mathbf{e}_3 + \mathbf{e}_2 \times (\delta\eta) = (\mathbf{e}_2 - \mathbf{e}_3) \times \delta\eta. \tag{83}$$

Hence, the Jacobian matrix of $\mathbf{Z}_1$ is

$$\mathbf{J}(\mathbf{Z}_1) = [\mathbf{e}_2]_\times - [\mathbf{e}_3]_\times = [\mathbf{d}_1]_\times, \tag{84}$$

where $[\mathbf{u}]_\times$ denotes the skew-symmetric matrix such that, for any vector $\mathbf{v}$, $[\mathbf{u}]_\times\mathbf{v} = \mathbf{u} \times \mathbf{v}$. An important property that will be utilized later is that $[\mathbf{u}]_\times^\top = -[\mathbf{u}]_\times$.

Another important observation is that the term $\mathbf{J}(P_t)^\top\mathbf{Z}_1$ vanishes when summed over the 1-ring neighborhood of $\mathbf{v}_{j_1}$. To see this, we first note that

$$\mathbf{J}(P_t)^\top\mathbf{Z}_1 = \nabla_\eta\nabla_\eta^\top(-\psi_t(\eta)) = \nabla_\eta\left(\int_t \nabla_\xi^\top G(\xi,\eta) \, d\sigma_\xi\right)\mathbf{Z}_1 = \left(-\int_t \nabla_\xi\nabla_\xi^\top G(\xi,\eta) \, d\sigma_\xi\right)\mathbf{Z}_1$$
$$= -\int_{t_j} \nabla_\xi(\mathbf{Z}_1 \cdot (\nabla_\xi G(\xi,\eta)) \, d\sigma_\xi = -\int_{\partial t_j} \big(\mathbf{Z}(\xi) \cdot \nabla_\xi G(\xi,\eta)\big)\nu(\xi) \, ds_\xi, \tag{85}$$

where $\partial t_j$ is the boundary of $t_j$, and $\nu(\xi)$ is the outward unit normal of $\partial t_j$.

Let $\Omega = \bigcup_{t \in \mathcal{N}(\mathbf{v}_{j_1})} t$ denote the patch formed by the union of the triangles in the 1-ring neighborhood of $\mathbf{v}_{j_1}$. Upon summing over all the 1-ring neighborhoods, each interior edge contributes once positively and once negatively, canceling out to zero. The only remaining term is the boundary integral over $\partial\Omega$ as

$$-\int_{\partial\Omega} \big(\mathbf{Z}(\xi) \cdot \nabla_\xi G(\xi,\eta)\big)\nu(\xi) \, ds_\xi. \tag{86}$$

As illustrated in Fig. 2, $\xi - \eta$ is perpendicular to $\mathbf{Z}_1$ when $\xi$ lies on the edge connecting $\mathbf{v}_{j_2}$ and $\mathbf{v}_{j_3}$. Therefore, $\xi - \eta$ is perpendicular to $\mathbf{Z}(\xi)$ on $\partial\Omega$ and the above boundary integral equals 0. Specifically,

$$\sum_{t \in \mathcal{N}(\mathbf{v}_{j_1})} \mathbf{J}(P_t)^\top\mathbf{Z}_{1,t} = 0, \tag{87}$$

and we conclude that

$$\boxed{\nabla\phi_{\mathbf{v}_{j_1}}(\eta) = -\sum_{t \in \mathcal{N}(\mathbf{v}_{j_1})} \frac{1}{2A_t}([\mathbf{d}_1]_\times P_t).} \tag{88}$$

**5. Derivation of $\mathbf{H}\psi_t(\eta)$.** The Hessian of $\psi_t(\eta)$ can be obtained by

$$\boxed{\mathbf{H}\psi_t(\eta) = -\mathbf{J}(P_t),} \tag{89}$$

where

$$\mathbf{J}(P_t) = \sum_{i=1}^{3} (\mathbf{n}_t \times \mathbf{d}_i)\nabla C_i^\top - \mathbf{n}_t \nabla\omega_t^\top. \tag{90}$$

We also have

$$\nabla C_i = \overline{C}_i(\overline{\mathbf{e}}_{i+1} + \overline{\mathbf{e}}_{i+2}). \tag{91}$$

Therefore,

$$\mathbf{J}(P_t) = [\mathbf{n}_t]_\times \left(\mathbf{d}_1, \mathbf{d}_2, \mathbf{d}_3\right)_{3\times3} \begin{pmatrix} \overline{C}_1(\overline{\mathbf{e}}_2 + \overline{\mathbf{e}}_3)^\top \\ \overline{C}_2(\overline{\mathbf{e}}_3 + \overline{\mathbf{e}}_1)^\top \\ \overline{C}_3(\overline{\mathbf{e}}_1 + \overline{\mathbf{e}}_2)^\top \end{pmatrix}_{3\times3} - \mathbf{n}_t \nabla \omega_t^\top. \tag{92}$$

To compute the derivative of $\omega_t$ with respect to $\eta$, we first calculate $\psi_t(\eta)$. Recall that

$$\psi_t(\eta) = -\sum_{i=1}^3 C_i(\mathbf{Z}_i \cdot \mathbf{n}_t) - \frac{3}{A_t} \omega_t \operatorname{vol}_t, \tag{93}$$

and notice that its gradient with respect to $\eta$ satisfies

$$\nabla\psi_t(\eta) = -P_t(\eta) = -\left(\mathbf{n}_t \times \sum_{i=1}^3 C_i\mathbf{d}_i - \omega_t\mathbf{n}_t\right). \tag{94}$$

Differentiating Eq. (93) with respect to $\eta$ and obtain

$$\nabla\psi_t(\eta) = -\sum_{i=1}^3 \left[ (\mathbf{Z}_i \cdot \mathbf{n}_t)\, \nabla C_i + C_i\, \nabla(\mathbf{Z}_i \cdot \mathbf{n}_t) \right] - \frac{3}{A_t}\left(\operatorname{vol}_t \nabla\omega_t + \omega_t \nabla\operatorname{vol}_t\right). \tag{95}$$

The signed volume of the tetrahedron is $\operatorname{vol}_t = \frac{1}{3}A_t H_t$. Consequently

$$\nabla\operatorname{vol}_t = \frac{1}{3}A_t \nabla H_t = -\frac{1}{3}A_t \mathbf{n}_t. \tag{96}$$

Rearranging above formulation and get

$$\nabla\omega_t = -\frac{1}{H_t}\sum_{i=1}^3 (\mathbf{Z}_i \cdot \mathbf{n}_t)\overline{C}_i(\overline{\mathbf{e}}_{i+1} + \overline{\mathbf{e}}_{i+2}), \tag{97}$$

where $H_t$ is the height of $\eta$ toward $t_j$, which can also computed as $H_t = \mathbf{e}_1 \cdot \mathbf{n}_t$. The formula for $\nabla\omega_t$ differs slightly from that in [Ben-Chen et al. 2009], but we have verified that both are correct.

**6. Derivation of $\mathbf{H}\phi_{\mathbf{v}_{j_1}}(\eta)$.** Differentiating Eq. (88) directly yields

$$\mathbf{H}\phi_{\mathbf{v}_{j_1}}(\eta) = -\sum_{t \in N(\mathbf{v}_{j_1})} \frac{1}{2A_t}([\mathbf{d}_1]_\times \mathbf{J}(P_t)). \tag{98}$$

The expression for $\mathbf{J}(P_t)$ can be found in Eq. (92).

## C  Supplementary Information for Experiments

### C.1  Anisotropic Matrices

Here, we provide some anisotropic matrices used in 2D deformation experiments. Note that $\mathcal{A}(\theta, 1, \lambda) = \mathcal{A}(\theta + \frac{\pi}{2}, \lambda, 1)$, so all rotation angles listed in the table are less than $\frac{\pi}{2}$.

| $\theta$ | $(\lambda_1, \lambda_2) = (1.0, 2.0)$ | $(\lambda_1, \lambda_2) = (1.0, 4.0)$ | $(\lambda_1, \lambda_2) = (2.0, 1.0)$ | $(\lambda_1, \lambda_2) = (4.0, 1.0)$ |
|---|---|---|---|---|
| $0$ | $\begin{pmatrix} 1.0 & 0 \\ 0 & 2.0 \end{pmatrix}$ | $\begin{pmatrix} 1.0 & 0 \\ 0 & 4.0 \end{pmatrix}$ | $\begin{pmatrix} 2.0 & 0 \\ 0 & 1.0 \end{pmatrix}$ | $\begin{pmatrix} 4.0 & 0 \\ 0 & 1.0 \end{pmatrix}$ |
| $\pi/6$ | $\begin{pmatrix} 1.25 & -0.4330127 \\ -0.4330127 & 1.75 \end{pmatrix}$ | $\begin{pmatrix} 1.75 & -1.2990381 \\ -1.2990381 & 3.25 \end{pmatrix}$ | $\begin{pmatrix} 1.75 & 0.4330127 \\ 0.4330127 & 1.25 \end{pmatrix}$ | $\begin{pmatrix} 3.25 & 1.2990381 \\ 1.2990381 & 1.75 \end{pmatrix}$ |
| $\pi/4$ | $\begin{pmatrix} 1.5 & -0.5 \\ -0.5 & 1.5 \end{pmatrix}$ | $\begin{pmatrix} 2.5 & -1.5 \\ -1.5 & 2.5 \end{pmatrix}$ | $\begin{pmatrix} 1.5 & 0.5 \\ 0.5 & 1.5 \end{pmatrix}$ | $\begin{pmatrix} 2.5 & 1.5 \\ 1.5 & 2.5 \end{pmatrix}$ |
| $\pi/3$ | $\begin{pmatrix} 1.75 & -0.4330127 \\ -0.4330127 & 1.25 \end{pmatrix}$ | $\begin{pmatrix} 3.25 & -1.2990381 \\ -1.2990381 & 1.75 \end{pmatrix}$ | $\begin{pmatrix} 1.25 & 0.4330127 \\ 0.4330127 & 1.75 \end{pmatrix}$ | $\begin{pmatrix} 1.75 & 1.2990381 \\ 1.2990381 & 3.25 \end{pmatrix}$ |

## C.2 Computation of distortion

We compute the isometric distortion $E_{iso}$, conformal distortion $E_{conf}$ and area distortion $E_{area}$ as follows:

$$E_{iso} = (\sigma_1 - 1)^2 + (\sigma_2 - 1)^2, \tag{99}$$

where $\sigma_1$ and $\sigma_2$ are the two singular values of the Jacobian matrix of the deformation map. The conformal distortion evaluates the deviation from angle preservation and is expressed as

$$E_{conf} = \frac{\sigma_1}{\sigma_2} + \frac{\sigma_2}{\sigma_1} - 2, \tag{100}$$

which vanishes if and only if $\sigma_1 = \sigma_2$. Finally, the area distortion quantifies the deviation from unit area scaling and is given by

$$E_{area} = |\sigma_1 \sigma_2 - 1|. \tag{101}$$

The metrics are computed as the average over all pixels in the input image.



\end{document}

%% file: sections/sec1.tex
\section{Introduction}
Space deformation is a basic problem in geometric modeling. Instead of directly modifying the object surface, space deformation treats the surface as an embedding in the ambient space and deforms the space through deformation handles~\cite{2009VariationalGC}.
Cage-based deformation is a typical example in which the embedded object deforms according to the manipulation of the cage~\cite{2024Survey}. For each point $\eta$ within the cage, a series of functions $\phi_{i}(\eta)$ is constructed for each cage vertex $\mathbf{v}_i$. These functions satisfy the partition of unity property, $\sum_{i=1}^{n}{\phi_{i}(\eta)}=1$, and the linear reproduction property $\eta=\sum_{i=1}^{n}{\phi_{i}(\eta)} \mathbf{v}_i$. The functions $\phi_{i}(\eta)$ are commonly called the Generalized Barycentric Coordinates (GBC)~\cite{2015ReviewBarycentric, 2017HormannBarycentric}. 

GBC is usually interpolatory and expresses the point inside the cage as an affine sum of the cage vertices. Although it demonstrates good consistency with the target cage, it introduces obvious shearing artifacts, particularly when the cage undergoes affine transformations~\cite{2008Green}. In practice, it is also desirable for the deformed object to maintain a certain degree of conformality with the original object. Based on the Green's third identity of harmonic functions,~\citet{2008Green} propose Green Coordinates (GC), which take into account both the cage vertices $\mathbf{v}_i$ and the cage normals $\mathbf{n}_{j}$. GC satisfy $\eta=\sum_{i=1}^{n}{\phi_{i}(\eta)} \mathbf{v}_i+\sum_{j=1}^{m}\psi_{j}(\eta)\mathbf{n}_{j}$ for $\eta$ inside the cage. These coordinates enable shape-preserving deformations, characterized by angle-preservation in 2D and quasi-conformality in 3D. However, conformality is a strict and absolute condition, which also restricts the deformation diversity of GC once the source and target cages are fixed. In practical applications and artistic design, there is often a demand for more flexible and versatile deformation styles, which can be achieved through adjusting various parameters. In the real world, materials such as bones, wood, and fibers exhibit anisotropic structures, prompting users to seek deformations that are more  ``flexible'' or more ``rigid'' along specific directions.

Based on previous observations, we propose to incorporate the anisotropic property into the traditional GC. The Laplacian operator is the composition of divergence and gradient, i.e., $\Delta = \nabla \cdot \nabla$. Therefore, we can introduce anisotropy within the divergence operator and consider the anisotropic Laplace equation $\nabla\cdot(\mathbf{A}\nabla u)=0$, where $\mathbf{A}$ is a symmetric positive definite (SPD) matrix. The matrix $\mathbf{A}$ introduces varying scales in different directions, thereby incorporating anisotropic properties in the formulation. Furthermore, the anisotropic Laplace equation also gives rise to corresponding boundary integral formulation and leads to the anisotropic Green’s identity, which form the foundation for constructing the Anisotropic Green Coordinates (AGC). We show that AGC satisfy linear reproduction and translation invariance properties, and produce versatile deformation results with varying control matrices $\mathbf{A}$. We present an important geometric interpretation that up to a proper scale factor, deforming with AGC is equivalent to first transforming the entire setup by $\mathbf{A}^{-1/2}$, applying isotropic GC, and then transforming back by $\mathbf{A}^{1/2}$. Moreover, we relate the distortion of the deformation mapping to the condition number of $\mathbf{A}$, and mathematically demonstrate that AGC can generate a quasi-conformal mapping. Through extensive experiments, we validate that our method provide artists with a broader range of deformation options and enhance the diversity of deformations choices. Both theoretically and empirically, we further find that when the normal of a cage face aligns with the eigenvector corresponding to the larger eigenvalue of $\mathbf{A}$, the resulting deformation is more pronounced and sensitive along that direction. This feature enables artists to intuitively control directional stiffness or flexibility, thereby facilitating anisotropic deformation effects that are essential for modeling materials with oriented structures and achieving application-driven behaviors such as controlled directional stretching.

To further demonstrate the broad applicability of the closed-form expressions of AGC, we also derive their gradients and Hessians and perform variational shape deformation~\cite{2009VariationalGC}, which minimizes energy functional of the deformation map. This approach enables flexible control through space constraints and supports the as-rigid-as-possible (ARAP) deformation strategy. The incorporation of anisotropy further extends the applicability of the variational shape deformation framework to a wider range of scenarios.

%% file: sections/sec2.tex
\section{Related Work}

\subsection{Cage-based Deformation}
Cage-based deformation is a fundamental topic in geometry processing, where users manipulate a coarse control structure to drive the deformation of complex objects within the enclosed space~\cite{2024Survey}. Two critical factors in this problem are the geometry of the cage structure and the deformation coordinates defined on the cage. The most commonly used cage structure is the oriented simplicial surface, such as closed polygonal chains in 2D and triangle meshes in 3D~\cite{2005MVC, 2008Green, 2009VariationalGC, 2012Biharmonic, 2019GeneralBarycentric, 2024Biharmonic3D}. It is the first-order approximation of co-dimensional one surfaces. Other methods utilize high-order surfaces, such as polynomial or Bézier surfaces, as the cage structure~\cite{2013CubicMVC, 2023PolyGreen, 2024GCpatych, 2024PolyCauchy, 2025VariationalBiharmonic, 2025BezierCage, 2025Polynomial3D}. Such structures enable more precise and flexible control over curved and bending deformations.

While the cage structure can influence the flexibility, the deformation effect is predominantly determined by the specific deformation coordinates established on the cage. The most widely utilized methods are Generalized Barycentric Coordinates (GBC), which formulate interpolatory coordinates as a partition of unity and can accurately represent an arbitrary point $\eta$ within the cage. As pioneering works, \citet{2003MVCPoly}, \citet{2005MVCPoly}, and \citet{2005MVC} construct closed-form Mean Value Coordinates (MVC) by projecting the enclosing cage onto a sphere centered at $\eta$ and calculating the contribution of each cage vertex based on this spherical projection. There are also other techniques for constructing GBC, such as using the Laplace equation~\cite{2007Harmonic}, Poisson integral formula~\cite{2013Poisson}, probability theory~\cite{2008MaxEntropy, 2023MaximumLikelihood}, or stochastic sampling and Monte Carlo integration~\cite{2024Stochastic}. Additionally, transfinite coordinates~\citep{2006Transfinite, 2009TransfiniteMVC, 2025TransfiniteArbitary} extend the concept of traditional discrete barycentric coordinates to continuous boundaries, offering a general theoretical framework.

In addition to GBC, some other deformation coordinates consider both the cage vertices and cage normals, and therefore achieve shape-preserving deformations. Shape preservation means that the deformed shape maintains a certain degree of conformality with the original shape and avoids large shearing artifacts. The typical methods are the Green Coordinates (GC)~\cite{2008Green, 2009DriveGreen, 2022QuadGreen, 2025Polynomial3D, 2025BezierCage}, which utilize Green’s third identity to establish the deformation formulation and naturally satisfy linear reproduction and transformation invariance. Other techniques involve Cauchy Coordinates~\citep{2009CBC, 2024PolyCauchy, 2025CauchyClose}, which is demonstrated to be geometrically equivalent to GC in the planar domain. However, the deformation effects of Green and Cauchy coordinates are fixed. Specifically, once the source and target cages are determined, the deformation is uniquely defined. To address this limitation, some approach solve the biharmonic Dirichlet problem, which incorporates gradient variations into the formulation~\cite{2012Biharmonic, 2024Biharmonic3D, 2025PolyBiharmonic}. Alternatively, Somigliana Coordinates (SC) ~\cite{2023Somigliana} generalize GC based on the Navier–Cauchy equation of linear elasticity. This approach enables physically plausible volume control. However, SC mainly rely on the scalar Poisson ratio to control the compressibility. It does not focus on the directional preference. Furthermore, ~\citet{2024Biharmonic3D} provide a closed-form expression for SC.

\subsection{Variational Shape Deformation}
Variational methods are widely used in formulating partial differential equations for practical problems, such as the minimal surface equation~\cite{1993MinimalSurface}. They address the most fundamental problem in functional analysis of finding a function $f$ that minimizes the energy functional $E(f)$. ~\citet{2009VariationalGC} introduce the variational harmonic map in space deformation, which defines $E(f)$ based on GC, enabling deformation control with space constraints and ARAP properties~\cite{2000ARAPInterpolate}. The optimization problem is solved through a local-global minimization strategy~\citep{2007ARAP, 2008LocalGlobal}, which requires computing the Jacobian and Hessian of GC. Subsequent research has extended this technique by incorporating various coordinate systems or employing higher-order cage structures~\citep{2012Biharmonic, 2024Biharmonic3D, 2025CauchyClose, 2025VariationalBiharmonic, 2025Polynomial3D}. Additionally,~\citet{2023VBC} introduce variational barycentric coordinates, a neural field-based framework that optimizes generalized barycentric coordinates by minimizing their total variation.

\subsection{Anisotropy in Geometry Processing}
Anisotropy is a fundamental phenomenon in both natural systems and physical processes, playing an important role in simulating directionally dependent materials and generating adaptive computational domains. Research in this field encompasses theoretical foundations, numerical methodologies, and practical applications.

At the mathematical fundamental level,~\citet{2007Upscaling} establish a metric-based upscaling framework for elliptic PDEs of the form $\mathcal{L}u=\nabla\cdot(\mathbf{A}(\mathbf{x})\nabla u(\mathbf{x}))$ and rigorously analyze its mathematical properties within the Sobolev space. ~\citet{2009InhomogeneousElastic} construct anisotropic elasticity tensors on coarse meshes through harmonic homogenization to preserve fine-scale directional properties.~\citet{2019AnisotropicElasticity} focus on anisotropic hyperelasticity and propose an anisotropic strain invariant for the inversion-safe energy model, complemented by a rehabilitation technique to address degenerate elements.~\citet{2024AnisotropicFiniteElement} derive the anisotropic ARAP energy for elastic materials and provide a robust formulation suitable for finite element simulations.

Anisotropy is also well introduced in mesh processing.~\citet{2022GoGreen} expand Green’s functions by leveraging the anisotropic characteristics of general linear elastic materials, which allows for real-time control of material deformation without volumetric discretization.~\citet{2018AnisotropicQuasiConformal} propose generating planar anisotropic meshes for a given metric tensor and establishing a quasi-conformal mapping to transform the anisotropic metric into a Euclidean metric.~\citet{2013ParticleSurfaceMeshing} embed the anisotropic Riemannian metrics into higher-dimensional isotropic spaces, achieving efficient particle-based anisotropic meshing. \citet{2025AnisotropicGauss} derive the anisotropic Gauss formulation and utilize this boundary integral formulation  for point cloud normal orientation and surface reconstruction. This work focuses on geometry topics that are distinct from our method and only considers the anisotropic equations in diagonal forms.

%% file: sections/sec3.tex
\section{Anisotropic Green Coordinates}
In this section, we present a detailed derivation of AGC. We begin by introducing the boundary integral formulation of the anisotropic Laplace equation. It is non-trivial that, despite the introduction of anisotropy, we can still use boundary integrals to determine the coordinates of any point within the cage. Subsequently, we discretize this identity to obtain the deformation coordinates that exhibit linear reproduction and translation invariance. Then, we analyze their properties.

\subsection{Anisotropic Laplace Equation}
Green's third identity can be derived from the Laplace equation $\Delta u=0$, which serves as the foundation for establishing GC. Since $\Delta=\nabla \cdot \nabla$, we can introduce anisotropy within the divergence operator. Our method starts with the anisotropic Laplace equation in $\mathbb{R}^{d}, d\ge2$, which is given by:
\begin{equation}
\nabla\cdot(\mathbf{A}\nabla u(\mathbf{x}))=0,
\end{equation}
where $\mathbf{A}$ is a $d \times d$ SPD matrix with constant coefficients. This equation is a special case of the second-order elliptic operator $\mathcal{L}u=\nabla\cdot(\mathbf{A}(\mathbf{x})\nabla u(\mathbf{x}))+ \mathbf{b}(\mathbf{x}) \cdot \nabla u(\mathbf{x}) + c(\mathbf{x})u(\mathbf{x})$, and shares many similar properties with the classical Laplacian operator. When $\mathbf{A}$ is not constant, properties of solutions are generally studied in Sobolev spaces~\cite{1064EvansPDE}, and obtaining explicit solutions is often challenging. Thus, we focus on the scenario where $\mathbf{A}$ is a constant-coefficient matrix, and both the convection-enhancing term $\mathbf{b}(\mathbf{x})$ and the zero-order term $c(\mathbf{x})$ are set to zero. The Malgrange-Ehrenpreis theorem states that every non-zero linear differential operator $\mathcal{L}$ with constant coefficients admits a fundamental solution~\cite[Theorem 1.56]{1995PDEbook}.

We first introduce the concept of the square root of a SPD matrix. It is known that every SPD matrix $\mathbf{A}$ can be diagonalized as $\mathbf{A}=\mathbf{P} \Lambda \mathbf{P}^{\top}$, where 
\begin{equation}
\Lambda = \mathrm{diag}(\lambda_1, \lambda_2,...,\lambda_d)
\end{equation}
is a $d$-dimensional diagonal matrix with $\lambda_i >0$ for $i=1,2,...,d$, and $\mathbf{P}$ is an orthogonal matrix satisfying $\mathbf{P}^{-1}=\mathbf{P}^{\top}$. By expressing 
\begin{equation}\Lambda^{1/2}=\mathrm{diag}(\sqrt{\lambda_1}, \sqrt{\lambda_2},...,\sqrt{\lambda_d}),
\end{equation}
we can define the square root of $\mathbf{A}$ as $\mathbf{A}^{1/2}=\mathbf{P} \Lambda^{1/2} \mathbf{P}^{\top}$. It is then straightforward to verify that $(\mathbf{A}^{1/2})^{2}=\mathbf{A}$. $\mathbf{A}^{1/2}$ is also SPD.

Next, we will present some theorems related to the anisotropic Laplace equation. While we believe these properties have been investigated in prior mathematical research, as they are closely related to the generalized Green’s function~\cite{2004GeneralizedGreen}, we are not aware of specific literature in the fields of cage or variational shape deformation that comprehensively addresses these results. Therefore, we provide a detailed discussion of them in the following.

\begin{theorem}
The fundamental solution of the anisotropic Laplace equation $\nabla_{\xi}\cdot(\mathbf{A}\nabla_{\xi} G_{\mathbf{A}}(\xi, \eta))=\delta(\xi-\eta)$ is:
\begin{equation}
\label{eq:anisotropic_fundamental_solution}
G_{\mathbf{A}}(\xi, \eta)=\begin{cases}
\frac{1}{2\pi\sqrt{\det{\mathbf {A}}}}
\log\sqrt{(\xi-\eta)^{\top} \mathbf{A}^{-1}(\xi-\eta)}, & d = 2,\\
-\frac{1}{(d - 2)\omega_{d}\sqrt{\det{\mathbf {A}}}}{{[(\xi-\eta)^{\top}\mathbf{A}^{-1}(\xi-\eta)]}^{\frac{2-d}{2}}}, & d\geq3,
\end{cases}
\end{equation}
where $\delta(\mathbf{x})$ is the Dirac delta distribution satisfying $\delta(\mathbf{x})=0$ for $\mathbf{x} \neq \mathbf{0}$ and $\int_{\mathbb{R}^{d}}{\delta(\mathbf{x})} \ \mathrm{d}\mathbf{x}=1$, and $\omega_{d}={2\pi}^{d/2}/\Gamma(d/2)$ represents the surface area of the unit sphere in $\mathbb{R}^{d}$.
\end{theorem}
\begin{proof} 
We begin with the variable substitutions $\mathbf{x}=\mathbf{A}^{-1/2}\xi$ and $\mathbf{y}=\mathbf{A}^{-1/2}\eta$. 
According to Lemma A.1 in the supplementary material, 
we have
\begin{equation}
\label{eq:laplacianxi_x}
\begin{gathered}
\nabla_{\xi} \cdot (\mathbf{A}\nabla_{\xi}u)=\nabla_{\xi} \cdot (\mathbf{A}\mathbf{A}^{-1/2}\nabla_{\mathbf{x}}u)=\nabla_{\xi} \cdot (\mathbf{A}^{1/2}\nabla_{\mathbf{x}}u) \\
=\nabla_{\mathbf{x}} \cdot (\mathbf{A}^{-1/2}\mathbf{A}^{1/2}\nabla_{\mathbf{x}}u) = \Delta_{\mathbf{x}}u.
\end{gathered}
\end{equation}
Since the Dirac delta distribution is a generalized function that violates the $\mathcal{C}^{2}$ condition of Lemma A.1, we introduce Lemma A.2 to establish the same variable substitution formula within the framework of distribution theory. Furthermore, within distribution theory, equality holds in the almost everywhere sense; thus, functions may differ on a set of measure zero. Nevertheless, for the fundamental solution, it is customary to seek a function that is smooth everywhere except at the singular point.

In the following, we denote by $G(\mathbf{u})=G(\mathbf{x}, \mathbf{y})$ the fundamental solution of the isotropic Laplace equation $\Delta_{\mathbf{x}}G(\mathbf{x}, \mathbf{y})=\delta(\mathbf{x}-\mathbf{y})$, where $\mathbf{u}=\mathbf{x}-\mathbf{y}$, and by $G_{\mathbf{A}}(\xi, \eta)$ the fundamental solution of the anisotropic Laplace equation $\nabla_{\xi}\cdot(\mathbf{A}\nabla_{\xi} G_{\mathbf{A}}(\xi, \eta))=\delta(\xi-\eta)$ with respect to $\mathbf{A}$. According to Eq.~\eqref{eq:laplacianxi_x}, we know that $G_{\mathbf{A}}$ satisfies
\begin{equation}
\Delta_{\mathbf{x}}G_{\mathbf{A}}(\mathbf{x},\mathbf{y})=\delta(\mathbf{A}^{1/2}(\mathbf{x}-\mathbf{y}))=\frac{1}{\sqrt{\det{\mathbf{A}}}}\delta(\mathbf{x}-\mathbf{y}).
\end{equation}
The last equality follows from Lemma A.3 together with the identity $\det {\mathbf{A}^{1/2}} = \det{\Lambda^{1/2}}=\sqrt{\det{\mathbf{A}}}$. Therefore, we conclude that $\sqrt{\det{\mathbf{A}}} \ G_{\mathbf{A}}(\mathbf{x}-\mathbf{y}) \ $ is the fundamental solution $G(\mathbf{u})$ of $d$-dimensional isotropic Laplace equation, which is given by
\begin{equation}
\label{eq:fundamental_solution_iso}
G(\mathbf{u})=\begin{cases}
\frac{1}{2\pi}\log\|\mathbf{u}\|, & d = 2,\\
-\frac{1}{(d - 2)\omega_{d}}\frac{1}{\|\mathbf{u}\|^{d-2}}, & d\geq3.
\end{cases}
\end{equation}
Substituting $\mathbf{u}=\mathbf{x}-\mathbf{y}=\mathbf{A}^{-1/2}(\xi - \eta)$ to $G(\mathbf{u})$ and we complete the proof.
\end{proof}

The gradient $\nabla_{\xi} G_{\mathbf{A}}(\xi, \eta)$ can be evaluated explicitly as:
{\footnotesize
\begin{equation}
\label{eq:nabla_G_A}
\nabla_{\xi}G_{\mathbf{A}}(\xi, \eta)=\begin{cases}
\frac{1}{2\pi\sqrt{\det{\mathbf {A}}}}{{[(\xi-\eta)^{\top}\mathbf{A}^{-1}(\xi-\eta)]}^{-1}\mathbf{A}^{-1}(\xi-\eta)}, & d = 2,\\
\frac{1}{\omega_{d}\sqrt{\det{\mathbf {A}}}}{{[(\xi-\eta)^{\top}\mathbf{A}^{-1}(\xi-\eta)]}^{-\frac{d}{2}}\mathbf{A}^{-1}(\xi-\eta)}, & d\geq3,
\end{cases}
\end{equation}
}
which will be used frequently in the subsequent analysis.

\subsection{Construction of Anisotropic Green Coordinates}
Before deriving AGC, we first revisit the construction of the classical GC, which are defined on an oriented simplicial surface cage $P=(\mathbb{V}, \mathbb{T})$~\cite{2008Green, 2009DriveGreen}, where $\mathbb{V}$ denotes the cage vertices and $\mathbb{T}$ represents the faces. GC are constructed based on Green’s third identity. Specifically, when $\Omega$ is a bounded open set in $\mathbb{R}^{d}$ with $\mathcal{C}^{2}$ boundary, and $u \in \mathcal{C}^{2}(\overline{\Omega})$, where $\overline{\Omega}=\Omega \cup \partial \Omega$ is the closure of $\Omega$. $u(\xi)$ is a harmonic function, i.e., $\Delta u = 0$ in $\Omega$. Then, for $\eta \in \Omega$, we have:
\begin{equation}
\label{eq:third_identity}
u(\eta)=\int_{\partial \Omega}{u(\xi) \frac{\partial G}{\partial \mathbf{n}}(\xi, \eta) \ \mathrm{d} \sigma_{\xi}} -\int_{\partial \Omega}{G(\xi,\eta)\frac{\partial u}{\partial \mathbf{n}}(\xi) \ \mathrm{d} \sigma_{\xi}},
\end{equation}
where $G(\xi, \eta)$ is the fundamental solution of the isotropic Laplace equation~\eqref{eq:fundamental_solution_iso}. The directional derivatives are taken along the outward normal. 

\begin{figure}[htb]
  \centering
  \includegraphics[width=1.0\linewidth]{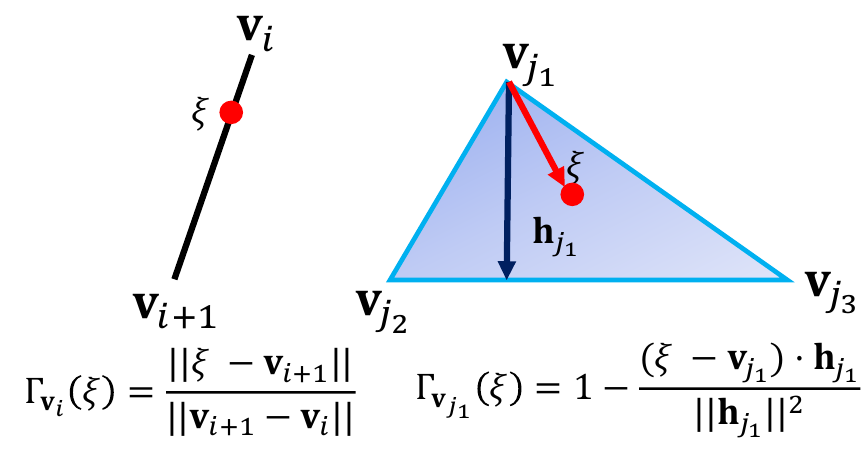}
  \caption{\label{fig:Figure_hat}
          Illustration of the hat function in 2D and 3D. }
\end{figure}

Given the cage $P=(\mathbb{V}, \mathbb{T})$, we can use above identity to reproduce any point $\eta$ inside the cage using the vertices $\{\mathbf{v}_{i} | i \in I_{\mathbb{V}}\}$ and outward normals $\{\mathbf{n}_j | j \in I_{\mathbb{T}}\}$ of the cage boundary as follows:
\begin{equation}
\label{eq:isotropic_GC_exp}
\eta = F(\eta, P)=\sum_{i \in I_{\mathbb{V}}}{\phi_{i}(\eta) \mathbf{v}_i} + \sum_{j \in I_{\mathbb{T}}}{\psi_{j}(\eta) \mathbf{n}_j}, \ \ \eta \in \Omega,
\end{equation}
where
\begin{align}
\phi_{i}(\eta) &= \int_{\xi \in \mathcal{N}\{\mathbf{v}_i\}}{\Gamma_{\mathbf{v}_i}(\xi) \frac{\partial G}{\partial \mathbf{n}}(\xi, \eta) \ \mathrm{d} \sigma_{\xi} , i \in I_{\mathbb{V}}}, \label{eq:isotropic_phi}\\
\psi_{j}(\eta) &= -\int_{\xi \in t_j}{G(\xi, \eta) \ \mathrm{d}\sigma_{\xi}, j\in I_{\mathbb{T}}}.\label{eq:isotropic_psi}
\end{align}
In the above formulations, $I_\mathbb{V}$ and $I_\mathbb{T}$ represent the index set of the cage vertices and faces, respectively; $\mathcal{N}\{\mathbf{v}_i\}$ represents the faces adjacent to $\mathbf{v}_i$, and $\Gamma_{\mathbf{v}_i}(\xi)$ is a piecewise linear hat function, which takes the value one at $\mathbf{v}_i$, zero at all other vertices, and varies linearly across each face within $\mathcal{N}\{\mathbf{v}_i\}$. 
As illustrated in Fig.~\ref{fig:Figure_hat}, the explicit form of $\Gamma_{\mathbf{v}_i}(\xi)$  in the 2D case is given by $\Gamma_{\mathbf{v}_i}(\xi)=\frac{\| \xi - \mathbf{v}_{i+1}\|}{\| \mathbf{v}_{i+1} - \mathbf{v}_i\|}$. In the 3D scenario, $\Gamma_{\mathbf{v}_{j_1}}(\xi)$ is expressed as $\Gamma_{\mathbf{v}_{j_1}}(\xi)=1-\frac{(\xi - \mathbf{v}_{j_1})\cdot \mathbf{h}_{j_1}}{\|\mathbf{h}_{j_1}\|^{2}}$, where $\mathbf{h}_{j_1}$ denotes the altitude corresponding to the edge opposite $\mathbf{v}_{j_1}$. Furthermore, $\Gamma_{\mathbf{v}_{j_1}}(\xi)$ essentially represents the barycentric coordinate of 
$\xi$ relative to $\mathbf{v}_{j_1}$ within $\triangle \mathbf{v}_{j_1}\mathbf{v}_{j_2}\mathbf{v}_{j_3}$. That is, $\xi=\Gamma_{\mathbf{v}_{j_1}}(\xi)\mathbf{v}_{j_1}+\Gamma_{\mathbf{v}_{j_2}}(\xi)\mathbf{v}_{j_2}+\Gamma_{\mathbf{v}_{j_3}}(\xi)\mathbf{v}_{j_3}$ and $\Gamma_{\mathbf{v}_{j_1}}(\xi)+\Gamma_{\mathbf{v}_{j_2}}(\xi)+\Gamma_{\mathbf{v}_{j_3}}(\xi)=1.0$.

When the source cage $P$ is deformed to the target cage $\tilde{P}$, the new location $\tilde{\eta}$ of $\eta$ is computed as:
\begin{equation}
\tilde{\eta} = F(\eta, \tilde{P})=\sum_{i \in I_{\mathbb{V}}}{\phi_{i}(\eta) \tilde{\mathbf{v}}_i} + \sum_{j \in I_{\mathbb{T}}}{\psi_{j}(\eta) s_j\tilde{\mathbf{n}}_j},
\end{equation}
where $\tilde{\mathbf{v}}_i$ and $\tilde{\mathbf{n}}_j$ denote the vertices and normals of the deformed cage, respectively. Throughout this work, the tilde symbol is used to represent quantities associated with the deformed value or target cage. The scale factor $s_j$ preserves scale invariance in 2D and quasiconformality in 3D, as detailed in Section 3 of ~\citep{2008Green}. 

Next, we derive the boundary integral formulation of the anisotropic Laplace equation, which establishes the mathematical foundation for AGC. 
\begin{theorem}
\label{th:anisotropic_identity}
If $\Omega$ is a bounded open set in $\mathbb{R}^{d}$ with $\mathcal{C}^{2}$ boundary, and $u \in \mathcal{C}^{2}(\overline{\Omega})$ satisfying $\nabla\cdot(\mathbf{A}\nabla u(\mathbf{\xi}))=0$ on $\Omega$, where $\mathbf{A}$ is symmetric positive definite, and $\overline{\Omega}=\Omega \cup \partial \Omega$ is the closure of $\Omega$. Then, for $\eta \in \Omega$, we have
\begin{equation}
\begin{gathered}
\label{eq:ani_identity}
u(\eta)=\int_{\partial \Omega}{u(\xi)\ (\mathbf{A} \nabla G_{\mathbf{A}}(\xi, \eta) \cdot \mathbf{n}(\xi)) \ \mathrm{d} \sigma_{\xi} } \\ -\int_{\partial \Omega}{G_{\mathbf{A}}(\xi,\eta)(\mathbf{A}\nabla u(\xi) \cdot \mathbf{n}(\xi)) \ \mathrm{d} \sigma_{\xi}},
\end{gathered}
\end{equation}
where $G_{\mathbf{A}}(\xi, \eta)$ is the fundamental solution of the anisotropic Laplace equation $\nabla_{\xi}\cdot(\mathbf{A}\nabla_{\xi} G_{\mathbf{A}}(\xi, \eta))=\delta(\xi-\eta)$, and $\mathbf{n}(\xi)$ the outward normal of $\partial \Omega$ at $\xi$. We also refer to such a function $u$ as \textbf{generalized harmonic with respect to} $\mathbf{A}$.
\end{theorem}

The proof of Theorem 2 is presented in Appendix B. We can derive AGC with the natural linear reproduction property by setting $u(\eta)=\eta$ in Eq.~\eqref{eq:ani_identity}. There is a remark that $u$ is a scalar function in Eq.~\eqref{eq:ani_identity}. However, we can apply the identity to each coordinate component separately. Specifically, let $u_{k}(\mathbf{\eta})=\eta_{k}$, which returns the $k$-th component of $\eta=(\eta_1,\eta_2,...,\eta_d)^{\top}$. Moreover, denote $\mathbf{A}=(\mathbf{a}_{1},\mathbf{a}_{2},...,\mathbf{a}_{d})$, where $\mathbf{a}_{k}$ is the $k$-th column of $\mathbf{A}$. Then, $\mathbf{A}\nabla u_{k}(\xi) \cdot \mathbf{n}(\xi)=(\mathbf{a}_k)^{\top}\mathbf{n}(\xi)$. We can then combine the $d$ components to obtain $\eta=(u_1(\eta),...,u_d(\eta))^{\top}$. Moreover, the composed normal term on the right-hand side is  $\mathbf{A}^{\top} \mathbf{n}_j=\mathbf{A}\mathbf{n}_j$. Considering the boundary integration region as the cage boundary, we have the following equations: 
\begin{equation}
\begin{aligned}
\label{eq:derive_ani_green}
\eta=&\sum_{j \in I_\mathbb{T}}\Big({\int_{t_j}{\xi\ (\mathbf{A} \nabla G_{\mathbf{A}}(\xi, \eta) \cdot \mathbf{n}(\xi)) \ \mathrm{d} \sigma_{\xi} } -\int_{t_j}{G_{\mathbf{A}}(\xi,\eta) \ \mathbf{A} \mathbf{n}(\xi) \ \mathrm{d} \sigma_{\xi}}}\Big) \\
=&\sum_{j \in I_\mathbb{T}}\sum_{\mathbf{v}_i \in \mathbb{V}(t_j)}{\mathbf{v}_i \Big(\int_{t_j}{\Gamma_{\mathbf{v}_i,t_j}(\xi) \mathbf{A} \nabla G_{\mathbf{A}}(\xi, \eta) \cdot \mathbf{n}(\xi) \ \mathrm{d} \sigma_{\xi}}\Big)} \\
&-\sum_{j \in I_\mathbb{T}}{\mathbf{A}\mathbf{n}_j}\Big(\int_{t_j}{G_{\mathbf{A}}(\xi, \eta) \ \mathrm{d} \sigma_{\xi}} \Big),
\end{aligned}
\end{equation}
where $I_\mathbb{T}$ denotes the index set of the cage faces, and $\mathbb{V}(t_j)$ represents the vertex set of $t_j$. By rearranging the above equation, AGC can be expressed as:
\begin{equation}
\label{eq:anisotropic_Green_coord}
\eta = \sum_{i \in I_{\mathbb{V}}}{\phi_{i}^{\mathbf{A}}(\eta) \mathbf{v}_i} + \sum_{j \in I_{\mathbb{T}}}{\psi_{j}^{\mathbf{A}}(\eta) \ \mathbf{A}\mathbf{n}_j}, \eta \in \Omega,
\end{equation}
where
\begin{align}
\phi_{i}^{\mathbf{A}}(\eta) &= \int_{\xi \in \mathcal{N}\{\mathbf{v}_i\}}{\Gamma_{\mathbf{v}_i}(\xi) \mathbf{A} \nabla G_{\mathbf{A}}(\xi, \eta) \cdot \mathbf{n}(\xi)\ \mathrm{d} \sigma_{\xi}, \ i \in I_{\mathbb{V}}}, \label{eq:anisotropic_phi}\\
\psi_{j}^{\mathbf{A}}(\eta) &= -\int_{\xi \in t_j}{G_{\mathbf{A}}(\xi, \eta) \ \mathrm{d}\sigma_{\xi}, \ j\in I_{\mathbb{T}}}. \label{eq:anisotropic_psi}
\end{align}
Compared to the traditional GC (Eqs~\eqref{eq:isotropic_GC_exp} to~\eqref{eq:isotropic_psi}), the integral expressions of $\phi_{i}^{\mathbf{A}}(\eta)$ and $\psi_{j}^{\mathbf{A}}(\eta)$ are now related to $G_{\mathbf{A}}(\xi, \eta)$ instead of $G(\xi, \eta)$. Moreover, the Neumann term of Eq.~\eqref{eq:anisotropic_Green_coord}, which is constructed based on the cage normals, now becomes $\mathbf{A} \mathbf{n}_j$.
When the source cage is deformed to $\tilde{P}$ with new vertices $\tilde{\mathbf{v}}_i$ and face normals $\tilde{\mathbf{n}}_j$, the deformed position $\tilde{\eta}$ of $\eta$ is given by:
\begin{equation}
\label{eq:deformed_position_ani}
\tilde{\eta} = \sum_{i \in I_{\mathbb{V}}}{\phi_{i}^{\mathbf{A}}(\eta) \tilde{\mathbf{v}}_i} + \sum_{j \in I_{\mathbb{T}}}{\psi_{j}^{\mathbf{A}}(\eta) s_j\mathbf{A}\tilde{\mathbf{n}}_j},
\end{equation}
where $s_j$ is the scale factor analogous to that in~\cite{2008Green}. A detailed discussion of this term will be provided in the following subsection.

\subsection{Properties}
To analyze the deformation properties of AGC, we first revisit the properties of the isotropic GC, as examined in~\citep{2008Green, 2009DriveGreen}. Specifically, these properties include: (1) Linear reproduction: $\eta=F(\eta, P)$ for $\eta \in \Omega$; (2) Translation invariance: $\sum_{i \in I_{\mathbb{V}}}{\phi_{i}(\eta)=1}$ for $\eta \in \Omega$; (3) Rotation and scale invariance: for an affine transformation $T$, $T\eta=F(\eta, TP)$; (4) Shape preservation: the deformation is angle-preserving in 2D and quasi-conformal in 3D; and (5) Smoothness: $\phi_{i}(\eta)$ and $\psi_{j}(\eta)$ are $C^{\infty}$ for $\eta \in \Omega$. We observe that AGC retain the properties of linear reproduction, translation invariance, scale invariance, quasi-conformality and smoothness. However, they does not satisfy the rotation invariance. Additionally, in 2D, these coordinates are not angle-preserving but exhibit quasi-conformality.

\textbf{Linear reproduction}:  The anisotropic boundary integral formula naturally satisfies the linear reproduction property. As shown in Eq.~\eqref{eq:deformed_position_ani}, when the source and target cages are identical and $s_j=1$, substituting the coordinate expressions back yields Eq.~\eqref{eq:derive_ani_green}, verifying that $\tilde{\eta}=\eta$.

\textbf{Translation invariance}:
Translation invariance follows from the fact that $\phi_{i}^{\mathbf{A}}(\eta)$ satisfies the partition of unity property, i.e., $\sum_{i \in I_{\mathbb{V}}}{\phi_{i}^{\mathbf{A}}(\eta)=1}$. This property can be proven by setting $u(\eta)=1$ in Eq.~\eqref{eq:ani_identity}, which yields: 
\begin{equation}
\int_{\partial \Omega}(\mathbf{A} \nabla G_{\mathbf{A}}(\xi, \eta) \cdot \mathbf{n}(\xi)) \ \mathrm{d} \sigma_{\xi}=1, \ \ \eta \in \Omega.
\end{equation}
Furthermore, $\Gamma_{\mathbf{v}_i, t_j}(\xi)$ in the expression for $\phi_{i}^{\mathbf{A}}(\eta)$ is actually the barycentric coordinates of $\mathbf{v}_i$ with respect to $t_j$. Therefore, 
\begin{equation}
\begin{aligned}
\sum_{i \in I_{\mathbb{V}}}\phi_{i}^{\mathbf{A}}(\eta)=&\sum_{j \in I_\mathbb{T}}\sum_{\mathbf{v}_i \in \mathbb{V}(t_j)}\Big({\int_{t_j}{\Gamma_{\mathbf{v}_i, t_j}(\xi)\ (\mathbf{A} \nabla G_{\mathbf{A}}(\xi, \eta) \cdot \mathbf{n}(\xi)) \ \mathrm{d} \sigma_{\xi}\Big) }} \\
=&\int_{\partial \Omega}(\mathbf{A} \nabla G_{\mathbf{A}}(\xi, \eta) \cdot \mathbf{n}(\xi)) \ \mathrm{d} \sigma_{\xi}=1.
\end{aligned}
\end{equation}

\textbf{Scale and rotation invariance}
AGC do not necessarily satisfy rotation invariance, as $\mathbf{T}$ and $\mathbf{A}$ are not necessarily commutative when $\mathbf{T}$ is an arbitrary rotation matrix. Regarding the selection of the scaling factor $s_j$ in Eq~\eqref{eq:deformed_position_ani}, we define $s_j$ as a scalar because scale invariance is not an anisotropic property. To determine its value, GC~\cite{2008Green} establish a simplex $S_j$ formed by $t_j$ and $\mathbf{v}_{j_1} + \mathbf{n}_j$, and $\tilde{S}_j$ formed by $\tilde{t}_j$ and $\tilde{\mathbf{v}}_{j_1} + s_j\tilde{\mathbf{n}}_j$, where $\mathbf{v}_{j_1}$ is a vertex of $t_j$. The scaling factor $s_j$ in Eq.~\eqref{eq:deformed_position_ani} is derived by analyzing the stretch of the face to ensure that $T(S_j)$ is similar to $\tilde{S}_j$, resulting in $s_j=\|T\|_2$, where $T$ represents the scale transformation. In the anisotropic scenario, the simplex is formed by $t_j$ and $\mathbf{v}_{j_1}+\mathbf{A}\mathbf{n}_j$, as well as $\tilde{t}_j$ and $\tilde{\mathbf{v}}_{j_1}+s_j\mathbf{A}\tilde{\mathbf{n}}_j$, respectively. To ensure that $T(S_j)$ approximates $\tilde{S}_j$ in shape, we can similarly conclude that $s_j=\|T\|_2$. Following the approach proposed by \citet{2008Green}, we define $s_j$ as follows. In the 2D case, $s_j$ is given by the ratio of the deformed edge length to the original edge length, i.e., $s_j=||\tilde{t}_j||/||t_j||$. For the 3D scenario, the scale factor is defined as:
\begin{equation}
s_j=
\sqrt{\frac{\sigma_1^2+\sigma_2^2}{2}}=\frac{\sqrt{\|\tilde{u}\|^2\|v\|^2 - 2(\tilde{u} \cdot\tilde{v})(u \cdot v)+\|\tilde{v}\|^2 \|u\|^2}}{\sqrt{8} \mathrm{area}(t_j)},
\end{equation}
where $\sigma_1$ and $\sigma_2$ denote the singular values of the linear map that transforms $t_j$ to $\tilde{t}_j$ when restricted to a plane. Moreover, $u, v$ and $\tilde{u}, \tilde{v}$ represent the edge vectors of $t_j$ and $\tilde{t}_j$, respectively.

\textbf{Quasi-conformality}
Although AGC are not strictly conformal in two dimensions, their quasi-conformality in both 2D and 3D can be established through the analysis of their relationship with the isotropic formulation presented in Section~\ref{sec:5}. Consequently, we also include the analysis of quasi-conformality in Section~\ref{sec:5}.

\textbf{Smoothness and generalized harmonicity} 
We also have the property that $\phi_{i}^{\mathbf{A}}(\eta)$ and $\psi_{j}^{\mathbf{A}}(\eta)$ are smooth for $\eta \in \Omega$. The reason is that $G_{\mathbf{A}}(\xi, \eta)$ is an elementary function with respect to $\eta$ and has no singularities within the integration domain as $\xi \neq \eta$. Consequently, $\phi_{i}^{\mathbf{A}}(\eta)$ and $\psi_{j}^{\mathbf{A}}(\eta)$ are also smooth with respect to $\eta$, as they are obtained through differentiation and integration of $G_{\mathbf{A}}(\xi, \eta)$ with respect to $\xi$, and $\Gamma_{i}(\xi)$ is independent of $\eta$.
Moreover, we also conclude that $\phi_{i}^{\mathbf{A}}(\eta)$ and $\psi_{j}^{\mathbf{A}}(\eta)$ are generalized harmonic with respect to $\mathbf{A}$, i.e., $\nabla_{\eta}\cdot(\mathbf{A}\nabla_{\eta} \phi_{i}^{\mathbf{A}}(\eta))=0$ and $\nabla_{\eta}\cdot(\mathbf{A}\nabla_{\eta} \psi_{j}^{\mathbf{A}}(\eta))=0$. This can be briefly proven as follows: Due to the smoothness of $G_{\mathbf{A}}(\xi, \eta)$ when $\xi \neq \eta$, differentiation with respect to $\eta$ can be interchanged with the integration with respect to $\xi$. Moreover, the gradient of $G_{\mathbf{A}}(\xi, \eta)$ with respect to $\eta$ is simply the negative of its gradient with respect to $\xi$. Given that $G_{\mathbf{A}}(\xi, \eta)$ is generalized harmonic with respect to $\mathbf{A}$ in $\xi$, it is also generalized harmonic with respect to $\eta$ in $\Omega$. It follows that $\phi_{i}^{\mathbf{A}}(\eta)$ and $\psi_{j}^{\mathbf{A}}(\eta)$ also inherit this property. 

%% file: sections/sec4.tex
\section{Derivation of Closed-form AGC}\label{sec:4}

We have derived the deformation formula for AGC and investigated its fundamental properties. In this section, we present closed-form expressions for $\phi_{i}^{\mathbf{A}}(\eta)$ and $\psi_{j}^{\mathbf{A}}(\eta)$ in both 2D and 3D. These analytical solutions are more accurate and computationally efficient compared to numerical methods for evaluating the integrals. In the following derivations, we frequently use formulations for the substitution of area elements on curves and surfaces embedded in $\mathbb{R}^{2}$ and $\mathbb{R}^{3}$. These formulations are detailed in Lemmas A.4, A.5, and A.6 of the supplementary material. While these results do not rely on advanced mathematical concepts, they are non-trivial and not immediately obvious. Additionally, Appendix C provides a detailed derivation of the closed-form expressions for classical isotropic GC and their gradients and Hessians, which serve as the foundation for the anisotropic formulation.

\subsection{Closed-form Expressions for 2D Scenario}\label{sec:4_1}
We begin by presenting the method in 2D. We assume that the 2D cage is a closed and oriented 1-manifold in $\mathbb{R}^2$ with vertices $I_{\mathbb{V}}=\{\mathbf{v_{i}}\}_{i=1}^{N}$ and faces $I_{\mathbb{T}}=\{f_j|f_j=\overrightarrow{\mathbf{v}_{j}\mathbf{v}_{j+1}}\}$ arranged in counter-clockwise order. We denote $\mathbf{a}_j=\mathbf{v}_{j+1}-\mathbf{v}_{j}$. Then, the unit outward normal $\mathbf{n}_j$ of $f_j$ is given by $\mathbf{n}_j={\mathbf{a}_j}^{\bot}/\|{\mathbf{a}_j}^{\bot}\|$, where the superscript ``$\bot$'' denotes the vector obtained by rotating the original vector 90 degrees clockwise. Then, $\psi_{j}^{\mathbf{A}}(\eta)$ has the following expression:
\begin{align}
\psi_{j}^{\mathbf{A}}(\eta) = &-\int_{\xi \in f_j} G_{\mathbf{A}}(\xi, \eta) \ \mathrm{d}\sigma_{\xi}, \\
= & \int_{\xi \in f_j}{-\frac{1}{2\pi\sqrt{\det{\mathbf {A}}}}
\log \sqrt{(\xi-\eta)^{\top} \mathbf{A}^{-1}(\xi-\eta)}} \ \mathrm{d}\sigma_{\xi}.
\end{align}
\begin{figure}[htb]
  \centering
  \includegraphics[width=1.0\linewidth]{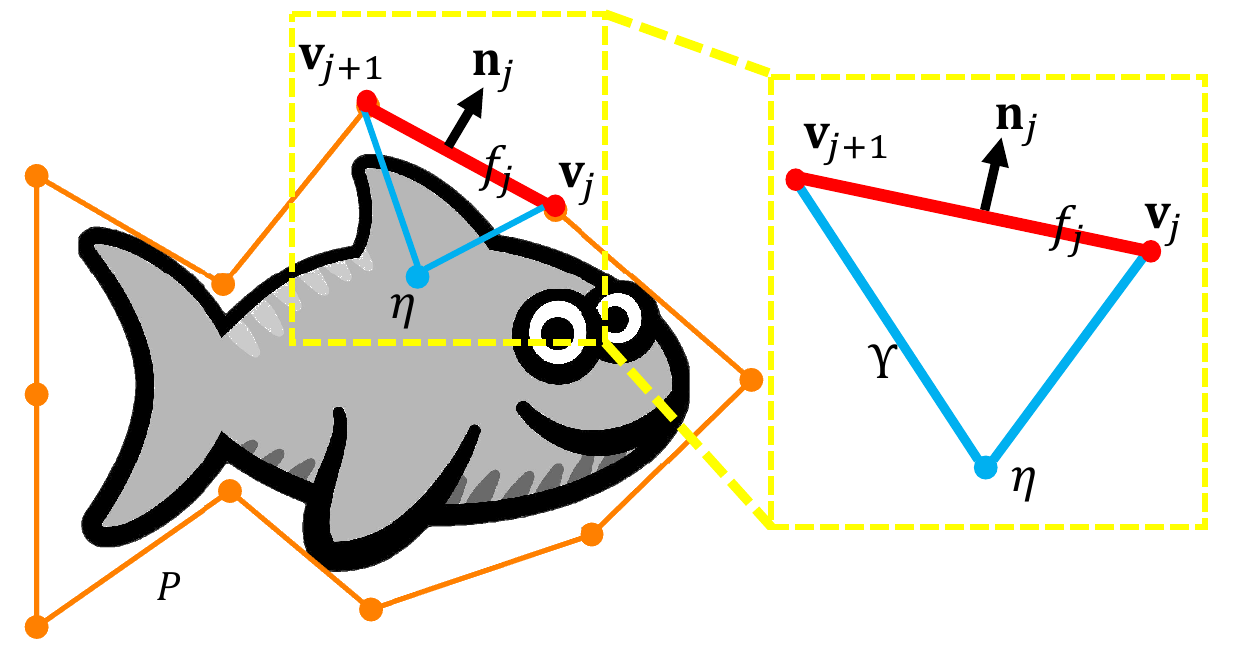}
  \caption{\label{fig:Figure1}
          The illustration of a 2D cage and its mathematical notations. }
\end{figure}

The key step is to transform the calculation of the anisotropic formulation into the isotropic one through the variable substitution. Specifically, let $\mathbf{x}=\mathbf{A}^{-1/2}\xi,\  \mathbf{y}=\mathbf{A}^{-1/2}\eta$, $\mathbf{v}^{\prime}_{j}=\mathbf{A}^{-1/2}\mathbf{v}_{j}$ and $\mathbf{v}^{\prime}_{j+1}=\mathbf{A}^{-1/2}\mathbf{v}_{j+1}$. According to Lemma A.4 and Lemma A.5, we have $\mathrm{d} \sigma_{\mathbf{x}}=|\det \mathbf{A}^{-1/2}|\|\mathbf{A}^{1/2} \mathbf{n}_j\| \ \mathrm{d} \sigma_{\xi}$, where $\mathbf{n}_j$ is the unit normal of $f_j$. Denoting $f_j^{\prime}=\overrightarrow{\mathbf{v}_{j}^{\prime}\mathbf{v}_{j+1}^{\prime}}$, we have
\begin{equation}
\begin{aligned}
\label{eq:closed_form_ani_2D_psi}
\psi_{j}^{\mathbf{A}}(\eta) =& \int_{\xi \in f_j}{-\frac{1}{2\pi\sqrt{\det{\mathbf {A}}}}
\log \sqrt{(\xi-\eta)^{\top} \mathbf{A}^{-1}(\xi-\eta)}} \ \mathrm{d}\sigma_{\xi}\\
= &\frac{1}{|\det{\mathbf{A}^{-1/2}}|\|\mathbf{A}^{1/2}\mathbf{n}_{j}\|}\int_{\mathbf{x} \in f_j^{\prime}}-{\frac{1}{2\pi\sqrt{\det{\mathbf {A}}}} \log\|\mathbf{x}-\mathbf{y}\|}  \ \mathrm{d} \sigma_{\mathbf{x}} \\
=& \frac{1}{\sqrt{\mathbf{n}_{j}^{\top}\mathbf{A}\mathbf{n}_{j}}}\int_{\mathbf{x} \in f_j^{\prime}}{-\frac{1}{2\pi} \log\|\mathbf{x}-\mathbf{y}\|} \ \mathrm{d} \sigma_{\mathbf{x}}.
\end{aligned}
\end{equation}
According to~\citep{2009DriveGreen}, we can define $\mathbf{a}_j=\mathbf{v}_{j+1}^{\prime}-\mathbf{v}_{j}^{\prime}$ and $\mathbf{b}_j=\mathbf{v}_{j}^{\prime} - \mathbf{y}$. Then, the term $-\int_{\xi \in f_j^{\prime}}{\frac{1}{2\pi} \log\|\mathbf{x}-\mathbf{y}\|} \ \mathrm{d} \sigma_{\mathbf{x}}$ is equal to $-\frac{||\mathbf{a}_{j}||}{4\pi}(Q(1)-Q(0))$, where
\begin{equation}
\begin{aligned}
&Q(t)=
\left(t+\frac{\mathbf a_j\cdot\mathbf b_j}{\|\mathbf a_j\|^2}\right)
\log\left(\|\mathbf a_j\|^2t^2+2t(\mathbf a_j\cdot\mathbf b_j)+\|\mathbf b_j\|^2\right)
-2t \\
&+\frac{2\sqrt{\|\mathbf a_j\|^2\|\mathbf b_j\|^2-(\mathbf a_j\cdot\mathbf b_j)^2}}{\|\mathbf a_j\|^2}
\arctan\left(
\frac{\|\mathbf a_j\|^2t+(\mathbf a_j\cdot\mathbf b_j)}
{\sqrt{\|\mathbf a_j\|^2\|\mathbf b_j\|^2-(\mathbf a_j\cdot\mathbf b_j)^2}}
\right)
\end{aligned}
\end{equation}
Please refer to Appendix C.1 for more details on this formula.

We now proceed to the Dirichlet term $\phi_{\mathbf{v}_j}^{\mathbf{A}}(\eta)$, which is obtained by summing the contributions over the edges associated with $\mathbf{v}_i$. We focus on the integral over a single edge $f_j$, denoted as $\phi_{\mathbf{v}_j, f_j}^{\mathbf{A}}(\eta)$. Recall that  $\mathbf{x}=\mathbf{A}^{-1/2}\xi,\  \mathbf{y}=\mathbf{A}^{-1/2}\eta$, $\mathbf{v}^{\prime}_{j}=\mathbf{A}^{-1/2}\mathbf{v}_{j}$, $\mathbf{v}^{\prime}_{j+1}=\mathbf{A}^{-1/2}\mathbf{v}_{j+1}$, $f_j=\overrightarrow{\mathbf{v}_{j}\mathbf{v}_{j+1}}$ and $f_j^{\prime}=\overrightarrow{\mathbf{v}_{j}^{\prime}\mathbf{v}_{j+1}^{\prime}}$. Let $\Gamma_{\mathbf{v}_j,f_j}(\xi)$ be the hat-function of $\xi$ with respect to $\mathbf{v}_j$ in $f_j$, and $\Gamma_{\mathbf{v}_j^{\prime},f_j^{\prime}}(\mathbf{x})$ be the hat-function of $\mathbf{x}$ with respect to $\mathbf{v}_j^{\prime}$ in $f_j^{\prime}$. We first show that $\Gamma_{\mathbf{v}_j,f_j}(\xi_0)=\Gamma_{\mathbf{v}_j^{\prime},f_j^{\prime}}(\mathbf{x}_0)$ for $\xi_{0} \in f_j$ and $\mathbf{x}_0=\mathbf{A}^{-1/2}\xi_{0}$. Recall that $\Gamma_{\mathbf{v}_j,f_j}(\xi_0)$ is the barycentric coordinates of $\xi_0$ with respect to $\mathbf{v}_j$ in $f_j$, we obtain
\begin{equation}
\xi_0 = \Gamma_{\mathbf{v}_j,f_j}(\xi_0) \mathbf{v}_j+\Gamma_{\mathbf{v}_{j+1},f_j}(\xi_0) \mathbf{v}_{j+1}
\end{equation}
and
\begin{equation}
\Gamma_{\mathbf{v}_j,f_j}(\xi_0)+\Gamma_{\mathbf{v}_{j+1},f_j}(\xi_0)=1.
\end{equation}
Then, 
\begin{equation}
\begin{aligned}
\mathbf{x}_0 = &\mathbf{A}^{-1/2} \xi_0 =\Gamma_{\mathbf{v}_j,f_j}(\xi_0) \mathbf{A}^{-1/2}\mathbf{v}_j+\Gamma_{\mathbf{v}_{j+1},f_j}(\xi_0) \mathbf{A}^{-1/2}\mathbf{v}_{j+1} \\
=&\Gamma_{\mathbf{v}_j,f_j}(\xi_0) \mathbf{v}_j^{\prime}+\Gamma_{\mathbf{v}_{j+1},f_j}(\xi_0) \mathbf{v}_{j+1}^{\prime}.
\end{aligned}
\end{equation}
Therefore, $\Gamma_{\mathbf{v}_j,f_j}(\xi_0), \Gamma_{\mathbf{v}_{j+1},f_j}(\xi_0)$ and $\Gamma_{\mathbf{v}_j^{\prime},f_j^{\prime}}(\mathbf{x}_0), \Gamma_{\mathbf{v}_{j+1}^{\prime},f_j^{\prime}}(\mathbf{x}_0)$ are both barycentric coordinates of $\mathbf{x}_0$ in $f_j^{\prime}$. As established in Lemma A.7, the barycentric coordinates of a simplex are uniquely determined due to the affine independence of its vertices. This property distinguishes them from barycentric coordinates in arbitrary polygonal domains, where uniqueness is generally not guaranteed. Therefore, we conclude that $\Gamma_{\mathbf{v}_j,f_j}(\xi_0)=\Gamma_{\mathbf{v}_j^{\prime},f_j^{\prime}}(\mathbf{x}_0)$ for any $\xi_{0} \in f_j$ and $\mathbf{x}_0=\mathbf{A}^{-1/2}\xi_{0}$. Moreover, we have $\mathrm{d} \sigma_{\xi}=|\det \mathbf{A}^{1/2}|\|\mathbf{A}^{-1/2} \mathbf{n}_{j}^{\prime}\| \ \mathrm{d} \sigma_{\mathbf{x}}$, where $\mathbf{n}_{j}^{\prime}$ is the unit normal of $\overrightarrow{\mathbf{v}_{j}^{\prime}\mathbf{v}_{j+1}^{\prime}}$. Then, we can derive the closed-from solution of $\phi_{\mathbf{v}_j, f_j}^{\mathbf{A}}(\eta)$ as follows: 
{\footnotesize
\begin{equation}
\begin{aligned}
\label{eq:closed_form_ani_2D_phi}
\phi_{\mathbf{v}_j, f_j}^{\mathbf{A}}(\eta) =& \frac{1}{2\pi\sqrt{\det{\mathbf {A}}}}\int_{\xi \in f_j}{\Gamma_{\mathbf{v}_j,f_j}(\xi) \frac{(\xi - \eta) \cdot \mathbf{n}_{j}}{(\xi - \eta)^{\top} \mathbf{A}^{-1}(\xi - \eta)}} \ \mathrm{d} \sigma_{\xi} \\
=& \frac{|\det \mathbf{A}^{1/2}|\|\mathbf{A}^{-1/2}\mathbf{n}_{j}^{\prime}\|}{2\pi\sqrt{\det{\mathbf {A}}}}\int_{\mathbf{x} \in f_j^{\prime}}{\Gamma_{\mathbf{v}_j^{\prime},f_j^{\prime}}(\mathbf{x})\frac{\mathbf{A}^{1/2}(\mathbf{x} - \mathbf{y}) \cdot (\frac{\mathbf{A}^{-1/2} \mathbf{n}_{j}^{\prime}}{\|\mathbf{A}^{-1/2} \mathbf{n}_{j}^{\prime}\|})}{(\mathbf{x} - \mathbf{y})^{\top}(\mathbf{x} - \mathbf{y})}} \ \mathrm{d}\mathbf{x} \\
=& \frac{1}{2\pi}\int_{\mathbf{x} \in f_j^{\prime}}{\Gamma_{\mathbf{v}_j^{\prime},f_j^{\prime}}(\mathbf{x}) \frac{(\mathbf{x} - \mathbf{y}) \cdot \mathbf{n}_{j}^{\prime}}{(\mathbf{x} - \mathbf{y})^{\top}(\mathbf{x} - \mathbf{y})}  \ \mathrm{d}\mathbf{x} }.
\end{aligned}
\end{equation}
}
Denote $\mathbf{a}_j=\mathbf{v}_{j+1}^{\prime}-\mathbf{v}_{j}^{\prime}$ and $\mathbf{b}_j=\mathbf{v}_{j}^{\prime} - \mathbf{y}$. The term 
\begin{equation}
\int_{\mathbf{x} \in f_j^{\prime}}{\Gamma_{\mathbf{v}_j^{\prime},f_j^{\prime}}(\mathbf{x}) \frac{(\mathbf{x} - \mathbf{y}) \cdot \mathbf{n}_{j}^{\prime}}{(\mathbf{x} - \mathbf{y})^{\top}(\mathbf{x} - \mathbf{y})}  \ \mathrm{d}\mathbf{x} }
\end{equation}
is equal to $-\frac{\|\mathbf{a}_{j}\|(\mathbf{b}_{j}\cdot \mathbf{n}_{j}^{\prime})}{2\pi}(W(1)-W(0))$, where
{\scriptsize
\begin{equation}
\begin{aligned}
W(t)=&
\frac{1}{2\|\mathbf{a}_{j}\|^{2}}\log(\|\mathbf{a}_{j}\|^{2}t^{2}+2t(\mathbf{a}_{j}\cdot \mathbf{b}_{j})+\|\mathbf{b}_{j}\|^{2})\\
-&\left(\frac{\|\mathbf{a}_{j}\|^{2}+(\mathbf{a}_{j} \cdot \mathbf{b}_{j})}{\|\mathbf{a}_{j}\|^{2} \sqrt{\|\mathbf{a}_{j}\|^{2}\|\mathbf{b}_{j}\|^{2}-(\mathbf{a}_{j} \cdot \mathbf{b}_{j})^2}}\right) \arctan\left(\frac{\|\mathbf{a}_{j}\|^{2}t+(\mathbf{a}_{j} \cdot \mathbf{b}_{j})}{ \sqrt{\|\mathbf{a}_{j}\|^{2}\|\mathbf{b}_{j}\|^{2}-(\mathbf{a}_{j} \cdot \mathbf{b}_{j})^2}}\right).
\end{aligned}
\end{equation}
}

\subsection{Closed-form Expressions for 3D Scenario}~\label{sec:4_2}
\begin{figure}[htb]
  \centering
  \includegraphics[width=1.0\linewidth]{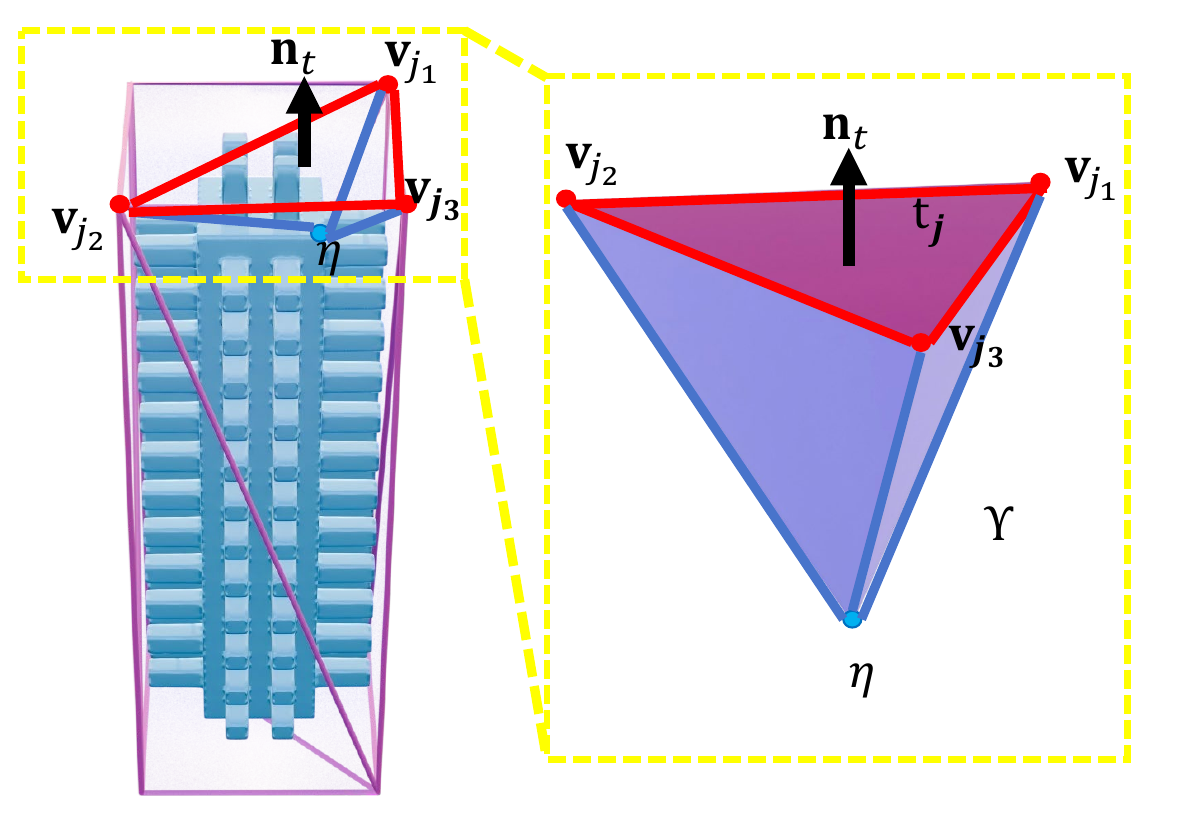}
  \caption{\label{fig:Figure2}
          A tetrahedron formed by a triangular face $t_j=\triangle \mathbf{v}_{j_1}\mathbf{v}_{j_2}\mathbf{v}_{j_3}$ of a cage and the query point $\eta$. }
\end{figure}

We can employ a similar methodology to derive the closed-form expression for the 3D case. Figure \ref{fig:Figure2} depicts a tetrahedron formed by a triangular face $t_j=\triangle\mathbf{v}_{j_1}\mathbf{v}_{j_2}\mathbf{v}_{j_3}$ of the cage and an interior point $\eta$ within the cage. Let $\mathbf{x}=\mathbf{A}^{-1/2}\xi,\  \mathbf{y}=\mathbf{A}^{-1/2}\eta$, $\mathbf{v}^{\prime}_{j_1}=\mathbf{A}^{-1/2}\mathbf{v}_{j_1}$, $\mathbf{v}^{\prime}_{j_2}=\mathbf{A}^{-1/2}\mathbf{v}_{j_2}$ and $\mathbf{v}^{\prime}_{j_3}=\mathbf{A}^{-1/2}\mathbf{v}_{j_3}$. According to Lemma A.6, we have $\mathrm{d} \sigma_{\mathbf{x}}=|\det \mathbf{A}^{-1/2}|\|\mathbf{A}^{1/2} \mathbf{n}_t\| \ \mathrm{d} \sigma_{\xi}$, where $\mathbf{n}_{t}$ denotes the unit outward normal vector of $t_j$. Furthermore, denoting $t_j^{\prime}=\triangle\mathbf{v}_{j_1}^{\prime}\mathbf{v}_{j_2}^{\prime}\mathbf{v}_{j_3}^{\prime}$, and its normal as $\mathbf{n}_{t}^{\prime}$. Then, the Neumann term $\psi_{j}^{\mathbf{A}}(\eta)$ on $t_j$ can be computed as follows:
\begin{equation}
\begin{aligned}
\label{eq:closed_form_ani_3D_psi}
\psi_{j}^{\mathbf{A}}(\eta) =& \int_{\xi \in t_j}{\frac{1}{4\pi\sqrt{\det{\mathbf {A}}}}
\frac{1}{[(\xi-\eta)^{\top} \mathbf{A}^{-1}(\xi-\eta)]^{1/2}}} \ \mathrm{d}\sigma_{\xi}\\
= &\frac{1}{|\det{\mathbf{A}^{-1/2}}|\|\mathbf{A}^{1/2}\mathbf{n}_{t}\|}\int_{\mathbf{x} \in t_j^{\prime}}{\frac{1}{4\pi\sqrt{\det{\mathbf {A}}}} \frac{1}{\|\mathbf{x}-\mathbf{y}\|}}  \ \mathrm{d} \sigma_{\mathbf{x}} \\
=& \frac{1}{\sqrt{\mathbf{n}_{t}^{\top}\mathbf{A}\mathbf{n}_{t}}}\int_{\mathbf{x} \in t_j^{\prime}}{\frac{1}{4\pi} \frac{1}{\|\mathbf{x}-\mathbf{y}\|}} \ \mathrm{d} \sigma_{\mathbf{x}}.
\end{aligned}
\end{equation}

For the Dirichlet term, we denote by $\phi_{\mathbf{v}_{j_1}, t_j}^{\mathbf{A}}(\eta)$ the contribution of $t_j$ to $\phi_{\mathbf{v}_{j_1}}^{\mathbf{A}}(\eta)$. According to Lemma A.6, we have $\mathrm{d} \sigma_{\xi}=|\det \mathbf{A}^{1/2}|\|\mathbf{A}^{-1/2} \mathbf{n}_{t}^{\prime}\| \ \mathrm{d} \sigma_{\mathbf{x}}$. Based on the uniqueness of the barycentric coordinates of a simplex (Lemma A.7), it holds that $\Gamma_{\mathbf{v}_{j_1},t_j}(\xi_0)=\Gamma_{\mathbf{v}_{j_1}^{\prime},t_j^{\prime}}(\mathbf{x}_0)$ for $\xi_0 \in t_j$ and $\mathbf{x}_0=\mathbf{A}^{-1/2}\xi_{0}$. Therefore,
{\footnotesize
\begin{equation}
\begin{aligned}
\phi_{\mathbf{v}_{j_1}, t_j}^{\mathbf{A}}(\eta) =& \frac{1}{4\pi\sqrt{\det{\mathbf {A}}}}\int_{\xi \in t_j}{\Gamma_{\mathbf{v}_{j_1},t_j}(\xi) \frac{(\xi - \eta) \cdot \mathbf{n}_{t}}{[(\xi - \eta)^{\top} \mathbf{A}^{-1}(\xi - \eta)]^{\frac{3}{2}}}} \ \mathrm{d} \sigma_{\xi} \\
=& \frac{|\det \mathbf{A}^{1/2}|\|\mathbf{A}^{-1/2}\mathbf{n}_{t}^{\prime}\|}{4\pi\sqrt{\det{\mathbf {A}}}}\int_{\mathbf{x} \in t_j^{\prime}}{\Gamma_{\mathbf{v}_{j_1}^{\prime},t_j^{\prime}}(\mathbf{x})\frac{\mathbf{A}^{1/2}(\mathbf{x} - \mathbf{y}) \cdot (\frac{\mathbf{A}^{-1/2} \mathbf{n}_{t}^{\prime}}{\|\mathbf{A}^{-1/2} \mathbf{n}_{t}^{\prime}\|})}{[(\mathbf{x} - \mathbf{y})^{\top}(\mathbf{x} - \mathbf{y})]^{\frac{3}{2}}}} \ \mathrm{d}\mathbf{x} \\
=& \frac{1}{4\pi}\int_{\mathbf{x} \in t_j^{\prime}}{\Gamma_{\mathbf{v}_{j_1}^{\prime},t_j^{\prime}}(\mathbf{x}) \frac{(\mathbf{x} - \mathbf{y}) \cdot \mathbf{n}_{t}^{\prime}}{[(\mathbf{x} - \mathbf{y})^{\top}(\mathbf{x} - \mathbf{y})]^{\frac{3}{2}}}  \ \mathrm{d}\mathbf{x} }.
\end{aligned}
\end{equation}
}
Then, we transform the computation of the anisotropic formulation into an isotropic one. The calculation of 3D isotropic GC is somewhat complex, and we provide the details in Appendix C of the supplementary material. These derivations are primarily based on the findings of ~\cite{2009VariationalGC} and ~\cite{Urago2000}, but we provide a more comprehensive derivation.

%% file: sections/sec5.tex
\section{Geometric Interpretation, Design Space, and Applications of AGC}~\label{sec:5}
In this section, we mainly derive the geometric interpretation of AGC in comparison to the isotropic counterparts, and demonstrate the quasi-conformality of our method. Furthermore, we emphasize the essential value of addressing problems from an anisotropic perspective. Additionally, we discuss the main applications of our method from both theoretical and experimental perspectives, with a particular focus on analyzing the influence of matrix $\mathbf{A}$ on the deformation results. The anisotropic matrix provides a new design space for artists. Based on its eigenvalues and eigenvectors, the sensitivity of cage faces to deformation can be analyzed.

We have the following observation when choosing the appropriate scale factor $s_j$: the two approaches below are equivalent: 
\begin{enumerate}[leftmargin=*]
\item Given an input, first apply a global multiplication by $\mathbf{A}^{-1/2}$ to the source cage, target cage, and the object; subsequently deform the object using the isotropic GC; and finally multiply globally by $\mathbf{A}^{1/2}$. 
\item Employ AGC for the source and target cages. 
\end{enumerate}
We will present the derivation of this equivalence subsequently. Despite the equivalence, we believe that there are still important points to illustrate the value of deriving the AGC. First, our method provides a geometric interpretation of Green’s identity derived from the anisotropic Laplace equation. Second, as deformation coordinates, they need to be expressed as functions relative to the original cage, rather than the cage has undergone a global transformation. Even if we know the equivalence in advance, expressing the coordinates as a closed-form expression of the source cage still requires the derivation presented in Section~\ref{sec:4}. Moreover, when performing the variational method for deformation control (which will be presented in Section~\ref{sec:6}), it is more intuitive to analyze directly from the closed-form anisotropic formulation as well as its gradient and Hessian. Moreover, analyzing the eigenvalues of $\mathbf{A}$ based on the anisotropic deformation formula demonstrates that the source cage face associated with the eigenvector with respect to the larger eigenvalue has a more pronounced effect on the target shape. The theoretical analysis of this will be provided at the end of this section. This facilitates a more targeted selection of the matrix. In addition, this study introduces a new perspective on cage-based and variational deformation. If the matrix $\mathbf{A}$ can be optimized, there is the potential to further enhance the practical value of the method, which remains a direction for our future work.

Next, we refer to the deformation effect obtained by ``globally multiplying by $\mathbf{A}^{-1/2}$, performing deformation using isotropic GC, and then globally multiplying by $\mathbf{A}^{1/2}$'', as \textit{similarity Green deformation}. The name stems from the fact that this process closely resembles a matrix similarity transformation. We denote the vertices and normals of the cage as $\{\mathbf{v}_i\}_{i=1}^{N}$ and $\{\mathbf{n}_j\}_{j=1}^{M}$. Moreover, we use the prime operator to denote the globally transformed values, e.g., $\mathbf{v}_i^{\prime}=\mathbf{A}^{-1/2}\mathbf{v}_i$, $\mathbf{n}_{j}^{\prime}=\frac{\mathbf{A}^{1/2}\mathbf{n}_{j}}{\|\mathbf{A}^{1/2}\mathbf{n}_{j}\|}$. Then, we can express the globally transformed values in terms of isotropic GC as:
\begin{equation}
\eta^{\prime}=\sum_{i \in I_{\mathbb{V}^{\prime}}}{\phi_{i}^{\prime}(\eta^{\prime}) \mathbf{v}}_{i}^{\prime} + \sum_{j \in I_{\mathbb{T}^{\prime}}}{\psi_{j}^{\prime}(\eta^{\prime}) \mathbf{n}_{j}^{\prime}}.
\end{equation}
The derivation of Section~\ref{sec:4_1} (Eqs.~\eqref{eq:closed_form_ani_2D_psi} and~\eqref{eq:closed_form_ani_2D_phi}) shows that $\phi_{i}^{\prime}(\eta^{\prime})=\phi_{i}^{\mathbf{A}}(\eta)$ and $\psi_{j}^{\prime}(\eta^{\prime})=\|\mathbf{A}^{1/2}\mathbf{n}_j\|\psi_{j}^{\mathbf{A}}(\eta)$, where $\mathbf{n}_j$ is the normal of $f_j$. Moreover, we have $\eta=\mathbf{A}^{1/2}\eta^{\prime}$. Then,
\begin{equation}
\begin{aligned}
\eta&=\mathbf{A}^{1/2}\eta^{\prime}=\mathbf{A}^{1/2}(\sum_{i \in I_{\mathbb{V}^{\prime}}}{\phi_{i}^{\prime}(\eta^{\prime}) \mathbf{v}}_{i}^{\prime} + \sum_{j \in I_{\mathbb{T}^{\prime}}}{\psi_{j}^{\prime}(\eta^{\prime}) \mathbf{n}_{j}^{\prime}}) \\
&=\mathbf{A}^{1/2}(\sum_{i \in I_{\mathbb{V}}}{\phi_{i}^{\mathbf{A}}(\eta) \mathbf{A}^{-1/2}\mathbf{v}}_{i} + \sum_{j \in I_{\mathbb{T}}}{\|\mathbf{A}^{1/2}\mathbf{n}_j\|\psi_{j}^{\mathbf{A}}(\eta) \frac{\mathbf{A}^{1/2}\mathbf{n}_{j}}{\|\mathbf{A}^{1/2}\mathbf{n}_j\|}}) \\
&=\sum_{i \in I_{\mathbb{V}}}{\phi_{i}^{\mathbf{A}}(\eta) \mathbf{v}}_{i} + \sum_{j \in I_{\mathbb{T}}}{\psi_{j}^{\mathbf{A}}(\eta) \mathbf{A}\mathbf{n}_{j}},
\end{aligned}
\end{equation}
and we show the equivalence between the AGC and the similarity Green deformation. It is worth noting that when the cage is deformed to the target configuration, $\|\mathbf{A}^{1/2}\mathbf{n}_j\|$ is generally not equal to $\|\mathbf{A}^{1/2}\tilde{\mathbf{n}}_j\|$. Nevertheless, this can be resolved by selecting a proper scaling factor. See Appendix D of the supplementary material for a detailed analysis.

\begin{figure}[htb]
  \centering
  \includegraphics[width=1.0\linewidth]{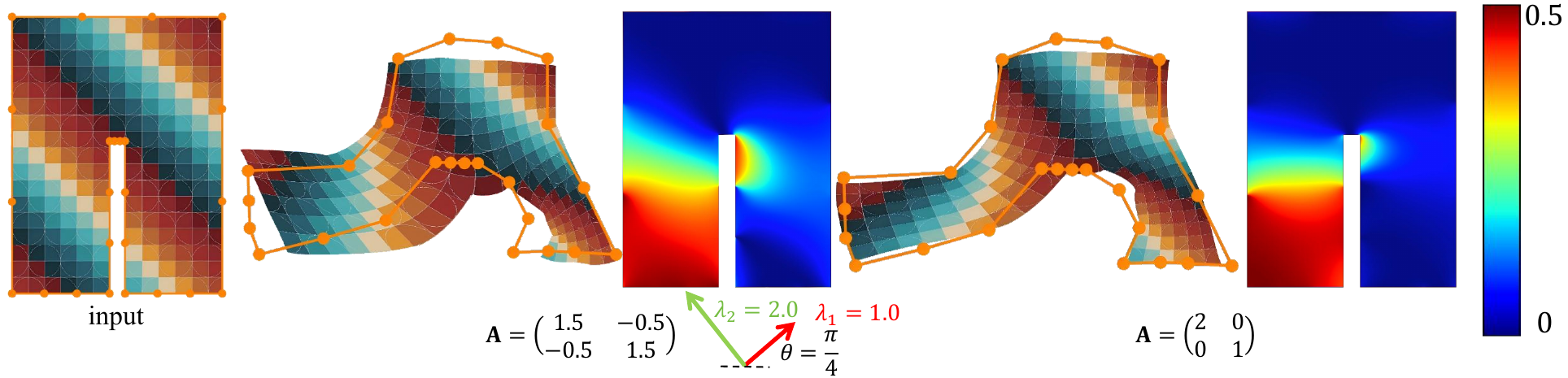}
  \caption{\label{fig:Figure24}
          Conformal distortion of our method with the anisotropic matrix where the ratio of the largest eigenvalue to the smallest eigenvalue is 2. $K_{\mathbf{A}}(\eta)=\sigma_{max}/\sigma_{min}\le 2$. The conformal distortion $\sigma_1/\sigma_2+\sigma_2/\sigma_1 -2 \le 0.5$.}
\end{figure}

Moreover, we can use the above analysis to demonstrate that our method can produce a quasi-conformal mapping. We denote the anisotropic deformation mapping as $f_{\mathbf{A}}(\eta)$, and the corresponding isotropic mapping as $f^{\prime}(\eta)$. Then,
\begin{equation}
f_{\mathbf{A}}(\eta)=\mathbf{A}^{1/2}f^{\prime}(\mathbf{A}^{-1/2}\eta).
\end{equation}
The Jacobian of $f_{\mathbf{A}}(\eta)$ is given by:
\begin{equation}
\mathbf{J}_{f_{\mathbf{A}}}(\eta)=\mathbf{A}^{1/2}\mathbf{J}_{f^\prime}(\mathbf{A}^{-1/2}\eta)\mathbf{A}^{-1/2},
\end{equation}
where $\mathbf{J}_{f^\prime}$ is the Jacobian of the isotropic formulation. Let $K_{\mathbf{A}}(\eta)$ denote the ratio of the maximum to the minimum singular values of the anisotropic deformation mapping, and $K^{\prime}(\eta)$ the corresponding isotropic ratio. Then,
\begin{equation}
\begin{aligned}
K_{\mathbf{A}}(\eta)=&\frac{\sigma_{max}(\mathbf{J}_{f_{\mathbf{A}}}(\eta))}{\sigma_{min}(\mathbf{J}_{f_{\mathbf{A}}}(\eta))}=\text{cond}(\mathbf{J}_{f_{\mathbf{A}}}(\eta))\\
&=\text{cond}(\mathbf{A}^{1/2}\mathbf{J}_{f^\prime}(\mathbf{A}^{-1/2}\eta)\mathbf{A}^{-1/2})\\
&\leq \text{cond}(\mathbf{A}^{1/2})\text{cond}(\mathbf{J}_{f^\prime}(\mathbf{A}^{-1/2}\eta))\text{cond}(\mathbf{A}^{-1/2})\\
&=\text{cond}(\mathbf{A})K^{\prime}(\mathbf{A}^{-1/2}\eta),
\end{aligned}
\end{equation}
where $\sigma_{max}(\mathbf{J}_{f_{\mathbf{A}}}(\eta))$ and $\sigma_{min}(\mathbf{J}_{f_{\mathbf{A}}}(\eta))$ represent the maximum and minimum singular values of $\mathbf{J}_{f_{\mathbf{A}}}(\eta)$, and $\text{cond}(\mathbf{A})$ represents the condition number of $\mathbf{A}$. Therefore, when GC are conformal in 2D, AGC are quasi-conformal with $K_{\mathbf{A}}(\eta)\leq \text{cond}(\mathbf{A})$. When GC are quasi-conformal in 3D with a distortion bound $M$, AGC are quasi-conformal in 3D with a bound $\text{cond}(\mathbf{A})\cdot M$. Moreover, since $\mathbf{A}=\mathbf{P}\Lambda \mathbf{P}^{\top}$, the condition number $\text{cond}(\mathbf{A})$ is precisely the ratio of the largest to the smallest eigenvalues of $\Lambda$. Fig.~\ref{fig:Figure24} illustrates the conformal distortion of our method using an anisotropic matrix where the ratio of the largest eigenvalue to the smallest eigenvalue is $2$. Therefore, $K_{\mathbf{A}}(\eta)=\sigma_{max}/\sigma_{min}\le \mathrm{cond}(\mathbf{A})=2$. The conformal distortion is calculated as $\sigma_1/\sigma_2+\sigma_2/\sigma_1 -2 \le 0.5$, which is also confirmed by the distortion map.

\begin{figure}[htb]
  \centering
  \includegraphics[width=1.0\linewidth]{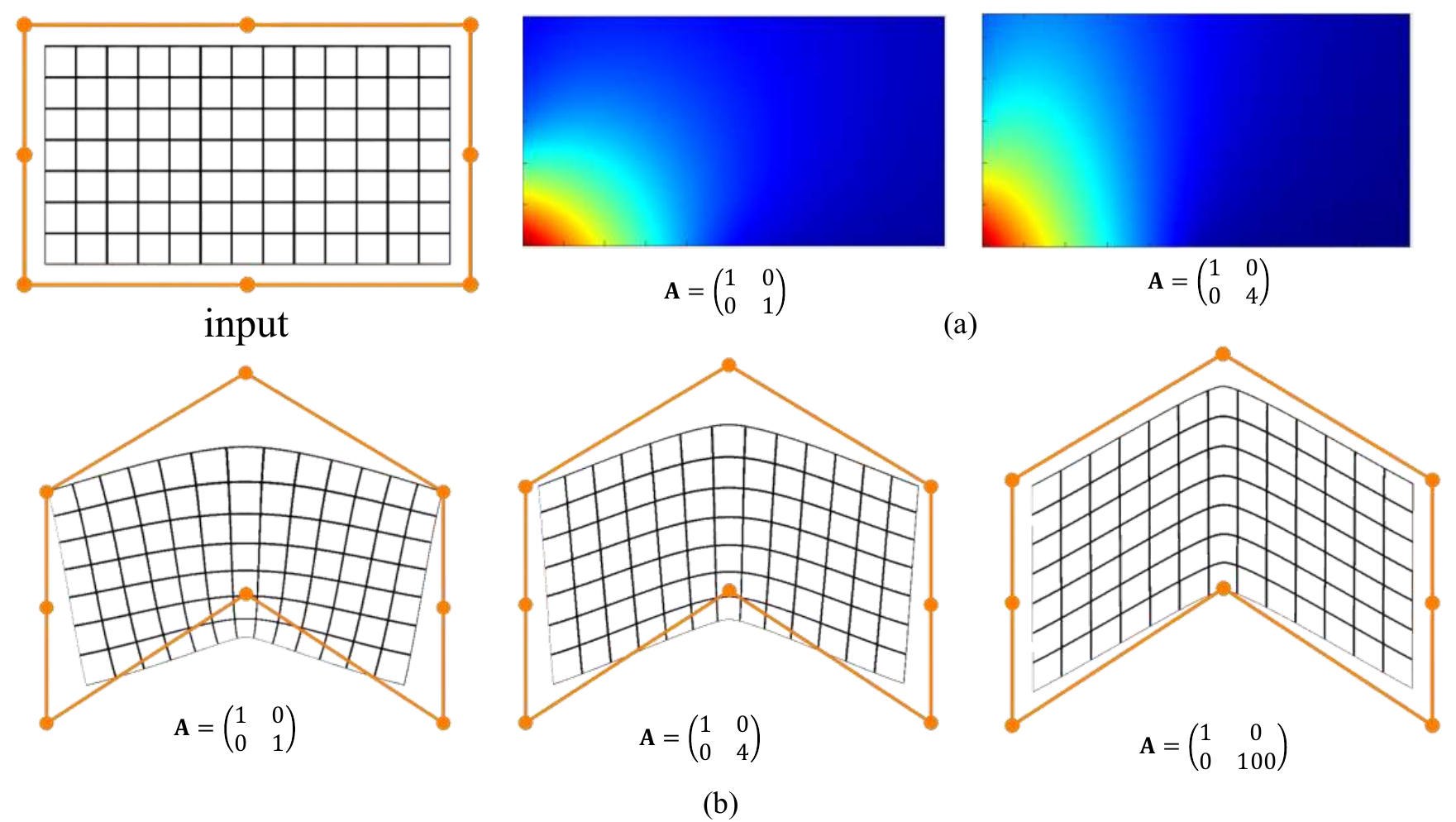}
  \caption{\label{fig:Figure20}
          Illustration of the effect of different anisotropic matrices on deformation. (a) Effect of different anisotropic matrices on the Dirichlet term $\phi_{i}^{\mathbf{A}}(\eta)$. (b) Deformation effect of moving the cage edges whose normals are aligned with the eigenvector of $\lambda_k$.}
\end{figure}

To further illustrate the practical applications of our method, we provide an analysis of the deformation effects resulting from different matrices $\mathbf{A}$, aimed at assisting artists in selecting the optimal matrix and designing the deformation cage. The matrix serves as a new design space, which mainly influences the shape (e.g., conformal or shearing) of the deformation. In the following, we analyze the influence of $\mathbf{A}$ on the deformation based on Eq.~\eqref{eq:anisotropic_Green_coord}. The Dirichlet term $\phi_{i}^{\mathbf{A}}(\eta)$ is primarily related to vertex coordinates and thus to the shape’s position, whereas the Neumann term $\psi_{j}^{\mathbf{A}}(\eta)$ is more associated with the cage normal and is more sensitive to local shape control. In Fig.~\ref{fig:Figure20}(a), we visualize the $\phi_{i}^{\mathbf{A}}(\eta)$ value for the bottom-left corner vertex. When the eigenvalue in the $y$-direction is larger, the influence of $\phi_{i}^{\mathbf{A}}(\eta)$ on this direction is relatively more significant, thereby increasing the influence of this vertex in the $y$-direction.

\begin{figure}[htb]
  \centering
  \includegraphics[width=1.0\linewidth]{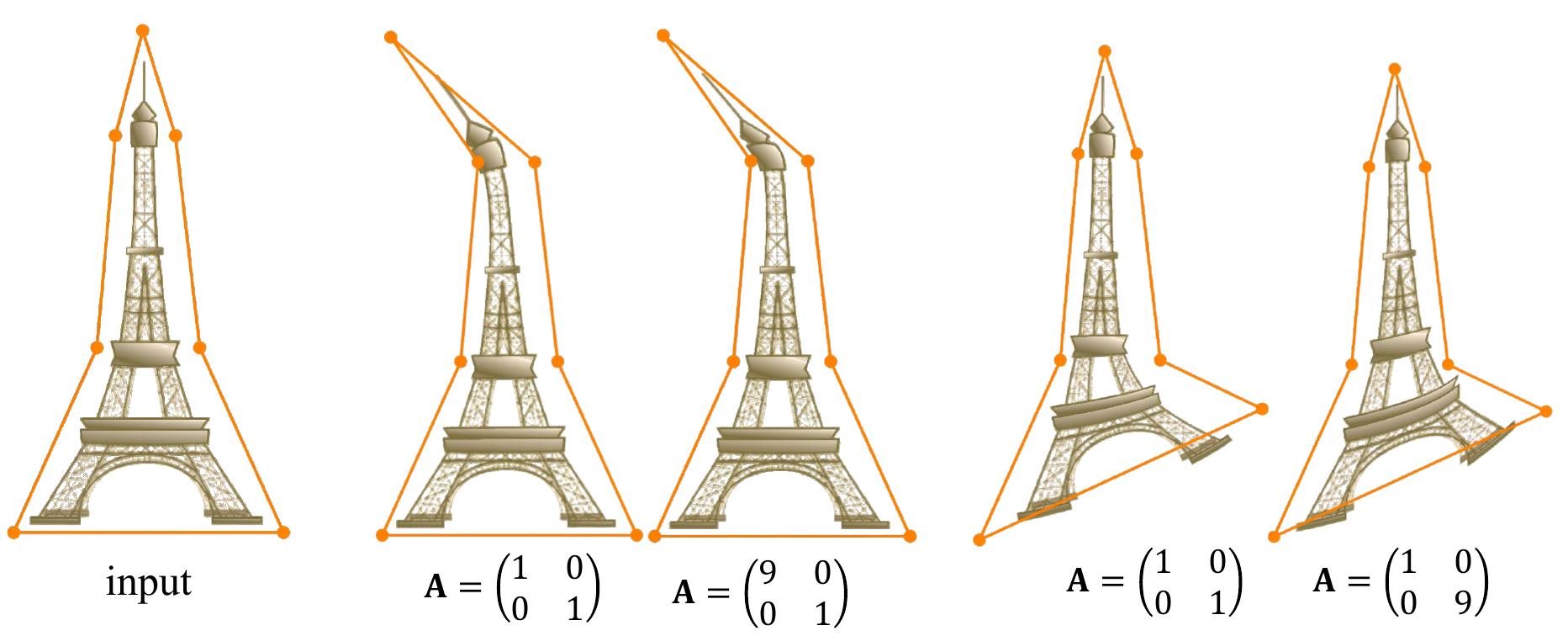}
  \caption{\label{fig:Figure21}
         Deformation result of a tower model. The deformation effect is more pronounced when the normal of a cage face aligns with the eigenvector corresponding to the larger eigenvalue of  $\mathbf{A}$.}
\end{figure}

Regarding the Neumann term, which is expressed as ${\psi_{j}^{\mathbf{A}}(\eta) \ \mathbf{A}\mathbf{n}_j}$, we can analyze it from a mathematical perspective. Based on the derivations in Section~\ref{sec:4_1}, we have
\begin{equation}
\psi_{j}^{\mathbf{A}}(\eta) \mathbf{A}\mathbf{n}_j=\frac{1}{\sqrt{\mathbf{n}_j^{\top}\mathbf{A}\mathbf{n}_j}} \psi^{\prime}_j(\mathbf{A}^{-1/2}\eta) \mathbf{A}\mathbf{n}_j,
\end{equation}
where $\psi^{\prime}_j(\mathbf{A}^{-1/2}\eta)$ is the isotropic coordinates at $\mathbf{A}^{-1/2}\eta$. If $\mathbf{n}_j$ is the eigenvector corresponding to the largest eigenvalue $\lambda_k$ of $\mathbf{A}$, then
\begin{equation}
\sqrt{\mathbf{n}_j^{\top}\mathbf{A}\mathbf{n}_j}=\sqrt{\lambda_k \mathbf{n}_j^{\top}\mathbf{n}_j}=\sqrt{\lambda_k}, \qquad \mathbf{A}\mathbf{n}_j=\lambda_k \mathbf{n}_j.
\end{equation}
Therefore, we obtain
\begin{equation}
\psi_{j}^{\mathbf{A}}(\eta) \mathbf{A}\mathbf{n}_j=\frac{1}{\sqrt{\lambda_k}}\psi^{\prime}_j(\mathbf{A}^{-1/2}\eta)\lambda_k \mathbf{n}_j=\sqrt{\lambda_k}\psi^{\prime}_j(\mathbf{A}^{-1/2}\eta)\mathbf{n}_j.
\end{equation}
It can be seen that as $\lambda_k$ increases, if the normal of a source edge is more consistent with the eigenvector corresponding to the eigenvalue $\lambda_k$ of $\mathbf{A}$, then the influence of the variation of this edge normal on the deformation result becomes more significant, with the magnitude being proportional to $\sqrt{\lambda_k}$. This phenomenon is also illustrated in Fig.~\ref{fig:Figure20}(b), where we set
$\mathbf{A}=\begin{pmatrix}
    1 & 0 \\
    0 & \lambda_k,
\end{pmatrix}$
and the corresponding eigenvector associated with $\lambda_k$ is $(0,1)^{\top}$. For $\lambda_k=4$ and more extreme values such as $\lambda_k=100$, the influence of cage edges with normals pointing in the $y$-direction is more pronounced (For all two-dimensional examples, the vertical direction is defined as the $y$-direction). Fig.~\ref{fig:Figure21} illustrates the results of moving the `horizontal' and `vertical' cage edges in a tower model, respectively. It provides further evidence that the deformation effect is more pronounced when the normal of a cage face aligns with the eigenvector corresponding to the larger eigenvalue of $\mathbf{A}$.

%% file: sections/sec6.tex
\section{Variational Shape Deformation of AGC} \label{sec:6}
In this section, we derive the expressions for the derivatives of AGC, specifically the gradients and Hessians of $\phi_{i}^{\mathbf{A}}(\eta)$ and $\psi_{j}^{\mathbf{A}}(\eta)$. These derivatives form the foundation for deformation control in the variational shape deformation framework introduced in~\citet{2009VariationalGC}. In this method, an energy functional is defined to determine the optimal deformation mapping that satisfies user-specified constraints while simultaneously balancing smoothness and ARAP transformations. Previous work also demonstrates that the introduction of anisotropy does not conflict with our desire for rigid transformations at specific positions~\cite{2024AnisotropicFiniteElement}.

For a simplicial surface cage with vertices $\mathbb{V}$ and faces $\mathbb{T}$, we can define the deformation map $f_{\mathbf{a},\mathbf{b}}(\eta)$ as a linear combination of basis functions $\phi_{i}^{\mathbf{A}}(\eta)$ and $\psi_{j}^{\mathbf{A}}(\eta)$ based on AGC as
\begin{equation}
\label{eq:ani_Green_eq_variational}
f_{\mathbf{a},\mathbf{b}}(\eta)=\sum_{v \in \mathbb{V}} \mathbf{a}_v \phi_{v}^{\mathbf{A}}(\eta)+\sum_{t \in \mathbb{T}} \mathbf{b}_t \psi_{t}^{\mathbf{A}}(\eta),
\end{equation}
where $\mathbf{b}_t=\mathbf{A}\mathbf{n}_t$. Here, $\mathbf{a}_v \in \mathbb{R}^d$ represents the piecewise-linear coefficients defined on the cage vertices, while $\mathbf{b}_t \in \mathbb{R}^d$ denotes the piecewise-constant coefficients defined on the cage faces. To enlarge the deformation space, $\mathbf{a}_v$ and $\mathbf{b}_t$ are treated as independent variables and optimized separately. Since $\mathbf{A}$ is invertible, we can directly optimize $\mathbf{b}_t$, as $\mathbf{n}_t$ can be easily computed through $\mathbf{n}_t = \mathbf{A}^{-1}\mathbf{b}_t$.

To perform variational shape deformation, it is necessary to compute the gradients and Hessians of $\phi_{i}^{\mathbf{A}}(\eta)$ and $\psi_{j}^{\mathbf{A}}(\eta)$. The core idea is to apply Lemma A.1 from the supplementary material to transform the formulation into an isotropic form. For the computation of the isotropic formulation, please refer to Appendix C.

Let $t_j$ be a face of the cage and $t_j^{\prime}$ its transformation via left-multiplication of vertices by $\mathbf{A}^{-1/2}$. For any point $\eta_0 \in \Omega$, let $\mathbf{y}_0=\mathbf{A}^{-1/2}\eta_0$. Moreover, let $\psi_{t_j}^{\mathbf{A}}(\eta_0)$ denote the Neumann term of $t_j$ for AGC at $\eta_0$, while $\psi_{t_j^{\prime}}(\mathbf{y}_0)$ represents the Neumann term of $t_j^{\prime}$ for the isotropic formulation at $\mathbf{y}_0$. Based on the derivations in Section~\ref{sec:4_1}, we obtain:
\begin{equation}
\label{eq:ani_psi_relation}
\psi_{t_j}^{\mathbf{A}}(\eta_0) = \frac{1}{\sqrt{\mathbf{n}_{j}^{\top}\mathbf{A}\mathbf{n}_{j}}}\psi_{t_j^{\prime}}(\mathbf{y}_0),
\end{equation}
Therefore, according to Lemma A.1, the gradient $\nabla_{\eta}\psi_{t_j}^{\mathbf{A}}(\eta_0)$ and Hessian $\mathbf{H}_{\eta}\psi_{t_j}^{\mathbf{A}}(\eta_0)$ can be calculated as:
\begin{equation}
\begin{gathered}
\nabla_{\eta}\psi_{t_j}^{\mathbf{A}}(\eta_0)=\frac{1}{\sqrt{\mathbf{n}_{j}^{\top}\mathbf{A}\mathbf{n}_{j}}} \mathbf{A}^{-1/2} \nabla_{\mathbf{y}}\psi_{t_j^{\prime}}(\mathbf{y}_0),  \\
\mathbf{H}_{\eta}\psi_{t_j}^{\mathbf{A}}(\eta_0)=\frac{1}{\sqrt{\mathbf{n}_{j}^{\top}\mathbf{A}\mathbf{n}_{j}}} \mathbf{A}^{-1/2} \mathbf{H}_{\mathbf{y}}\psi_{t_j^{\prime}}(\mathbf{y}_0)\mathbf{A}^{-1/2}.
\end{gathered}
\end{equation}
For the Dirichlet term, we denote $\phi_{i, t_j}^{\mathbf{A}}(\eta_0)$ as the contribution of the integration over $t_j$ to $\phi_{i}^{\mathbf{A}}(\eta_0)$. Based on the derivations in Section~\ref{sec:4_1}, we have:
\begin{align}
\phi_{i, t_j}^{\mathbf{A}}(\eta_0) =\phi_{i, t_j^{\prime}}(\mathbf{y}_0).
\end{align}
Then, the gradient $\nabla_{\eta}\phi_{i, t_j}^{\mathbf{A}}(\eta_0)$ and Hessian $\mathbf{H}_{\eta}\phi_{i, t_j}^{\mathbf{A}}(\eta)$ can be calculated as:
\begin{equation}
\begin{gathered}
\nabla_{\eta}\phi_{i, t_j}^{\mathbf{A}}(\eta_0)=\mathbf{A}^{-1/2} \nabla_{\mathbf{y}}\phi_{i, t_j^{\prime}}(\mathbf{y}_0)\\
\ \mathbf{H}_{\eta} \phi_{i, t_j}^{\mathbf{A}}(\eta_0)=\mathbf{A}^{-1/2} \mathbf{H}_{\mathbf{y}}\phi_{i, t_j^{\prime}}(\mathbf{y}_0)\mathbf{A}^{-1/2}.
\end{gathered}
\end{equation}
Thus, the computation of the derivatives of AGC is transformed into the isotropic formulation.

We now formulate the objective function for variational deformation. We express the deformation mapping  $f_{\mathbf{a},\mathbf{b}}(\eta)$ in matrix form as:
\begin{equation}
f_{\mathbf{a},\mathbf{b}}(\eta)_{1\times d}=\begin{pmatrix}\Phi_{1\times n}&\Psi_{1\times m}\end{pmatrix}\begin{pmatrix}\mathbf{a}_{n\times d}\\\mathbf{b}_{m\times d}\end{pmatrix},
\end{equation}
where $n = |\mathbb{V}|$ denotes the number of vertices, and $m = |\mathbb{T}|$ represents the number of faces.  The vector $\Phi$ is a row vector composed of all Dirichlet terms $\phi_{v}^{\mathbf{A}}(\eta)$, and $\Psi$ is a row vector composed of all Neumann terms $\psi_{t}^{\mathbf{A}}(\eta)$. 

The Jacobian matrix $\mathbf{J}_f(\eta)$ and the Hessian $\mathbf{H}_f(\eta)$ of the deformation mapping $f_{\mathbf{a},\mathbf{b}}(\eta)$ are also linear in $\mathbf{a}$ and $\mathbf{b}$, and are expressed as follows:
\begin{equation}
(\mathbf{J}_f(\eta))_{d \times d}^{\top}=\begin{pmatrix}(\mathbf{G}_{\Phi})_{d \times n}&(\mathbf{G}_{\Psi})_{d \times m}\end{pmatrix}\begin{pmatrix}\mathbf{a}_{n\times d}\\\mathbf{b}_{m\times d}\end{pmatrix},
\end{equation}
where each column of $\mathbf{G}_{\Phi}$ captures the gradients of $\phi_{v}^{\mathbf{A}}(\eta)$, and each column of $\mathbf{G}_{\Psi}$ records the gradients of $\psi_{t}^{\mathbf{A}}(\eta)$. Moreover,
{\small
\begin{equation}
(\mathbf{H}_f(\eta))_{[d(d+1)/2]\times d}=\begin{pmatrix}(\mathbf{H}_{\Phi})_{[d(d+1)/2]\times n}&(\mathbf{H}_{\Psi})_{[d(d+1)/2]\times m}\end{pmatrix}\begin{pmatrix}\mathbf{a}_{n\times d}\\\mathbf{b}_{m\times d}\end{pmatrix},
\end{equation}
}
where each column of $\mathbf{H}_{\Phi}$ records the elements of the Hessian matrix of $\phi_{v}^{\mathbf{A}}(\eta)$, comprising $d(d+1)/2$ values. This vector is derived by flattening the symmetric Hessian matrix, retaining the diagonal elements and the upper-right components. Similarly, $\mathbf{H}_{\Psi}$ contains the elements of the Hessian matrix of $\psi_{t}^{\mathbf{A}}(\eta)$.  

Based on the above formulation, we can now define an optimization framework. The energy functional $E(f_{\mathbf{a},\mathbf{b}})$ is designed to enforce ARAP behavior at specific points (typically medial axis points), ensure global smoothness (by minimizing the Hessian magnitude), and satisfy user constraints at designated points (i.e., deforming specific points to the user-defined positions). For discrete sampling, this problem is formulated as:
\begin{equation}
\label{eq:variational_opt_for}
\begin{gathered}
\min_{\mathbf{a},\mathbf{b},\mathbf{R}_i} E(f_{\mathbf{a},\mathbf{b}})=\sum_{i = 1}^p\left\|\mathbf{J}_f(\mathbf{m}_i)-\mathbf{R}_i\right\|_F^2+\lambda_1\sum_{i = 1}^r\|f_{\mathbf{a},\mathbf{b}}(\mathbf{q}_i)-\mathbf{f}_i\|_{2}^{2} \\
+\lambda_2\sum_{i = 1}^k\left\|\mathbf{H}_f(\mathbf{w}_i)\right\|_F^2 ,\\
\mathrm{s.t.}, \mathbf{R}_i^{\top} \mathbf{R}_i=\mathbf{I}, \forall i = 1,\dots,p.
\end{gathered}
\end{equation}
Unlike the earlier work \citep{2009VariationalGC}, which imposes hard
positional constraints, this formulation follows \citep{2025VariationalBiharmonic}
and uses soft positional constraints of the form $\|f_{\mathbf{a},\mathbf{b}}(\mathbf{q}_i)-\mathbf{f}_i\|_{2}^{2}$. Further details regarding parameter and sampling point selection will be provided in the experiment section. We apply the local/global strategy~\cite{2007ARAP, 2008LocalGlobal, 2009VariationalGC} to solve Eq.~\eqref{eq:variational_opt_for}. Specifically, in the local step, we fix $\mathbf{a}$ and $\mathbf{b}$ to solve for the optimal rotation matrices $\mathbf{R}_i$. This step formulates a well-known orthogonal Procrustes problem, which can be efficiently solved via singular value decomposition. In the global step, we fix $\mathbf{R}_i$ to solve for  $\mathbf{a}$ and $\mathbf{b}$, resulting in an unconstrained least-squares problem that can be directly reduced to a linear system. To avoid optimization instability, we also incorporate a Tikhonov regularization term in each global step, defined as the L2-norm of the difference between the current values ($\mathbf{a}^{(n)}$, $\mathbf{b}^{(n)}$) and those obtained in the previous iteration ($\mathbf{a}^{(n-1)}$, $\mathbf{b}^{(n-1)}$).

%% file: sections/sec7.tex
\section{Experiments}
This section presents the experimental results of our proposed method. We first focus on cage-based deformation and demonstrate the results obtained with different coefficient matrices $\mathbf{A}$. Subsequently, we showcase variational deformation outcomes subject to the same constraints using varying $\mathbf{A}$. All experiments are on a standard laptop equipped with an Intel CPU running at 2.40 GHz and 16GB of RAM.

We begin by describing how to obtain diverse matrices $\mathbf{A}$ for our experiments, focusing initially on the 2D scenario. By the spectral theorem, $\mathbf{A}$ can be decomposed into a diagonal matrix $\Lambda=\text{diag}(\lambda_1, \lambda_2)$ with $\lambda_1,\lambda_2>0$ and an orthogonal matrix $\mathbf{P}$ such that $\mathbf{A}=\mathbf{P}\Lambda\mathbf{P}^{\top}$. Moreover, since flipping the sign of an eigenvector leaves $\mathbf{P}\Lambda\mathbf{P}^{\top}$ unchanged, we may choose $\mathbf{P} \in SO(2)$, i.e., a rotation matrix. Therefore, we parameterize
$\mathbf{P}=\begin{pmatrix}
    \cos{\theta} & -\sin{\theta} \\
    \sin{\theta} & \cos{\theta}
\end{pmatrix}$ and denote $\mathbf{A}=\mathcal{A}(\theta, \lambda_1, \lambda_2)$. Moreover, since for any $c \in \mathbb{R}^{+}$, the equation $\nabla \cdot (\mathbf{A}\nabla u)=0$ is equivalent to $\nabla \cdot (c\mathbf{A}\nabla u)=0$ and leads to the same deformation mapping, we can normalize one of the diagonal entries of the matrix to $1.0$ in our formulation. Appendix E provides some anisotropic matrices used in 2D deformation experiments together with the $\lambda_1, \lambda_2, \theta$ values.

The construction of 3D matrices follows a similar approach. The 3D diagonal matrix $\Lambda$ is defined by three positive parameters $\lambda_1, \lambda_2$ and $\lambda_3$. The 3D rotation matrix $\mathbf{P}$ is parameterized by three Euler angles $\alpha, \beta, \gamma$ and can be expressed as $\mathbf{P}=\mathbf{R}_z(\alpha)\mathbf{R}_y(\beta)\mathbf{R}_x(\gamma)$, where
\begin{equation}
\begin{aligned}
& \mathbf{R}_z(\alpha)=\begin{pmatrix}
    \cos{\alpha} & -\sin{\alpha} & 0 \\
    \sin{\alpha} & \cos{\alpha} & 0 \\
    0 & 0 & 1
\end{pmatrix}, \ \mathbf{R}_y(\beta)=\begin{pmatrix}
    \cos{\beta} & 0 & -\sin{\beta} \\
     0 & 1 & 0 \\
    \sin{\beta} & 0 & \cos{\beta} 
\end{pmatrix}, \\
& \mathbf{R}_x(\gamma)=\begin{pmatrix}
    1 & 0 & 0 \\
     0 & \cos{\gamma} & -\sin{\gamma} \\
    0 & \sin{\gamma} & \cos{\gamma} 
\end{pmatrix}.
\end{aligned}
\end{equation}
We can denote $\mathbf{A}=\mathcal{A}(\lambda_1, \lambda_2, \lambda_3, \alpha, \beta, \gamma)$ for the 3D scenario. Therefore, in practical applications, artists do not need to design a precise matrix; instead, they can utilize tools such as sliders to adjust these angles and eigenvalues.

\subsection{Cage-based Deformation}
First, we present the results of cage-based deformation using AGC with various control matrices. When $\mathbf{A}=\mathbf{I}$, our results align with those of the traditional isotropic GC~\cite{2008Green}. We start with several representative examples to demonstrate the effects generated by different matrices $\mathbf{A}$, highlighting the diversity of deformation styles and options offered by our method. We will also analyze the design and selection of $\mathbf{A}$ based on its eigenvalues. Subsequently, we will compare our method with Mean Value Coordinates (MVC)~\citep{2003MVCPoly}, Harmonic Coordinates (HC)~\citep{2007Harmonic}, Biharmonic Coordinates (BiC)~\citep{2012Biharmonic} and Somigliana Coordinates (SC)~\citep{2023Somigliana}.
\begin{figure}[htb]
  \centering
  \includegraphics[width=1.0\linewidth]{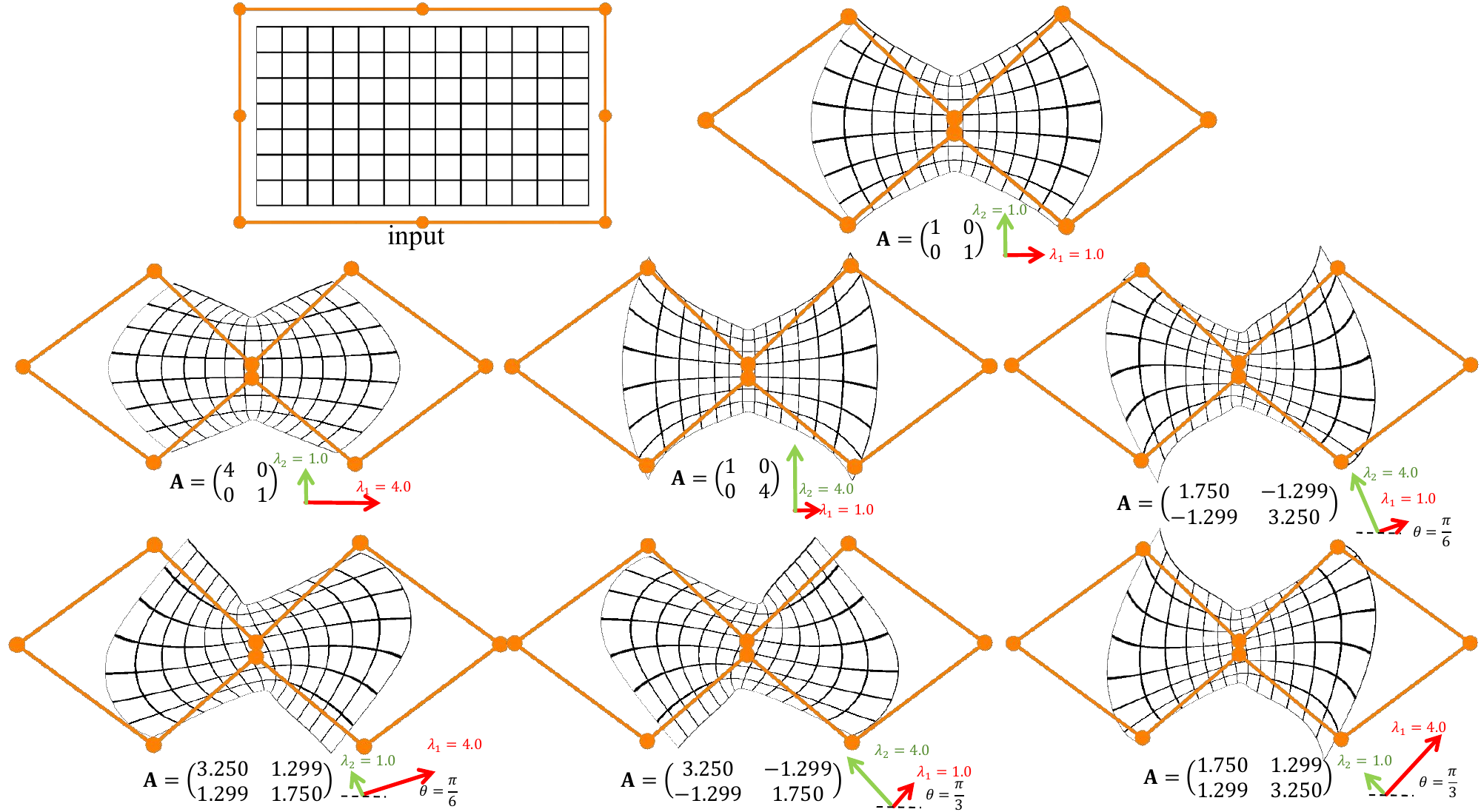}
  \caption{\label{fig:Figure17}
          Cage-based deformation of AGC in 2D. Different rotation matrices lead to varying ``stretching'' effects along different directions.}
\end{figure}

We first conduct experiments on 2D images composed of square and circular shapes. The vertical direction is defined as the $y$-direction. In Fig.~\ref{fig:Figure17}, we present the deformation results of a grid-like image. The eigenvalues of the matrix are $1$ and $4$, while the rotation angles are $\pi/6$ and $\pi/3$. Detailed parameter settings can be found in the table of Appendix E. A schematic coordinate frame is placed to the right of each matrix for which $\theta$ is not zero, and the associated values of $\lambda_1$, $\lambda_2$ and $\theta$ are also labeled on this schematic. We use the angle between the axis corresponding to $\lambda_1$ and the positive $x$-axis to represent $\theta$. We observe that different rotation matrices lead to varying degrees of ``stretching'' along different directions, which intuitively demonstrates the effects of anisotropy. For instance, when $\mathbf{A}=\begin{pmatrix}
    4 & 0 \\
    0 & 1
\end{pmatrix}$, the eigenvalue $4$ corresponds to the eigenvector $(1, 0)^{\top}$. Therefore, the stretching is more pronounced in the $x$-direction. Conversely, when $\mathbf{A}=\begin{pmatrix}
    1 & 0 \\
    0 & 4
\end{pmatrix}$, the stretching is more obvious in the $y$-direction. 

\begin{figure}[htb]
  \centering
  \includegraphics[width=1.0\linewidth]{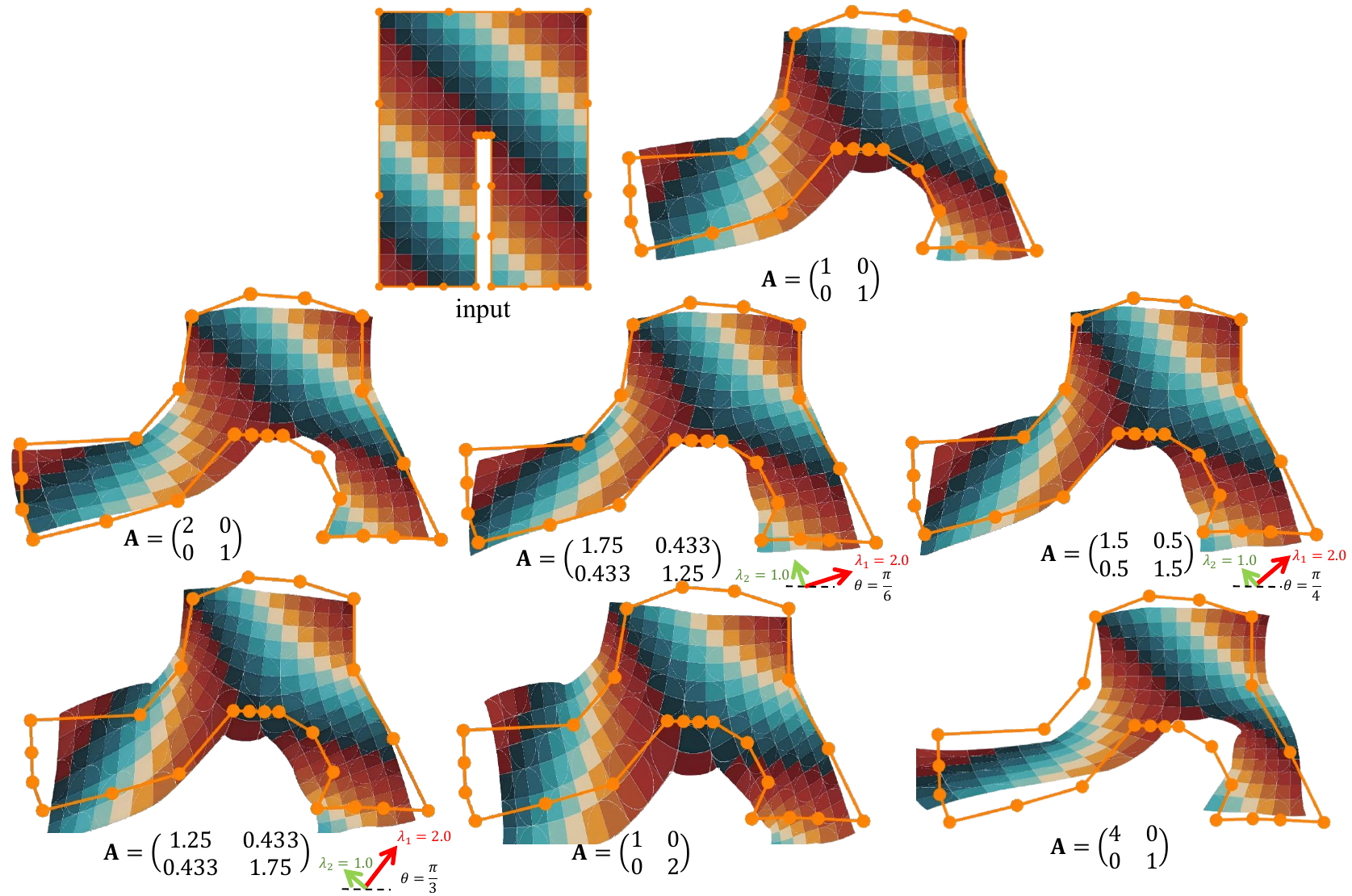}
  \caption{\label{fig:Figure5}
          Cage-based deformation results of an image containing circles and squares. AGC brings diverse deformation effects. }
\end{figure}

\begin{figure}[htb]
  \centering
  \includegraphics[width=1.0\linewidth]{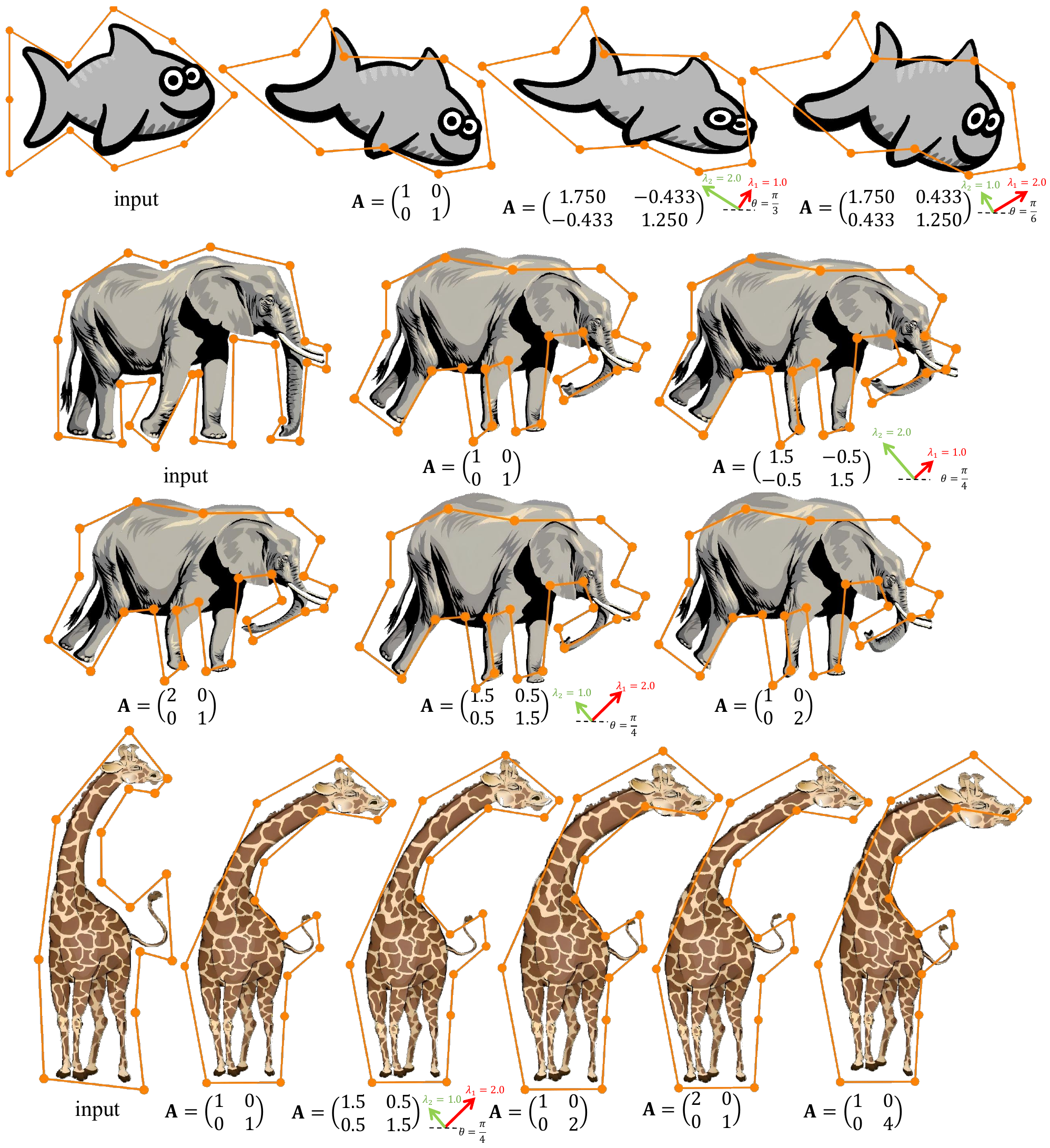}
  \caption{\label{fig:Figure14}
          Cage-based deformation results of a fish, an elephant, and a giraffe model, respectively.}
\end{figure}

Fig.~\ref{fig:Figure5} illustrates another example involving basic shapes such as squares and circles. The eigenvalues of the non-diagonal matrix are typically $1$ and $2$ (except in the last case). While isotropic GC are angle-preserving and map infinitesimal circles to infinitesimal circles, their deformation effect is fixed, which limits its flexibility and diversity. In contrast, our method enables a wide variety of deformation styles. Furthermore, this example also shows that parameter settings more sensitive to the $x$-direction (e.g., the results in the bottom-right corner) will also compress the ``fatness'' of the $y$-direction. Fig.~\ref{fig:Figure14} shows the cage-based deformation results of a fish, an elephant, and a giraffe model, respectively. It can be observed that different matrices control the ``fatness'' or ``thinness'' of the deformed objects, thereby affording artists and animators a more versatile range of deformation choices.

\begin{figure*}[htb]
  \centering
  \includegraphics[width=1.0\linewidth]{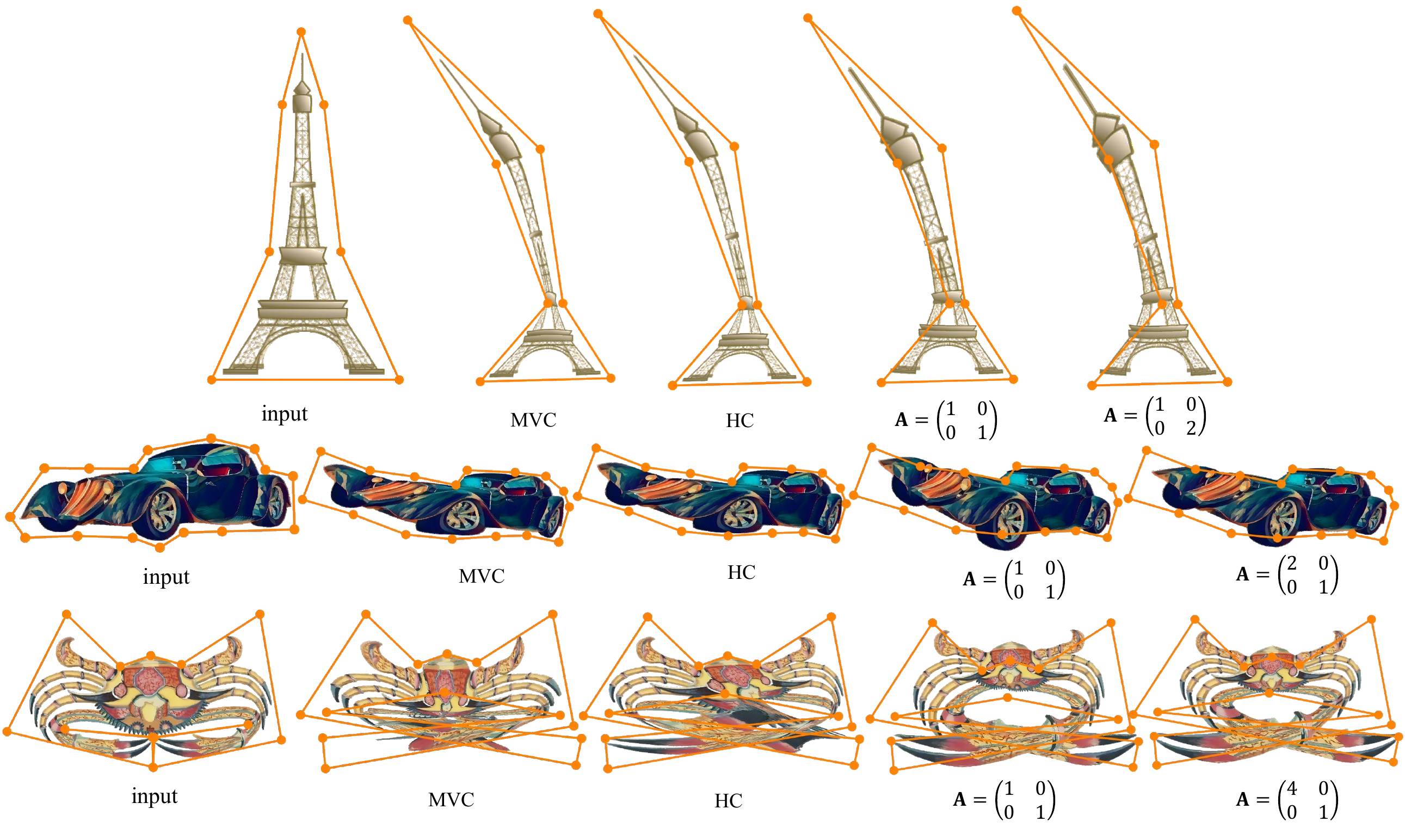}
  \caption{\label{fig:Figure4}
          Comparison of cage-based deformation results using our method, MVC \citep{2003MVCPoly}, and HC \citep{2007Harmonic}.}
\end{figure*}

\begin{figure*}[htb]
  \centering
  \includegraphics[width=1.0\linewidth]{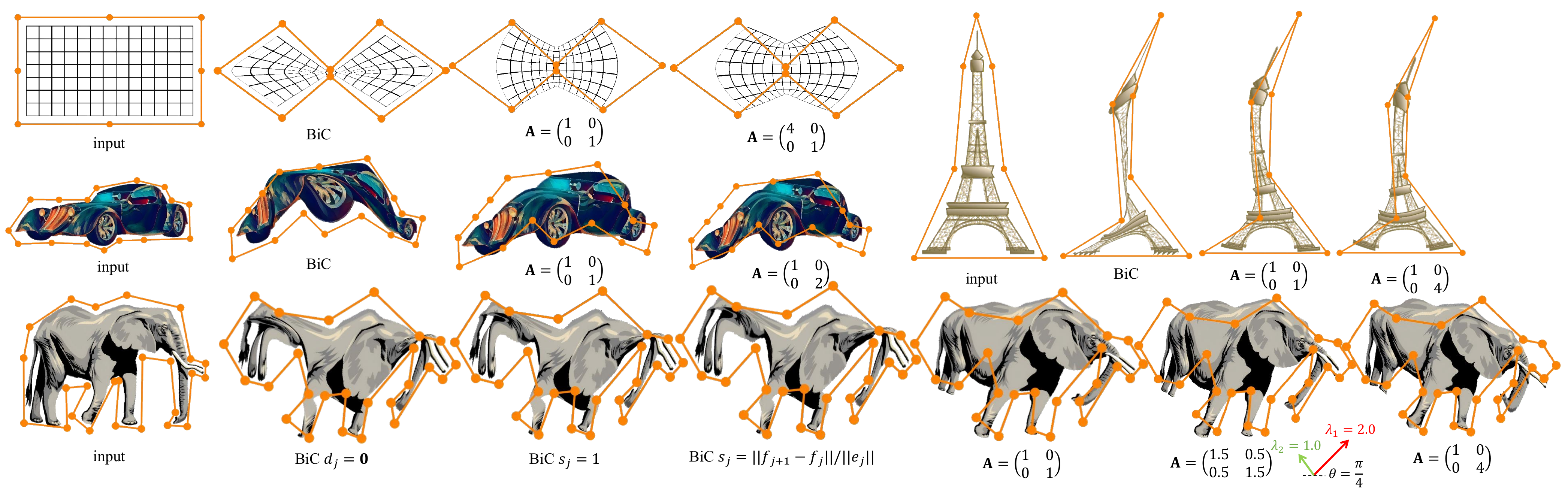}
  \caption{\label{fig:Figure22}
          Comparison of cage-based deformation results using our method and BiC~\citep{2012Biharmonic}}.
\end{figure*}

\begin{figure*}[htb]
  \centering
  \includegraphics[width=1.0\linewidth]{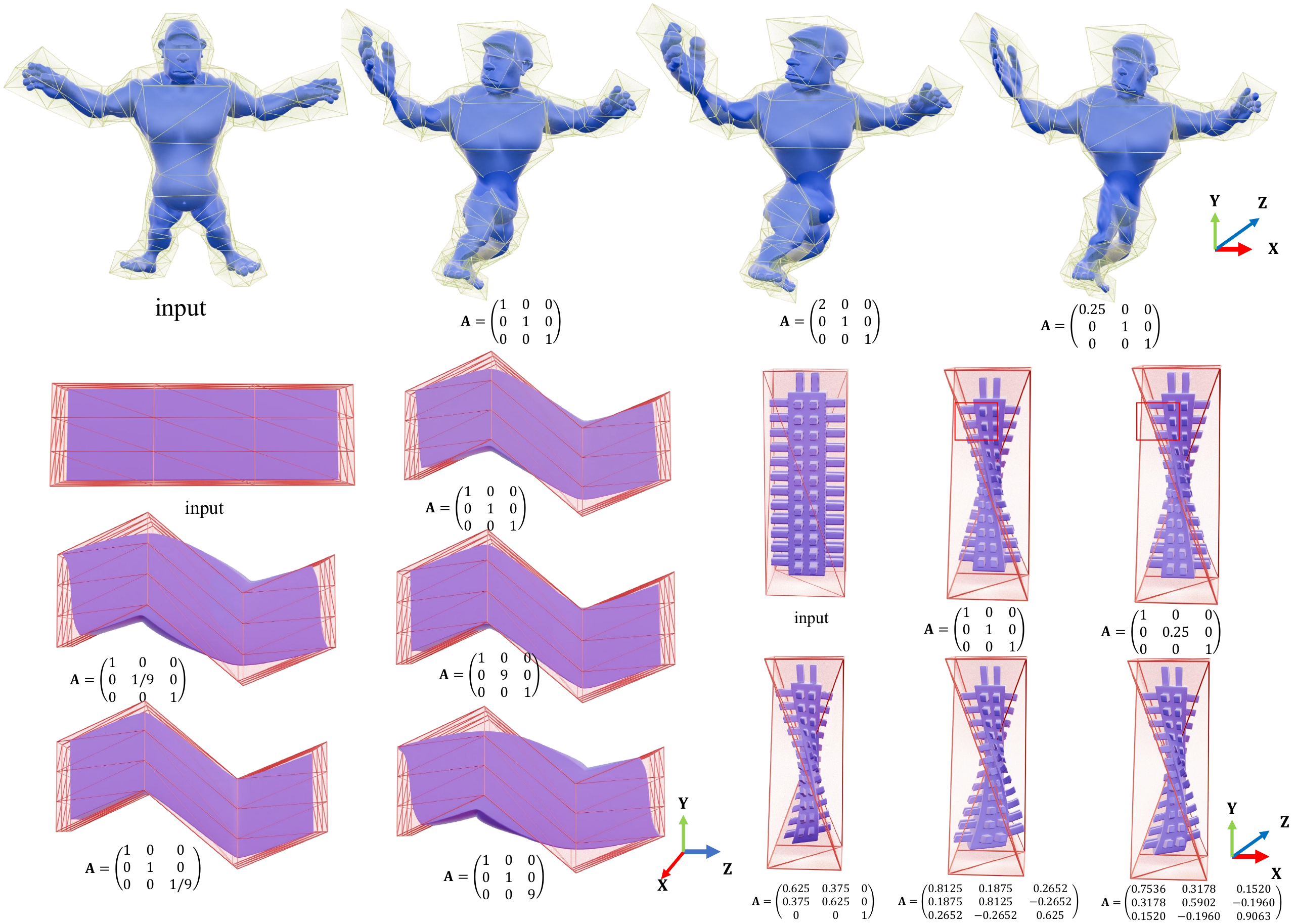}
  \caption{\label{fig:Figure8}
          Cage-based deformation results of a character, a bar and a skybox model in 3D, respectively.} 
\end{figure*}

\begin{figure*}[htb]
  \centering
  \includegraphics[width=1.0\linewidth]{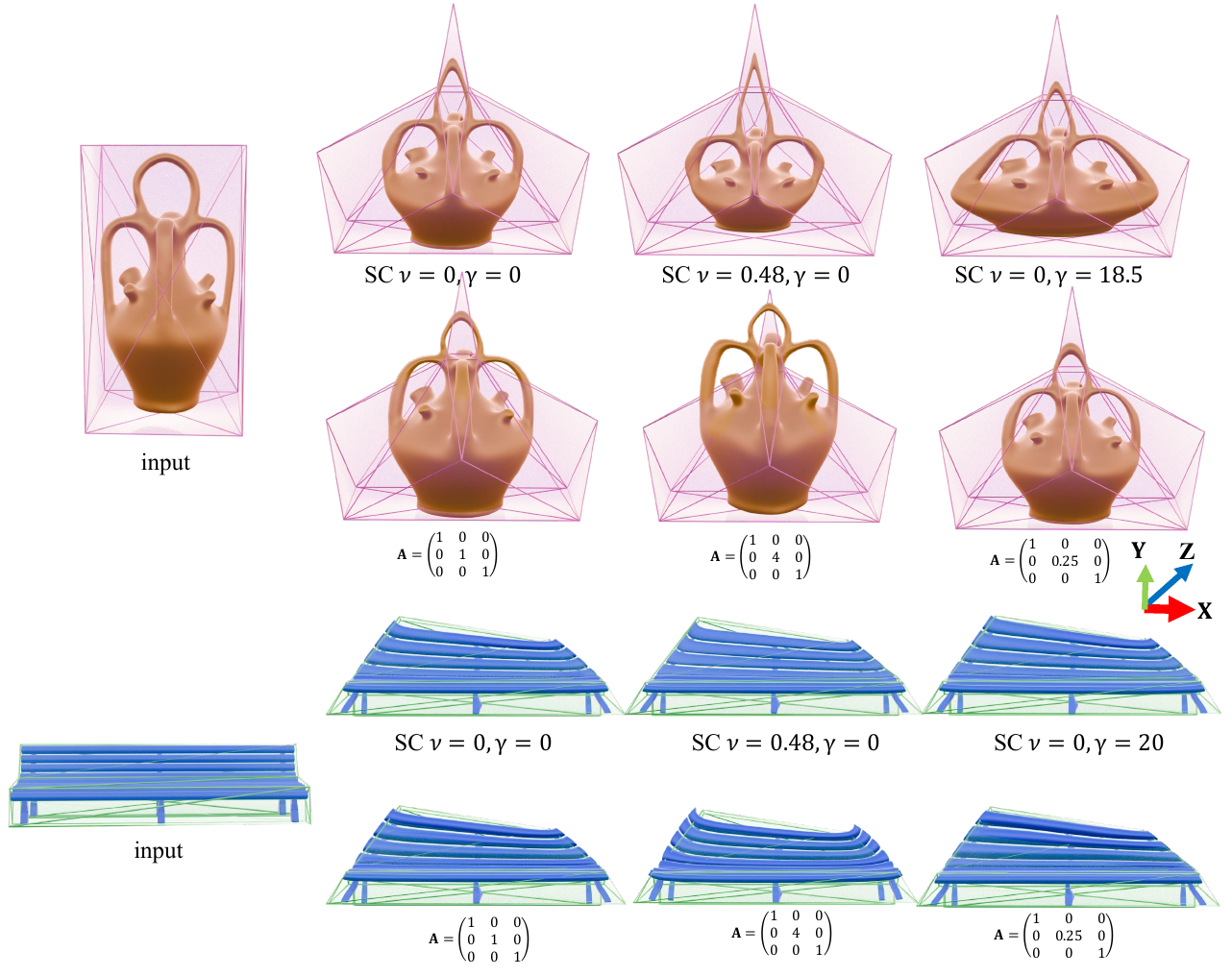}
  \caption{\label{fig:Figure23}
          Comparison of cage-based deformation results using our method and SC~\citep{2023Somigliana}}.
\end{figure*}

In Fig.~\ref{fig:Figure4}, we compare our method with MVC and HC. It is evident that MVC and HC lead to noticeable shearing and unnatural deformations. Our method yields different deformation results compared to the isotropic GC. Meanwhile, the quasi-conformal property of our method also guarantees a certain degree of shape preservation. Furthermore, in the crab model, to ensure the closure of the cage, there are some overlapping edges. The results demonstrate that MVC and HC are less robust than the boundary integral method. 

In Fig.~\ref{fig:Figure22}, we compare our method with BiC. BiC can achieve various deformation effects through different settings of the normal derivative $\mathbf{d}_j$. According to Section 5 of ~\citep{2012Biharmonic}, there are multiple configuration schemes. Specifically, when $\mathbf{d}_j$ is set to $\mathbf{0}$, it becomes a class of GBC. Alternatively, $\mathbf{d}_j$ can be set to $s_j\tilde{\mathbf{n}}_j$, where $\tilde{\mathbf{n}}_j$ is the deformed cage normal. The scalar $s_j$ can be chosen freely; it can be set to $1.0$, as described in Section 5.1 of~\citep{2012Biharmonic}, or to the ratio of the deformed edge length to the source edge length, as in Section 5.2 of~\citep{2012Biharmonic}. We evaluated all methods on the elephant example. Careful observation reveals slight variations in the shapes produced by different settings. Unlike GC and AGC, which only approximates the target cage shape, BiC can satisfy the Hermite property and conforms to the target cage more faithfully. At the same time, BiC achieves different deformation effects by controlling the normal derivative. However, when the target cage undergoes significant changes compared to the source cage, BiC may generate unnatural deformation results. This observation is further corroborated by other examples.

\begin{figure*}[htb]
  \centering
  \includegraphics[width=1.0\linewidth]{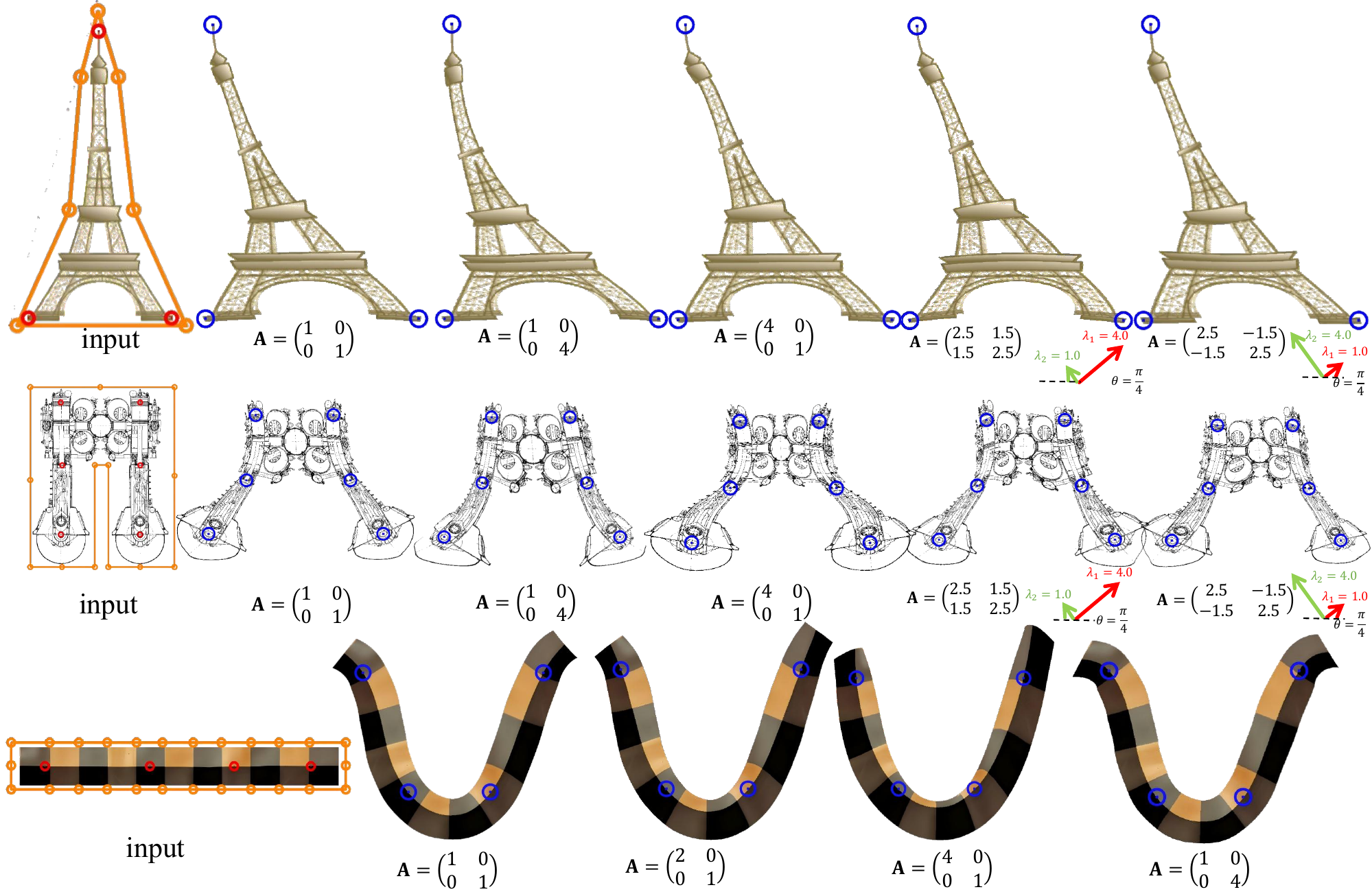}
  \caption{\label{fig:Figure7}
          Results of variational shape deformation in 2D. The red dots indicate the user-specified pre-deformation positions, while the blue dots mark the target positions.}
\end{figure*}

\begin{figure*}[htb]
  \centering
  \includegraphics[width=1.0\linewidth]{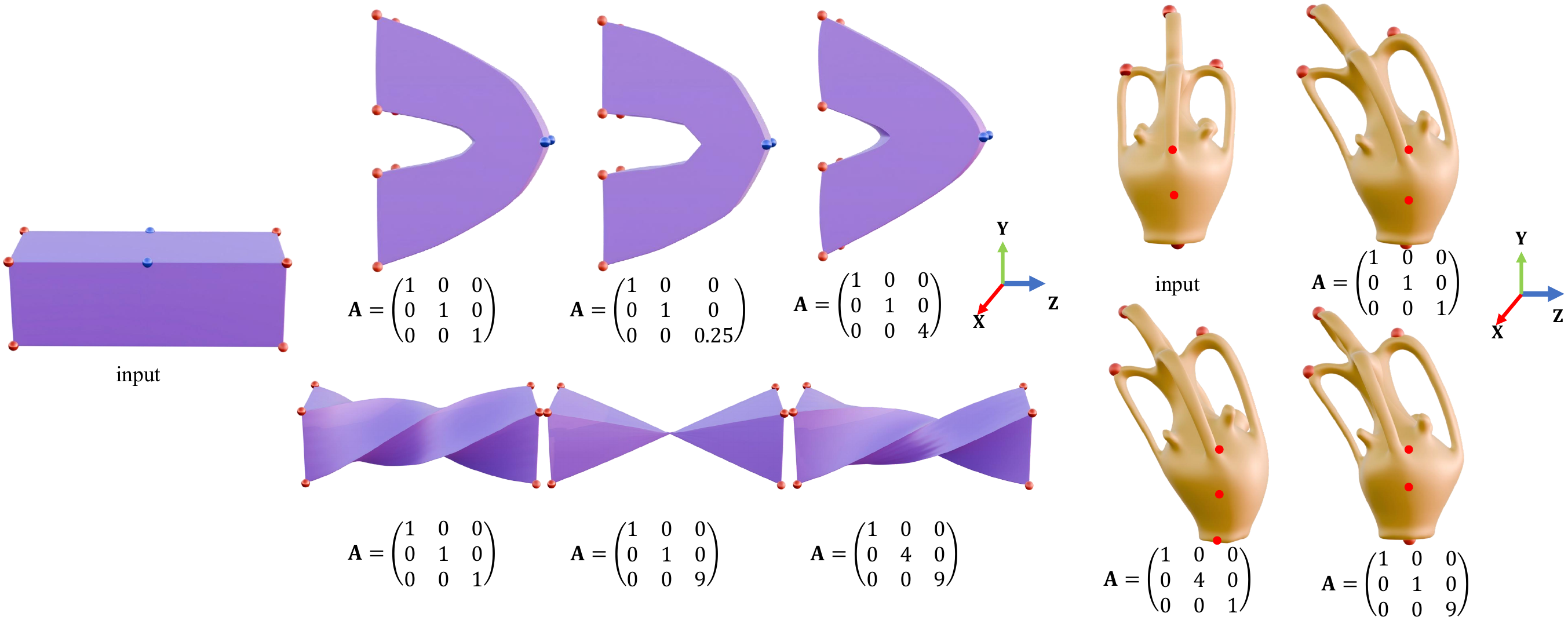}
  \caption{\label{fig:Figure11}
          Results of variational shape deformation in 3D. The dots mark the spatial constraints. Blue dot constraints are not apply when twisting the bar model.}
\end{figure*}

We then present the results of 3D deformation. In Fig.~\ref{fig:Figure8}, we show the results of a character, a bar and a skybox model, respectively. We also mark the spatial coordinate axes in the figure. In the bar model, the deformation is mainly observed along the $y$-axis, while the long side is oriented along the $z$-axis. It can be observed that different matrices produce deformation effects with varying degrees of sharpness. When $\mathbf{A}=\begin{pmatrix}
    1 & 0 & 0\\
    0 & 9 & 0\\
    0 & 0 & 1 \\
\end{pmatrix}$, the two eigenvalues are $\lambda_1=9$ and $\lambda_2=1$, with $\lambda_1=9$ corresponding to the eigenvector $(0,1,0)^{\top}$. Therefore, in the source cage, the face with a normal pointing in the $y$-direction has a greater impact on the deformation, resulting in a sharper deformed shape. When $\mathbf{A}=\begin{pmatrix}
    1 & 0 & 0\\
    0 & 1 & 0\\
    0 & 0 & 1/9 \\
\end{pmatrix}$, the two eigenvalues are $\lambda_1=1$ and $\lambda_2=1/9$, and the eigenvectors corresponding to $\lambda_1=1$ are $(1,0,0)^{\top}$ and $(0,1,0)^{\top}$. Thus, this also results in a sharper deformation in the $y$-direction. In the skybox model, the first row displays results obtained using diagonal matrices, while the second row presents outcomes for non-diagonal matrices. Specifically, from left to right, the matrices are $\mathbf{A}=\mathcal{A}(1, 0.25, 1, \pi/4,0,0)$, $\mathbf{A}=\mathcal{A}(1, 0.25, 1, \pi/4,0,\pi/4)$ and $\mathbf{A}=\mathcal{A}(1, 0.25, 1, \frac{\pi}{3},\frac{\pi}{4},\frac{\pi}{6})$. The character model has complex shapes, further demonstrating the diverse deformation effects achieved by our approach.

In Fig.~\ref{fig:Figure23}, we compare our method with SC~\citep{2023Somigliana}. SC is also a generalization of GC based on linear elasticity and typically controls the shape in two ways: 1. varying the Poisson’s ratio $\nu$ to control local volume, and 2. updating the corotations of the cage faces, which can be achieved via adjusting a bulging parameter $\gamma$. We show two examples in the figure. SC can generate physically plausible 3D deformations that follow the cage trend better. However, for the chair model, even with changes to $\nu$ and $\gamma$ such that the thickness and position of the chair edge undergo certain changes, the edge remains `uplifted' if carefully observed. In contrast, our method with
$\mathbf{A}=\begin{pmatrix}
    1 & 0 & 0\\
    0 & 0.25 & 0\\
    0 & 0 & 1 \\
\end{pmatrix}$ successfully straightens the chair edge. For the vase model, our method can adjust the height of the model while preserving the shape.

\subsection{Variational Shape Deformation}
In this subsection, we conduct experiments on variational shape deformation proposed in Section~\ref{sec:6}. Specifically, we establish the deformation mapping based on AGC as Eq.~\eqref{eq:ani_Green_eq_variational}, and optimize $\mathbf{a}$ and $\mathbf{b}$ by enforcing spatial constraints and ARAP constraints, as formulated in Eq.~\eqref{eq:variational_opt_for}. The spatial constraints dictate the deformation of $\mathbf{q}_i$ to $\mathbf{f}_i$ for $i=1,2,...,r$. The ARAP constraints aim to preserve rigidity during deformation as much as possible for some selected positions. Similar to previous methods, we sample a set of points $\mathbf{m}_i$ for $i=1,2,...,p$ along the medial axis. The Hessian energy term and the regularization term ensure the smoothness of the deformation and the stability of optimization, respectively. Since harmonic functions satisfy the extremum principle at the boundary, this property is inherited by our approach, in the sense that boundary points map to boundary points under linear transformation. Consequently, we sample the boundary points $\mathbf{w}_i (i=1,2,...,k)$ near the cage faces.

Figures~\ref{fig:Figure7} and~\ref{fig:Figure11} illustrate the 2D and 3D results, respectively, which are obtained using the same input and keeping all parameters constant except for the anisotropic matrix. Red dots denote the user-specified pre-deformation positions, while blue dots mark the target positions. The 2D input cages are illustrated in orange lines. The parameters are configured as $\lambda_1=\lambda_2=100$. In the first two examples, we sample $10$ boundary points near each cage face to compute the Hessians, whereas for the third example, we sample $2$ points near each face. The results demonstrate that varying the anisotropic matrices yields different deformation outcomes, thereby expanding the range of achievable deformation effects. In the 3D scenario, we also use dots to mark the points of spatial constraints. Blue dot constraints are not apply when twisting the bar model. Furthermore, we manually annotate the constraints that are occluded during rendering. The initial cages are dense, subdivided rectangular triangle meshes enclosing the input object. We sample two points near each face to compute the Hessians. The parameters are set to $\lambda_1=100$ and $\lambda_2=0.05$ for the bar model, and to $\lambda_1=100$ and $\lambda_2=0.005$ for the vase model. These results further demonstrate the variety of deformation effects achieved by our method.

%% file: sections/sec8.tex
\section{Conclusion}
In this work, we introduce Anisotropic Green Coordinates (AGC), which incorporate a SPD coefficient matrix into the divergence operator of the Laplace equation. This approach enables a variety of deformation styles for both cage-based deformation and variational shape deformation. We derive the Green’s identity based on the anisotropic Laplace equation and demonstrate that our method satisfies both linear reproduction and translation invariance. Furthermore, AGC, along with their gradients and Hessians, admit closed-form expressions. This facilitates the efficient computation of deformation mappings and supports variational shape deformation using ARAP energy, while accommodating user-specified positional constraints.

The primary limitation of our method is that the matrix $\mathbf{A}$  remains globally invariant and is not optimized based on deformation constraints, which we leave to future work. Additionally, GC cannot satisfy both Dirichlet and Neumann boundary conditions simultaneously. We adopt GC as the foundation of our approach because of their theoretical elegance and strong compatibility with the concept of anisotropy. We anticipate that this framework will offer novel insights for anisotropic space deformation and inspire further research in this field.